\documentclass[pre,superscriptaddress,twocolumn,sort&compress,round,
footinbib,amsmath,amssymb,aps]{revtex4-1}

\usepackage{graphicx}
\usepackage{dcolumn}
\usepackage{bm}
\usepackage{hyperref}
\usepackage[mathlines]{lineno}
\usepackage{float}
\usepackage{mathrsfs}
\usepackage{xcolor}
\usepackage{verbatim}

\begin{document}

\preprint{APS/123-QED}

\title{Transmission of a Pressure Signal through a Confined Bubble Array}

\author{Edgar Ortega-Roano}
\author{Devaraj van der Meer}
\affiliation{Department of Applied Physics and J.M. Burgers Centre for Fluid Dynamics, University of Twente, P.O. Box 217,
7500 AE Enschede, The Netherlands}

\date{\today}

\begin{abstract}
Pressure changes travel at an infinite sound speed in an incompressible ideal fluid, opposite to what happens in a bubbly liquid, where the presence of bubbles adds compressibility such that the sound speed becomes finite. Here, the transmission of a pressure signal through a confined bubble array is studied numerically. An axisymmetric Boundary Integral (BI) code was used to simulate a horizontal array of spherical bubbles inside a cylindrical container. On one end of the container, a piston is able to move in either a sinusoidal or impulsive way to excite the bubbles, whereas the other end of the cylinder is fixed. A one-dimensional model, hereafter called the Multiple Bagnold Problem (MBP) model, was developed to predict the behavior of the 3D system. A good agreement between the model and the simulations was found for initial excitation times and qualitatively good agreement for later times. The MBP model was subsequently used to obtain an expression for the sound speed in monodisperse systems, which depends on the ratio between the driving and the natural bubble frequency. Two regimes were found, one where the pressure signal travels unattenuated and the other where attenuation is present. Subsequently we turn to more complex, bidisperse systems, where the influence of the presence of bubbles of two different sizes is explored. Analytical expressions to predict the eigenfrequencies were obtained for monodisperse and bidisperse alternating systems. For bidisperse stacked systems the eigenfrequencies were computed numerically giving a good agreement with the MBP simulations. 
Finally, we discuss the propagation of a pressure signal through a polydisperse system.
\end{abstract}

\maketitle

\section{Introduction}

Pressure changes inside a fluid travel at different speeds depending on the compressibility of the medium. In the linear limit, when the pressure disturbance is not too large, the signal moves at the sound speed of the material, which is defined as the square root of the ratio of a change in pressure and the corresponding change in density inside the medium \cite{kieffer1977sound,kundu1990fluid,pozrikidis2016fluid}. The exact thermodynamical definition can then be expressed as
\begin{equation}
c = \sqrt{ \left( \dfrac{\partial P}{\partial \rho} \right) }_{S}
= \sqrt{\dfrac{K}{\rho}}, \nonumber
\end{equation}
where $c$ is the sound speed, $P$ the pressure, $\rho$ the density of the medium, and $K$ the bulk modulus of the material (reciprocal of the isentropic
compressibility). The partial derivative is taken at a constant entropy $S$ since the process is considered to be adiabatic due to the lack of heat exchange with the surroundings \cite{kieffer1977sound}. If the pressure disturbance is very large as compared to the ambient pressure, a shock wave traveling faster than the sound speed is developed.

From the expression of the sound speed, we see that an incompressible fluid can be defined as one in which the sound speed is infinite or, equivalently, where the density inside the fluid does not vary with changes in pressure. On the other hand, we can define a compressible fluid as the one where the sound speed is finite, or where the density varies with pressure changes.

For many practical purposes, most liquids can be regarded as incompressible. Although the sound speed is not infinite, it is usually much larger than other relevant velocities in a system. In particular it is much larger than that of any gas. As an example, the sound speed in water and air at ambient conditions are around 1500 m/s and 340 m/s respectively. Nonetheless, if air bubbles are present inside water, the sound speed of the mixture may decrease drastically to values as low as 20 m/s, even considerably lower than the sound speed in pure air. The reason is because the mixture now has a small bulk modulus $K$, dominated by that of the gas, and a high density $\rho$, close to the one of the liquid phase \cite{kieffer1977sound}.

Incompressible liquids have been researched since the origin of hydrodynamics and it is now a subject of study covering most of any introductory fluid mechanics book \cite{kundu1990fluid,pozrikidis2016fluid}. Compressible fluids, in particular gases, have also been researched extensively, but scientists started paying more attention to bubbles and bubbly liquids from the first half of the last century \cite{rayleigh1917viii,minnaert1933xvi,plesset1949dynamics,
wijngaarden1972one,plesset1977bubble,prosperetti1986bubble}. The first study of a bubble inside an infinite incompressible liquid can be dated back to 1917 \cite{rayleigh1917viii}, when Lord Rayleigh studied the pressure in a liquid during the collapse of a spherical bubble and derived what is now known as the Rayleigh equation for spherical bubbles. Another seminal article about bubbles is the one of Minnaert called \textit{On musical air-bubbles and the sounds of running water} \cite{minnaert1933xvi}, where he obtained an expression for the natural frequency of a spherical bubble inside an infinite quiescent water pool, assuming that the bubbles follow a polytropic equation of state:
\begin{equation}
\omega_{M} = \sqrt{\dfrac{3 \gamma P_{0}}{\rho_{l} r^{2} } }. \nonumber
\end{equation} 
Here, $\gamma$ is the polytropic exponent, $P_{0}$ and $\rho_{l}$ are the ambient pressure and density of the liquid respectively, and $r$ the bubble radius.

By the second half of the last century, \textcite{plesset1949dynamics} and \textcite{plesset1977bubble} worked on bubble dynamics and cavitation by solving an extended version of the Rayleigh equation, which includes surface tension and viscosity effects, an equation now known as the Rayleigh-Plesset equation. Later on, \textcite{prosperetti1986bubble} studied the dynamics of a bubble in a compressible liquid. For reference, most of the theory of the last century about cavitation and individual bubbles can be found in the books of \textcite{brennen2014cavitation} and \textcite{leighton2012acoustic}.

Research about the collective behaviour of many bubbles can be traced back to the end of the 1980's, when \textcite{smereka1988dynamics} and \textcite{d1989linearized} studied driven bubble clouds using a linearized average flow approach, disregarding the exact interaction between bubbles. More recently, \textcite{ida2004investigation,ida2005avoided} investigated the coupling between two, and later three, acoustically interacting spherical bubbles in a liquid pool. The coupling occurs due to the secondary Bjerknes forces, appearing in each of the Rayleigh-Plesset equations that describe the radial evolution of the bubbles. In the present day, these non-linear coupled equations can be integrated numerically even for a large number of bubbles. \textcite{zeravcic2011collective} solved this way a linearized version of the coupled Rayleigh-Plesset equation for systems of $\sim 1,000$ and up to $\sim 10,000$ bubbles. The bubbles were driven with an acoustic field in order to study the effects of viscous damping, distance between bubbles, polydispersity, geometric disorder, size of the bubbles and size of the cloud.

Bubbly liquids are of particular interest to the cryogenic industry since cryogenics have low boiling points and (vapour) bubbles will inevitably appear during their transportation. Furthermore, cryogenic liquids, like Liquid Natural Gas (LNG), are shipped across the world in huge containers, where other processes like sloshing or bubble entrapment are present. Therefore, it is important to analyze the effect that bubbly liquids have on the container, when they slosh back and forth colliding with the walls. The effect of gas pockets during the impact has been a problem of interest since the first half of the last century. In fact, in 1939, Bagnold carried out experiments to explore the impact of pressure shocks on vertical sea-walls due to breaking waves \cite{bagnold1939interim}. He described the phenomenon using a simplified model of a water slab colliding a vertical wall with a gas pocket in between. He found that the presence of air bubbles, trapped between the wall and the wave, cushions the pressure exerted on the wall, being greater the buffering effect the bigger the bubble. More recently, \textcite{braeunig2010effect} used a Bagnold-like model to study the effect of phase transition on the impact pressures. Subsequently, \textcite{ancellin2012influence}, followed this idea and used a generalized Bagnold model to study phase transition when a vapour pocket is trapped between a liquid pocket and a moving piston.

In the context of solid impact onto a liquid with bubbles, one important question that surfaces is how the liquid responds to pressure and sound signals that are transmitted to the bubbly liquid. How do these signals travel to the liquid? Are they attenuated or transmitted? And how does the response depend on the size, or more specifically the size distribution, of the bubbles? And finally, how does the load experienced by the solid depend on the characteristics of the bubbly liquid? The answer to the last question clearly depends crucially on the answers to the previous ones, and doesn't seem likely addressed in a single article. Therefore, we decided to take a first step by studying a highly simplified system, namely the transmission of a pressure signal in a one-dimensional, confined array of bubbles.

More specifically, in this article we will study the effect of bubbles trapped in an incompressible liquid from a different perspective, namely by looking at the resonant behavior of a bubble column array when driven by a pressure signal. For this, we make use of an in-house built axisymmetric Boundary Integral (BI) code that has proven to give satisfactory results simulating bubbles inside an incompressible fluid. A similar code was first used and described by \textcite{oguz1993dynamics} when doing research on the dynamics of bubble growth and detachment from a needle. The in-house code was used to simulate the impact of a solid disk on a water pool \cite{bergmann2009controlled,gekle2009approach,gekle2011compressible} as well as impacting, oscillating and levitated drops \cite{bouwhuis2012maximal,bouwhuis2013oscillating,hendrix2015universal}. Furthermore, we will develop a model, called the Multiple Bagnold Problem (MBP) model, that predicts satisfactorily the resonant behavior of the bubbles and from which we subsequently generalize to more complex systems, containing many bubbles of different sizes.

In Section \ref{sec:1_2} we describe the three-dimensional axisymmetric BI setup and derive the one-dimensional MBP model. For the latter, we will derive the equations of motion for the bubbles, or rather for the liquid slabs that separate them, and arrive to an expression for the bubble natural frequency and sound speed in this 1D model. In Section \ref{sec:1_3} we compare the results of the MBP model and those of the BI simulations and find an overall good agreement between them. In Section \ref{sec:1_4} we will study distinct types of systems using the MBP model. We explore the effect of different driving frequencies, bubble number, bidispersity, polydispersity and impulsive driving on the resonant response of the bubbles, using inspiration from \textcite{van2007microbubble} and \textcite{versluis2010microbubble} to characterize the various systems by their power spectra and response. We conclude in Section \ref{sec:1_5} by giving an overview of the most important results of each section.

\section{Multiple Bagnold Problem model}\label{sec:1_2}

Here we briefly describe the basics of the Boundary Integral (BI) code and the characteristics of the three-dimensional axisymmetric setup. Then, we will develop a one-dimensional model that successfully predicts the behavior of a pressure signal traveling through an array of bubbles. We call this the Multiple Bagnold Problem (MBP) model. 

For potential flow, BI codes work by solving the Laplace equation for the scalar velocity potential and its normal derivative along the fluid boundaries, then updating the potential in time using the time-dependent Bernoulli equation. This has the advantage that the problem is reduced by one dimension, thereby making the simulations faster than when using other numerical methods, such as Volume of Fluid for instance. In particular, the BI code used in this work is an axisymmetrical one, which makes simulations even faster than a fully three-dimensional BI code. However, when the system increases to a considerable size, bigger than ten bubbles for example, BI simulations also tend to become slow, thus indicating the need to present a model that can predict the behavior of confined bubble arrays while being faster computationally speaking. For more information about how the BI code works and its capabilities, we will refer to Refs. \cite{oguz1993dynamics,
bergmann2009controlled,gekle2009approach,gekle2011compressible,
bouwhuis2012maximal,bouwhuis2013oscillating,hendrix2015universal}. At this point it is good to note that, for the sake of simplicity, we neglect surface tension in this work, although it could in principle be included in both the BI simulations and the MBP model without significant difficulties. 

\begin{figure}
\includegraphics[width=\linewidth]{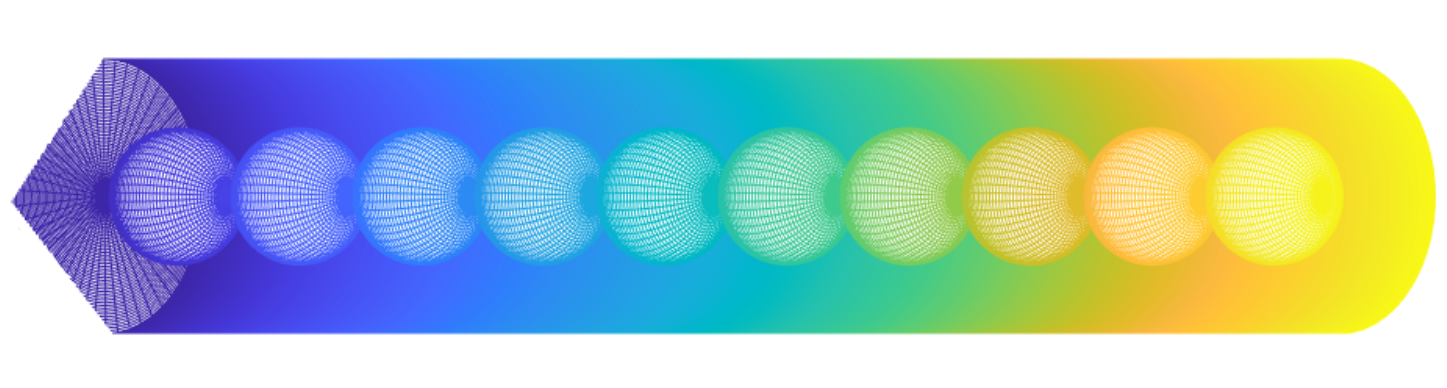}
\centering{a)}
\includegraphics[width=\linewidth]{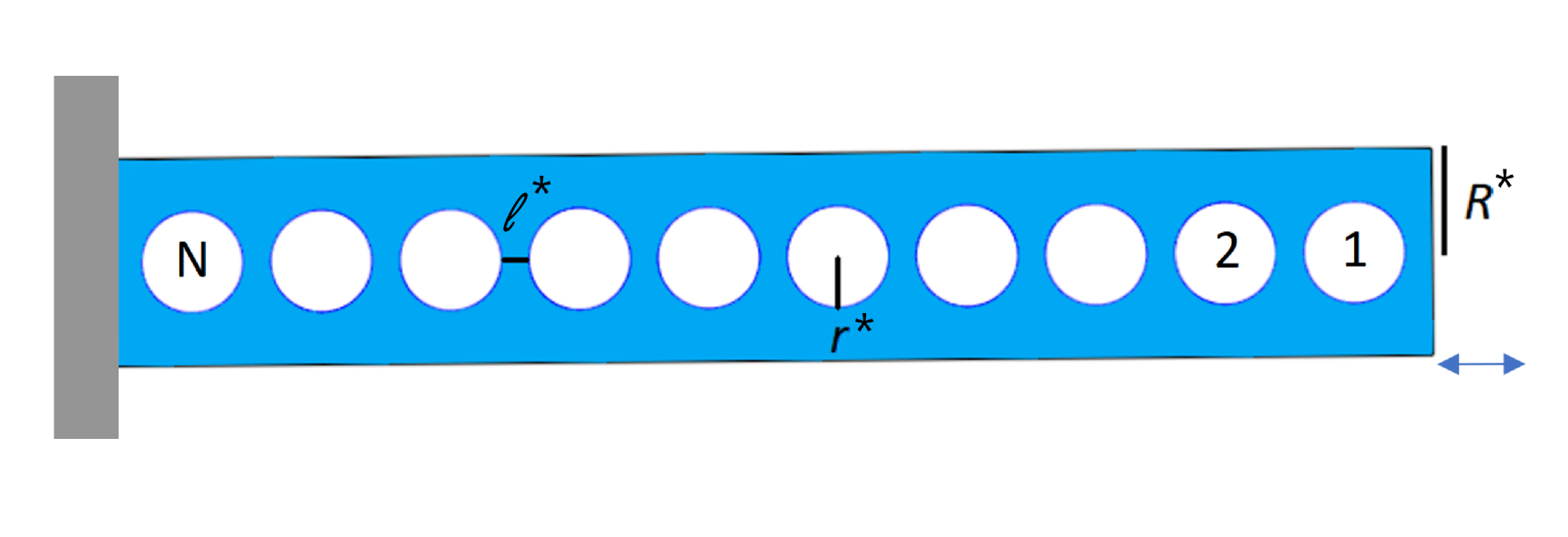}
\centering{b)}
\includegraphics[width=\linewidth]{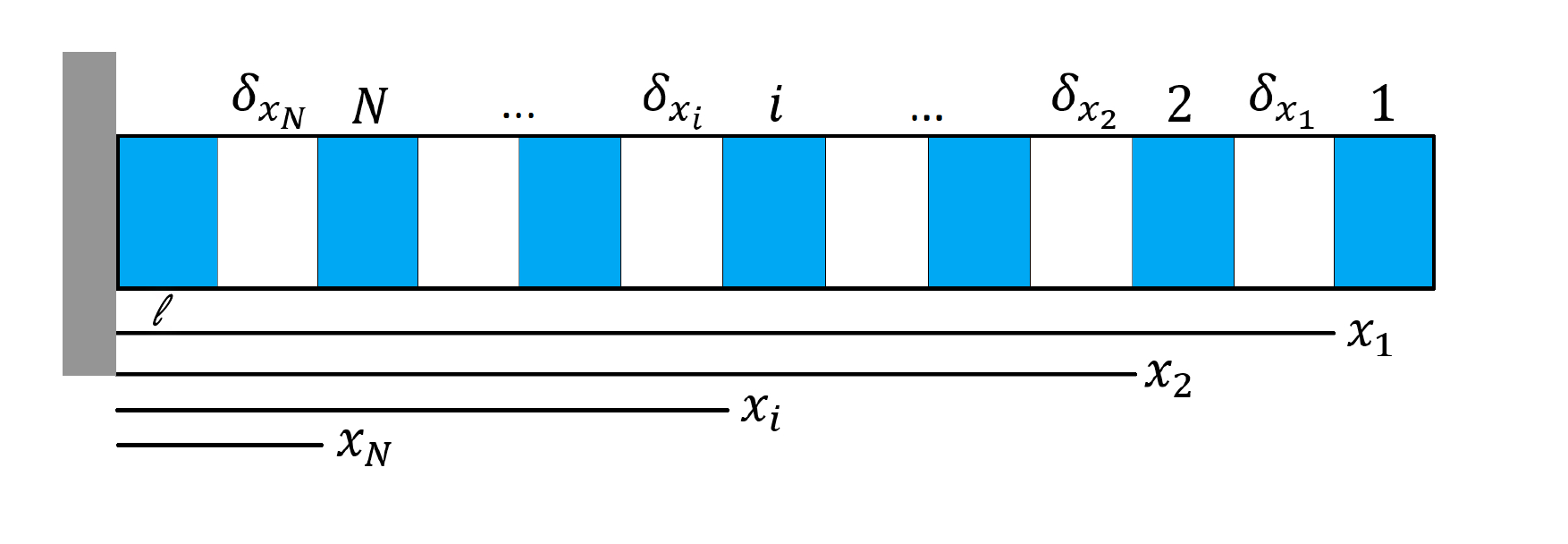}
\centering{c)}
\caption{a) Confined axisymmetric array of bubbles as simulated by the Boundary Integral code. b) Two-dimensional cross section of the axisymmetric setup; $N$ is the number of bubbles, $r^*$ the radius of the bubbles, $R^*$ the radius of the confining cylinder and $\ell^*$ is the minimum distance between bubbles. c) Multiple Bagnold Problem model, the blue color represents liquid patches with width $\ell$, the white color are the bubbles with width $\delta x_{i}$ and $x_{i}$ is the distance from the left wall to the right side of every bubble.}\label{fig:1_fig1}
\end{figure}

To develop the Multiple Bagnold Problem (MBP) model, we start by considering a three-dimensional, axisymmetric bubble array as it is simulated by the BI code, see Fig. \ref{fig:1_fig1} (a). The spherical objects inside the cylinder represent the bubbles arranged in a horizontal manner. Gravity is neglected in order to avoid buoyancy effects. Fig. \ref{fig:1_fig1} (b) shows a horizontal cut of the 3D representation, where $R^*$ is the radius of the cylinder. The right wall is connected to a \textit{piston} that is able to move left and right in a prescribed way, as represented by the double-headed arrow. The left wall, represented by the grey bar, is fixed. We label the bubble closest to the right wall as number one, to its left is bubble number two and so on until we fill the cylinder with $N$ bubbles. Each bubble has a radius $r^*$ and we define $\ell^*$ as the minimum distance between consecutive bubbles. The blue area represents an incompressible ideal fluid surrounding the bubbles. In particular, Figs. \ref{fig:1_fig1} (a) and \ref{fig:1_fig1} (b) show ten bubbles, but the simulations and model can at least in principle be extended to any number.

In order to simplify the setup shown in Fig. \ref{fig:1_fig1} (b), we present the pseudo one-dimensional model shown in Fig \ref{fig:1_fig1} (c). The bubbles are no longer spherical but now are represented in white as rectangular gas patches with width $\delta x_{i}$ where $i$ is the bubble number. The system now has an arbitrary cross-sectional dimension $R^*$ and all motion is confined horizontally.  We show in blue $N$ liquid patches with constant width $\ell$. Since the leftmost liquid patch is fixed to the left wall, it will not appear in the equations of motion and therefore does not have a number. The origin of our reference frame is fixed at the left wall. From it, we quantify the motion of the liquid patches by using the coordinates $x_{i}$, that go from the origin to the left side of each liquid patch.  

Given the aforementioned definitions, we relate $\delta x_{i}$ to $x_{i}$ as
\begin{equation}
\begin{array}{ll}
    \delta x_{1} &= x_{1} - ( x_{2} + \ell ) , \\
    &\vdots \\ 
    \delta x_{i} &= x_{i} - ( x_{i+1} + \ell ) , \\
    &\vdots \\
    \delta x_{N} &= x_{N} - \ell .
\end{array}
\label{eq:1_1}
\end{equation}
The pressure $P_{i}$ inside each bubble follows a polytropic equation of state
\begin{equation}
P_{i}(t) = P_{0} \left(  \dfrac{\delta x_{i}(t)}{\delta x_{i}(0)}  \right)^{-\gamma} ,
\label{eq:1_2}
\end{equation}
where $\gamma$ is the polytropic exponent and $P_{0}$ the pressure at time $t=0$, when the system is at rest (i.e., the ambient pressure). For the BI simulations, the bubbles follow a similar equation of state dependent on their volumes $V(t)$ instead of $\delta x_{i}(t)$.

Using Eq.~\eqref{eq:1_2}, we are able to write an equation of motion (Newton's second law) for the liquid patches, by balancing the pressure that a patch $i$ feels on each side: 
\begin{equation}
\begin{array}{ll}
\rho_{l} \ell \ddot{x}_{1} &=
P_{0} \left[ 
\left( \dfrac{\delta x_{1}(t)}{\delta x_{1}(0)} \right)^{-\gamma} - 1
\right] , \\
&\vdots \\ 
\rho_{l} \ell \ddot{x}_{i} &=
P_{0} \left[ 
\left( \dfrac{\delta x_{i}(t)}{\delta x_{i}(0)} \right)^{-\gamma} - 
\left( \dfrac{\delta x_{i-1}(t)}{\delta x_{i-1}(0)} \right)^{-\gamma}
\right] , \\
&\vdots \\ 
\rho_{l} \ell \ddot{x}_{N} &=
P_{0} \left[ 
\left( \dfrac{\delta x_{N}(t)}{\delta x_{N}(0)} \right)^{-\gamma} - 
\left( \dfrac{\delta x_{N-1}(t)}{\delta x_{N-1}(0)} \right)^{-\gamma}
\right] .
\end{array}
\label{eq:1_3}
\end{equation}
On the left hand side of Eqs. \ref{eq:1_3}, we see the liquid density $\rho_{l}$, the liquid width $\ell$, together with the second time derivative of the horizontal coordinate $\ddot{x}_{i}$. The pressure balance is on the right hand side of each equation. Note that the equation for $x_{1}$ assumes that the first liquid patch is open to the atmosphere and therefore an atmospheric pressure $P_{0}$ is pushing towards the left. In order to model a \textit{piston} driving the system in a sinusoidal way, one needs to impose the motion of $x_1$ as 
\begin{equation}
x_{1}(t) = x_1(0) - A \sin (\omega_{D} t ) ,\label{eq:1_4B}
\end{equation}
which implies that the first equation of motion in~\eqref{eq:1_3} must be modified to 
\begin{equation}
\ddot{x}_{1} = A \omega_{D}^{2} \sin (\omega_{D} t ) ,\label{eq:1_4}
\end{equation}
where $A$ is the amplitude and $\omega_{D}$ the angular frequency of the driving.

Considering this, Eqs. \ref{eq:1_3} can be solved numerically. However, to arrive at an expression for the natural frequency of the one-dimensional bubbles and the sound speed inside the system, it is necessary to look at the linearized version of Eqs. \ref{eq:1_3}. For this, let us assume for the moment that each liquid patch $i$ moves from its equilibrium position $x_{i}(0)$ a small perturbation $\epsilon_{i}(t)$, such that $\epsilon_{i}(t) \ll x_{i}(0)$. Then we can write the position of the liquid patches in time as function of the perturbations as
\begin{equation}
x_{i}(t) = x_{i}(0) + \epsilon_{i}(t) .\label{eq:1_5}
\end{equation}
Using Eqs. \ref{eq:1_1}, we rewrite the bubble widths in terms of $\epsilon_{i}(t)$
\begin{equation}
\begin{array}{ll}
\delta x_{i} &= \delta x_{i}(0) + \epsilon_{i} - \epsilon_{i+1} , \qquad \mathrm{for} \; i = 1, ..., N-1 \\ \\
\delta x_{N} &= \delta x_{N}(0) + \epsilon_{N} .
\end{array}\label{eq:1_6}
\end{equation}
Let us assume, for the time being, that all bubbles are of the same size. Then $\delta x_{i}(0) = \delta x_{0}$. After substituting Eqs. \ref{eq:1_6} into \ref{eq:1_3} and linearizing to first order in $\epsilon_{i}(t)$, we arrive at
\begin{equation}
\begin{array}{ll}
\ddot{\epsilon}_{2} + \dfrac{1}{2} \omega_{0}^{2} \epsilon_{2} &= \tfrac{1}{2} \omega_{0}^{2} \left(\epsilon_{1} + \epsilon_{3}\right), \\
&\vdots \\ 
\ddot{\epsilon}_{i} + \omega_{0}^{2} \epsilon_{i} &=
\tfrac{1}{2} \omega_{0}^{2} \left( \epsilon_{i-1} + \epsilon_{i+1}
\right) , \\
&\vdots \\ 
\ddot{\epsilon}_{N} + \omega_{0}^{2} \epsilon_{N} &=
\tfrac{1}{2} \omega_{0}^{2} \epsilon_{N-1} ,
\end{array}\label{eq:1_7}
\end{equation}
where $\epsilon_{1}$ is provided by Eq.~\eqref{eq:1_4B}, i.e., $\epsilon_{1} = -A\sin (\omega_{D} t )$, and where we have defined the natural frequency of the bubbles as
\begin{equation}
\omega_{0} = \sqrt{\dfrac{2 \gamma P_{0}}{\rho_{l} \ell \delta x_{0} } } .\label{eq:1_8}
\end{equation}
Equation \eqref{eq:1_8} is the one-dimensional equivalent of Minnaert's frequency for spherical bubbles. The same derivation can be done if the bubbles are of different sizes, then the natural frequency of the bubbles will depend on their corresponding initial size $\delta x_{i}(0)$. As a remark, Eqs. \ref{eq:1_7} have the same form as other classical problems such as the one of coupled harmonic oscillators in series \cite{florencio1985exact} or a one-dimensional granular chain \cite{taghizadeh2021stochastic}. Note however, that the natural frequency of the system changes depending on the physical parameters of the problem at hand.

We propose a solution to the linearized MBP equations \ref{eq:1_7} of the form
\begin{equation*}
\epsilon_k(t) = A_k e^{i(\omega_D t + \phi_k)}\qquad\textrm{for}\,\,k \in \mathbb{Z}\,,
\end{equation*}
where $A_{k}$ and $\phi_k$ are the amplitude and phase of the $k$-th position. After substitution and using translational invariance, it is possible (see Appendix \ref{1_A_1}) to derive expressions for the ratio between amplitudes $\alpha \equiv A_{k+1}/A_{k}$ and  phase shift $\Delta \phi \equiv \phi_{k+1} - \phi_k$ given by
\begin{equation}
\alpha =
    \begin{cases}
      1 & \text{if $\tilde{\omega} \leq \sqrt{2}$}\\
      \tilde{\omega}^2\left(1 - \sqrt{1-2\tilde{\omega}^{-2}}\right) -1 & \text{if $\tilde{\omega} \geq \sqrt{2}$}
    \end{cases}\qquad 
    \label{eq:1_9}
\end{equation}
and
\begin{equation}
\Delta\phi =
    \begin{cases}
      \arccos(1-\tilde{\omega}^2) & \text{if $\tilde{\omega} \leq \sqrt{2}$}\\
      \pi & \text{if $\tilde{\omega} \geq \sqrt{2}$}
    \end{cases}\qquad .
    \label{eq:1_10}
\end{equation}

Both expressions depend on the dimensionless angular frequency ratio $\tilde{\omega} \equiv \omega_{D}/\omega_{0}$. It is interesting to see that there are two regimes depending on the value of $\tilde{\omega}$. On the one hand, when $\tilde{\omega} \leq \sqrt{2}$, all oscillation amplitudes are equal, therefore the pressure signal is transmitted without attenuation but with a time-dependent phase. On the other hand, when $\tilde{\omega} \geq \sqrt{2}$, the phase shift becomes constant but there is a strong attenuation. As $\tilde{\omega} \rightarrow \infty$ the ratio between amplitudes tends to zero so no signal is transmitted, in agreement with the behavior of a driven harmonic oscillator at 
very high driving frequencies (i.e., well above the natural frequency).

Since a phase shift $\Delta\phi$ corresponds to a time shift $\Delta t = \Delta\phi/\omega_D$ occurring at a distance $\Delta x = \ell + \delta x_0$, we obtain the wave speed as
\begin{equation}
c = \frac{\Delta x}{\Delta t} = \omega_0\,(\ell + \delta x_0)\frac{\tilde{\omega}}{\Delta\phi}\,. \label{eq:1_11}
\end{equation}
Then if we define the typical sound speed $c_0$ as
\begin{equation}
c_0 \equiv \omega_0\,(\ell + \delta x_0) = \sqrt{\frac{2\gamma P_0}{\rho_l\nu(1-\nu)}}\,, \label{eq:1_12}
\end{equation}
with $\nu = \delta x_0/(\ell + \delta x_0)$ the gas fraction in the system. We can use Eq.~\eqref{eq:1_10} to write the dimensionless frequency-dependent sound speed for both regimes
\begin{equation}
\tilde{c} =
    \begin{cases}
      \frac{\tilde{\omega}}{\arccos(1-\tilde{\omega}^2)} & \text{if $\tilde{\omega} \leq \sqrt{2}$}\\
      \frac{\tilde{\omega}}{\pi} & \text{if $\tilde{\omega} \geq \sqrt{2}$}
    \end{cases}\qquad , 
    \label{eq:1_13}
\end{equation}
where we have defined $\tilde{c} \equiv c/c_{0}$.

\section{Validating the 1D Model with 3D Simulations}\label{sec:1_3}

\begin{figure*}
\centering
\includegraphics[width=0.49\textwidth]{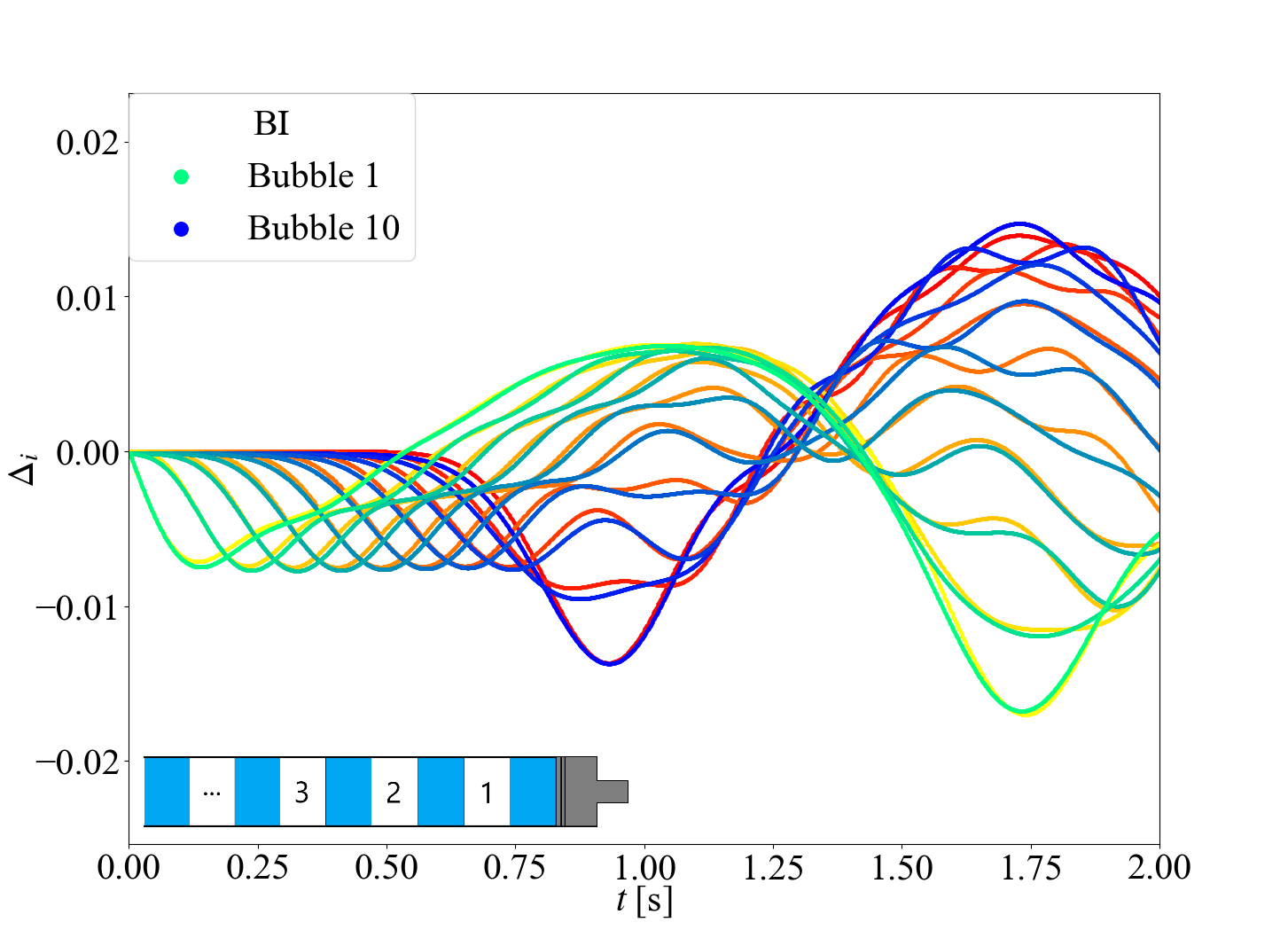}
\includegraphics[width=0.49\textwidth]{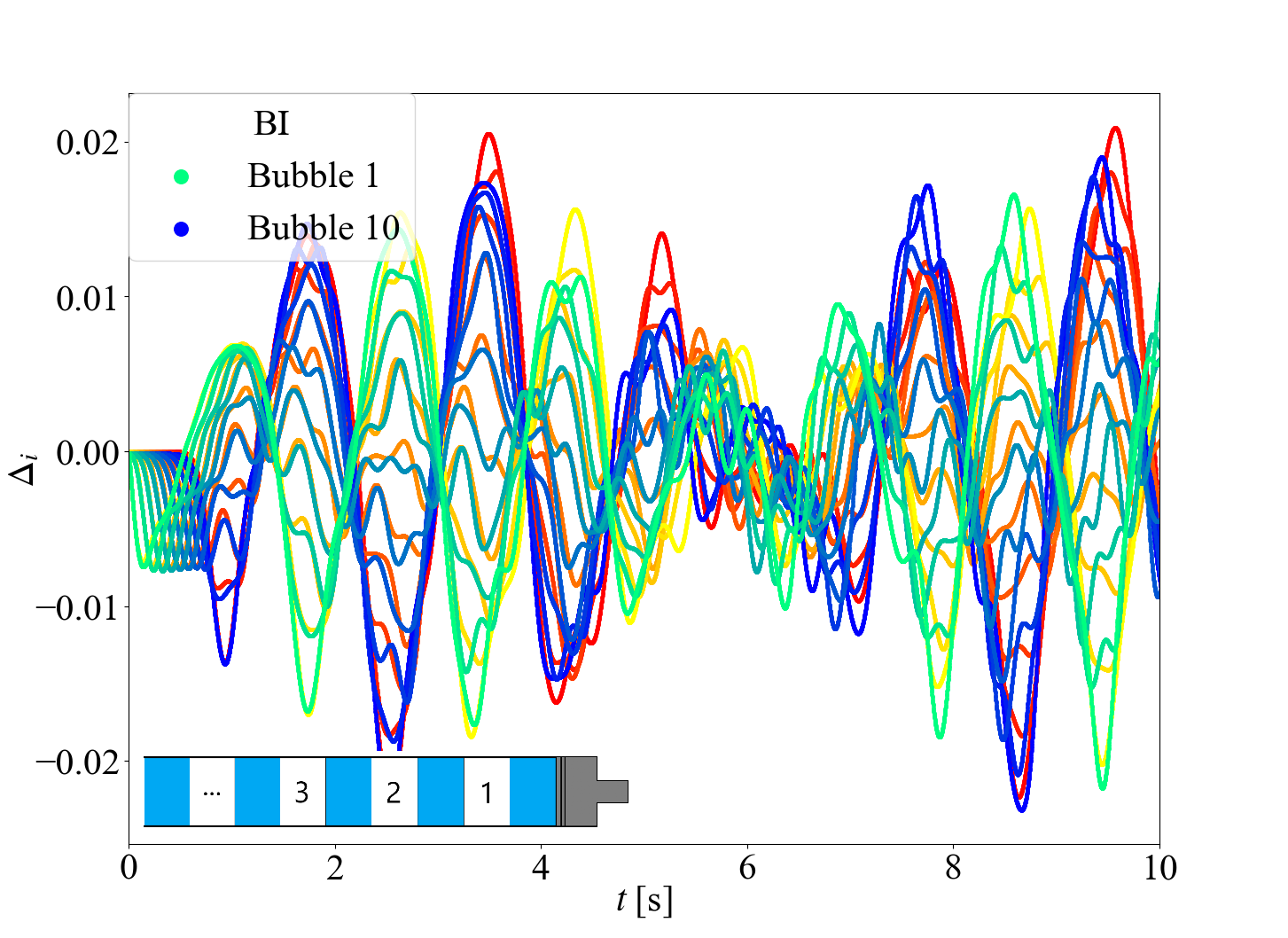}
\makebox[0.49\textwidth]{a)}
\makebox[0.49\textwidth]{b)}
\caption{Signal of the relative bubble volume change of a monodisperse 10 bubble system. In green-bluish colors are the results from the BI  simulations. In the background, in yellow-redish colors, the results from the MBP model. In order to acquire a good fit, the sound speed and signal amplitude from the MBP results had to be matched with those of the BI simulations. The driving frequency is $f_{D}=0.5$ Hz. Figure a) shows a good agreement between model and simulations for initial times. Figure b) shows a qualitative agreement between the two results for longer times.}\label{fig:1_fig2}
\end{figure*}

We simulated a monodisperse system with 10 bubbles using the BI code. The radius of the cylinder was set (rather arbitrarily) to $R^{*} = 2$ m, the radius of the bubbles is $r^{*} = 0.5$ m and the separation between bubbles $\ell^{*} = 0.5$ m. The piston driving the system moves at a frequency $f_{D} = 0.5$ Hz with amplitude $A^{*} = 0.01$ m. For all the BI and MBP simulations described in the article, the polytropic exponent was set to $\gamma = 1$, which corresponds to isothermal bubbles. Simulations with $\gamma = 1.4$, corresponding to air bubbles undergoing an adiabatic process, were also carried out and lead to similar results.

Here it is good to note that for definiteness we use dimensional units in spite of the fact that (of course) our numerical codes use dimensionless quantities. Since we restrict ourselves to atmospheric pressures ($P_0 = 101$ kPa) and use water as the working liquid ($\rho = 1.01\cdot10^3$ kg/m$^3$), the only relevant scales are time and length, which are related by the one-dimensional Minnaert frequency Eq.~\eqref{eq:1_8}. That is, rescaling the bubble size and distance down by a factor $1,000$ to the millimeter range implies scaling up the frequency by a factor $1,000$ into the kHz range.

Figure \ref{fig:1_fig2} shows the relative bubble volume change $\Delta$ defined as
\begin{equation}
\Delta_{i} = \dfrac{\delta x_{i}(t)-\delta x_{i}(0)}
{\delta x_{i}(0) } = \dfrac{\delta V_{i}(t)-\delta V_{i}(0)}
{\delta V_{i}(0) } , \label{eq:1_14}
\end{equation}
where $i$ denotes the individual bubbles. 
In green-blue colors are the results from the BI simulations. In the background, in yellow-red colors, are the results from the MBP model. Figure \ref{fig:1_fig2} (a) compares the results for initial driving times. The agreement between simulations and model is best between 0 and 0.75 s, when the bubbles first deform and the system behaves almost linearly. Figure \ref{fig:1_fig2} (b) compares both results for longer times. The agreement is still qualitatively good since it captures the major features like the envelope of the signal. 

The driving amplitude and sound speed of the model had to be fitted to the BI simulations to obtain a good agreement. The values that were found to give the best match were $A = 0.0335$ m and $c_0 = 21.9$ m/s respectively. The amplitude can be easily changed, whereas the corresponding sound speed was achieved by setting $\ell = 0.5$ m and $\delta x (0) = 1.16$ m, so that $\alpha \approx 0.7$, in the MBP model.  \footnote{Note that bubbles are made large to minimize the influence of surface tension, which is present in the BI simulations for stability reasons and set to the value of air-water. With the exception of surface tension, all results are scalable.}

From Fig. \ref{fig:1_fig2} it is hard to 
determine the resonant frequencies of the system. To visualize them, we apply a Fast Fourier Transform ($\mathscr{F}\left\lbrace \right\rbrace$) to $\Delta$ for each bubble signal
\begin{equation}
\mathit{fft}_{i} = \mathscr{F} \left\lbrace  \Delta_{i} \right\rbrace
\label{eq:1_15}
\end{equation}
and compute the power spectrum $E$ defined as:
\begin{equation}
E (f,f_{D}) = \dfrac{2}{N N_{t}} \left[ 
\vert \mathit{fft}_{1} \vert^2 + \ldots + \vert \mathit{fft}_{N} \vert^2
\right] , \label{eq:1_16}
\end{equation}
where $N_{t}$ is the number of time steps used for the numerical integration.

\begin{figure}
\includegraphics[width=\columnwidth]{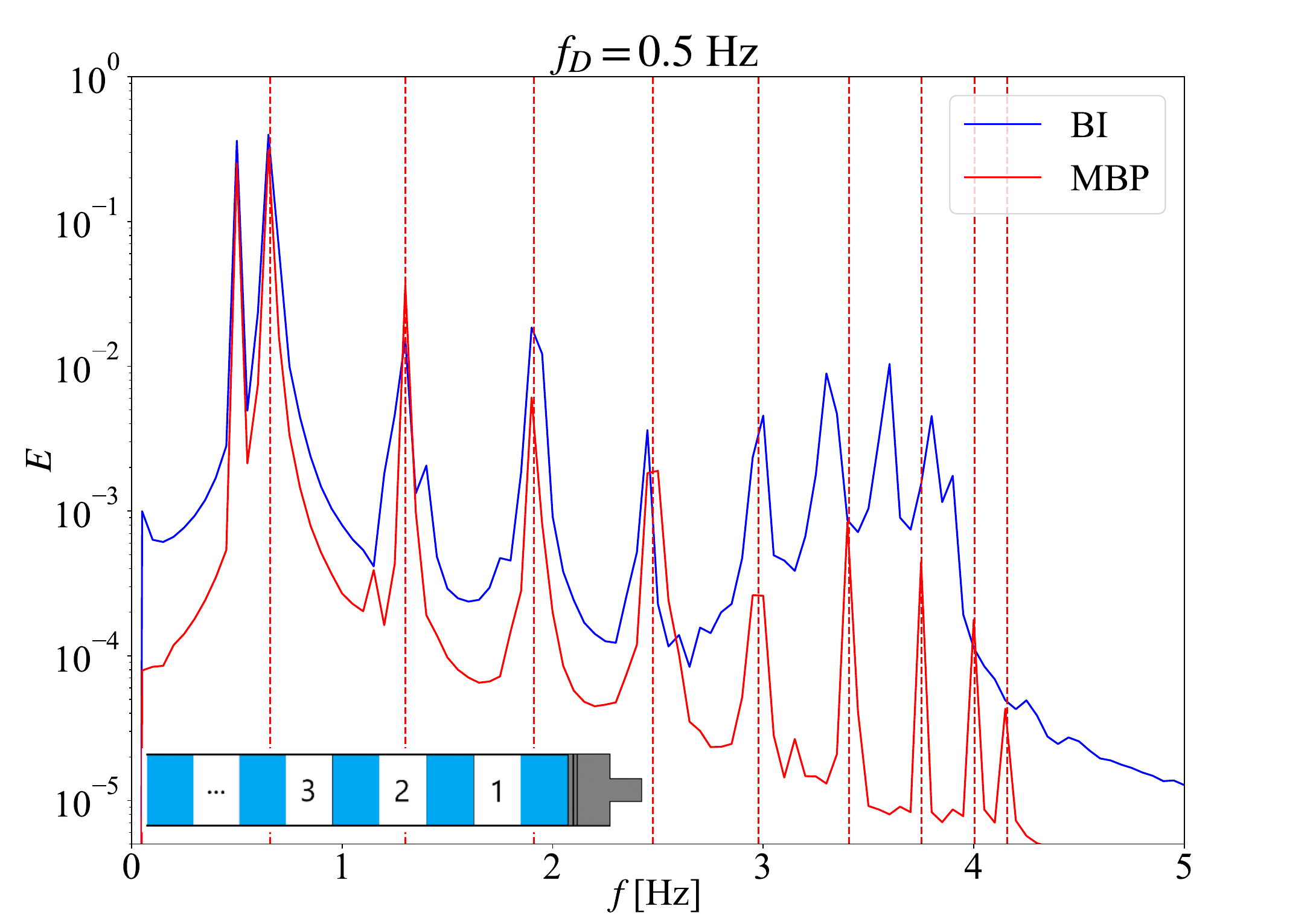}
\caption{Power spectrum of the relative bubble volume signal showing the resonant frequencies of the bubbles. In blue the results from the BI simulations and in red those of the MBP model. The leftmost peak is the driving frequency $f=0.5$ Hz. To its right, the peaks arise at the eigenfrequencies of the system. In red dotted vertical lines are the predictions of the eigenfrequencies obtained from Eq.~\eqref{eq:1_17}. The agreement between the MBP simulations peaks and the prediction is excellent. In blue, the BI simulation peaks agree perfectly up to the fifth eigenfrequency at $f=2.9$ Hz, which also corresponds to the natural frequency of the bubbles.
}\label{fig:1_fig3}
\end{figure}

In Fig. \ref{fig:1_fig3}, we compare the power spectra of the BI simulations (blue curve) and MBP model (red curve). The first peak appearing at $f = 0.5$ Hz coincides with the driving frequency. After it, several resonant peaks arise, corresponding to the eigenfrequencies of the system given by the expression
\begin{equation}
f_{n} = \dfrac{\omega_{0}}{2 \pi} \sqrt{1- \cos\left( \dfrac{n \pi}{N} \right)} , \label{eq:1_17}
\end{equation}
where $n = 1, ... , N-1$. 
For the derivation of Eq.~\eqref{eq:1_17} we 
refer to Appendix \ref{1_A_2}. Thus, Eq.~\eqref{eq:1_17} predicts the appearance of the second peak at $f_{1} = 0.66$ Hz, in agreement to what we observe in Fig. \ref{fig:1_fig3}. Each vertical red dotted line marks the place where an eigenfrequency peak should arise according to Eq.~\eqref{eq:1_17}. We observe that the MBP simulation peaks coincide perfectly with the prediction, whereas for the BI peaks, there is agreement for the first five eigenfrequencies only.

The natural frequency of the bubble, as given by Eq.~\eqref{eq:1_8}, is $f = 2.9$ Hz which is approximately the same value as the fifth eigenfrequency $f_{5}$. After this value, the BI peaks are shifted to the left and do not agree with the theoretical prediction. It is speculated that three-dimensional bubble deformations cause this deviation from the theory.

\section{A variety of systems}\label{sec:1_4}

Using BI simulations to explore the effect of different driving frequencies on the confined bubble array is timewise prohibitively expensive. The MBP model has the advantage to be computationally much faster. Therefore, since the model has shown to satisfactorily predict the main resonant frequencies, we use it to study the effect of a range of driving frequencies on different bubble systems. In Sections \ref{sec:1_4_1}, \ref{sec:1_4_2}, \ref{sec:1_4_3} and \ref{sec:1_4_4} we integrate numerically the MBP equations of motion defined in Eqs. \ref{eq:1_3} and \ref{eq:1_4} for driven monodisperse, bidisperse and polydisperse bubble systems, consisting of a varying number of bubbles. In Section \ref{sec:1_4_3} we explore an impulsively kicked system, i.e., a system where the first liquid patch is displaced instantaneously left from its initial position by a distance $A$. For all following simulations, a liquid patch length of $\ell = 1$ m was used.

For driven motion, when the driving frequency matches one of the eigenfrequencies or the natural bubble frequency, the response is enhanced. This can be understood as more energy is going into the oscillations as compared to off-resonance driving. To quantify this effect we follow the idea of Sander van der Meer et al. \cite{van2007microbubble} and compute the area under the power spectrum for each driving frequency $f_{D}$. We call this the \textit{response} of the system $R$ defined as
\begin{equation}
R(f_{D}) = \int E (f,f_{D}) \; \mathrm{d}f . \label{eq:1_18}
\end{equation}
Then, higher $R$ corresponds to more energetic resonant oscillations, i.e, larger displacements from the bubbles equilibrium position.

\subsection{Monodisperse systems}\label{sec:1_4_1}

\begin{figure}
\includegraphics[width=\columnwidth]{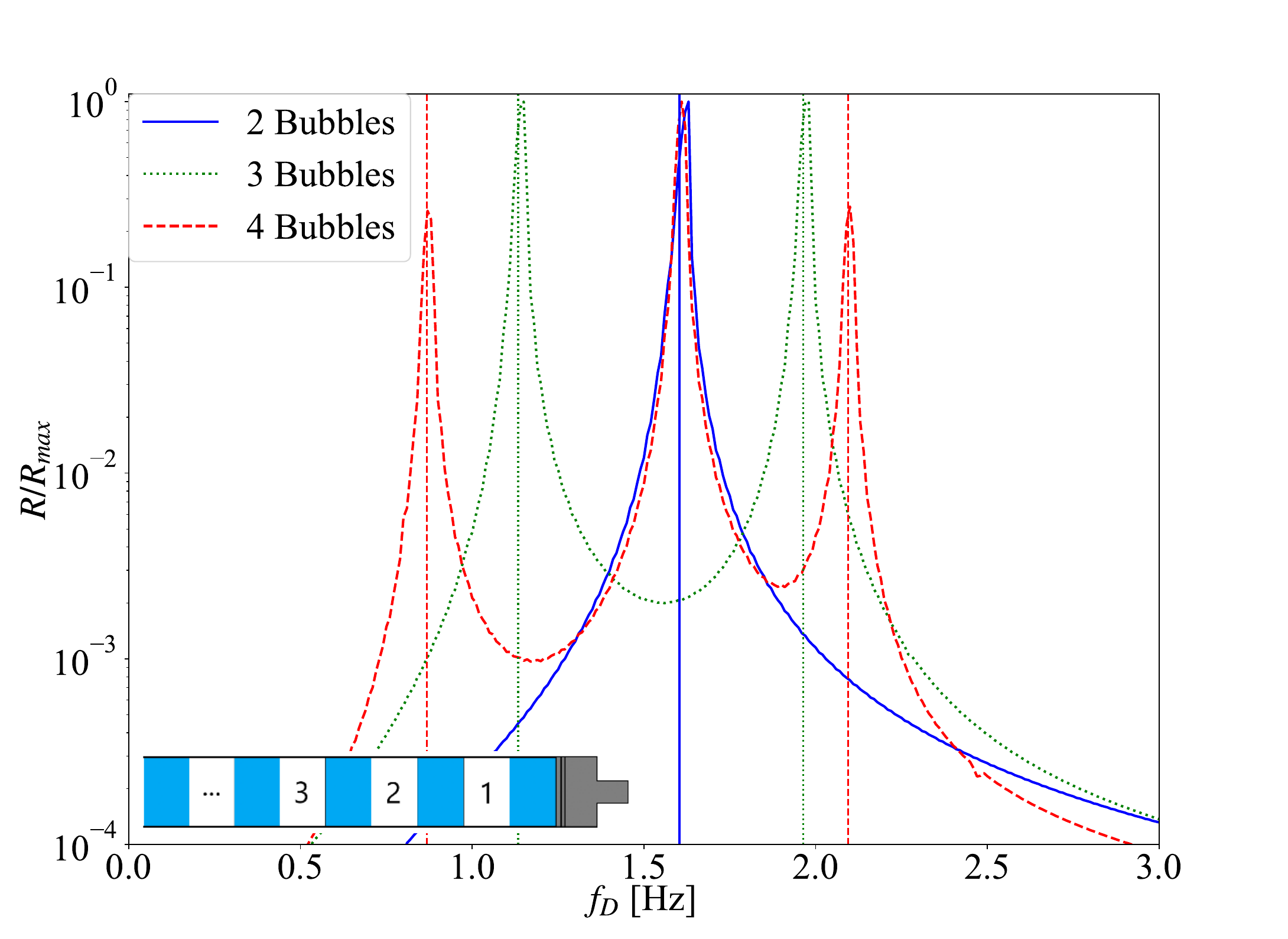}
\caption{Response curves normalized by the maximum response for different monodisperse systems. The blue solid line corresponds to a two-bubble system, the green dotted line to three bubbles and the red dashed line to four bubbles. The vertical lines of the corresponding colors are the predictions for the eigenfrequencies from Eq.~\eqref{eq:1_17} for $N = 2, 3$ and 4 respectively. When the number of bubbles is odd an eigenfrequency matches the natural frequency of the bubbles and the response is enhanced. If the number is even, the highest peaks emerge to the left and right of the bubble natural frequency.}\label{fig:1_fig4}
\end{figure}

Now that we found a way to characterize the bubble oscillations by looking at the response, we start by exploring monodisperse systems with bubble sizes of $\delta x (0) = 2$ m. Figure \ref{fig:1_fig4} shows $R$ for three monodisperse systems of 2, 3 and 4 bubbles with driving frequencies ranging from $f_{D} = 0.05$ to $3.0$ Hz and driving amplitude $A = 0.01$ m. The response is normalized to the maximum value such that the highest peak is at 1. The solid blue line represents the result 
for a two-bubble system, the 
dotted green line that for three bubbles, and the 
red dashed line for four bubbles. It is interesting to notice that the number of peaks is equal to the number of bubbles minus one. This is no coincidence since the number of eigenfrequencies $n$ for monodisperse systems follows the condition $n < N$, as derived in Appendix \ref{1_A_2}, and, generally speaking, $N-1$ is the number of degrees of freedom of the system.

In general, following Eq.~\eqref{eq:1_17}, the peaks arise centered around the bubble resonant frequency $f_{D} \approx 1.6$ Hz, with $N = 2, 3, 4$ for each bubble number respectively. When the number of bubbles is even, the highest peak emerges exactly at the natural bubble frequency. When the number is odd, there is no peak at the bubble frequency but two tall peaks to the left and right of it. The vertical lines in Fig. \ref{fig:1_fig4}  are the eigenfrequencies obtained from Eq.~\eqref{eq:1_17} for each bubble number. It can be seen that there is a perfect agreement with the MBP simulations for the three cases.

\begin{figure}
\includegraphics[trim={30 20 10 30},clip,width=0.84\linewidth]{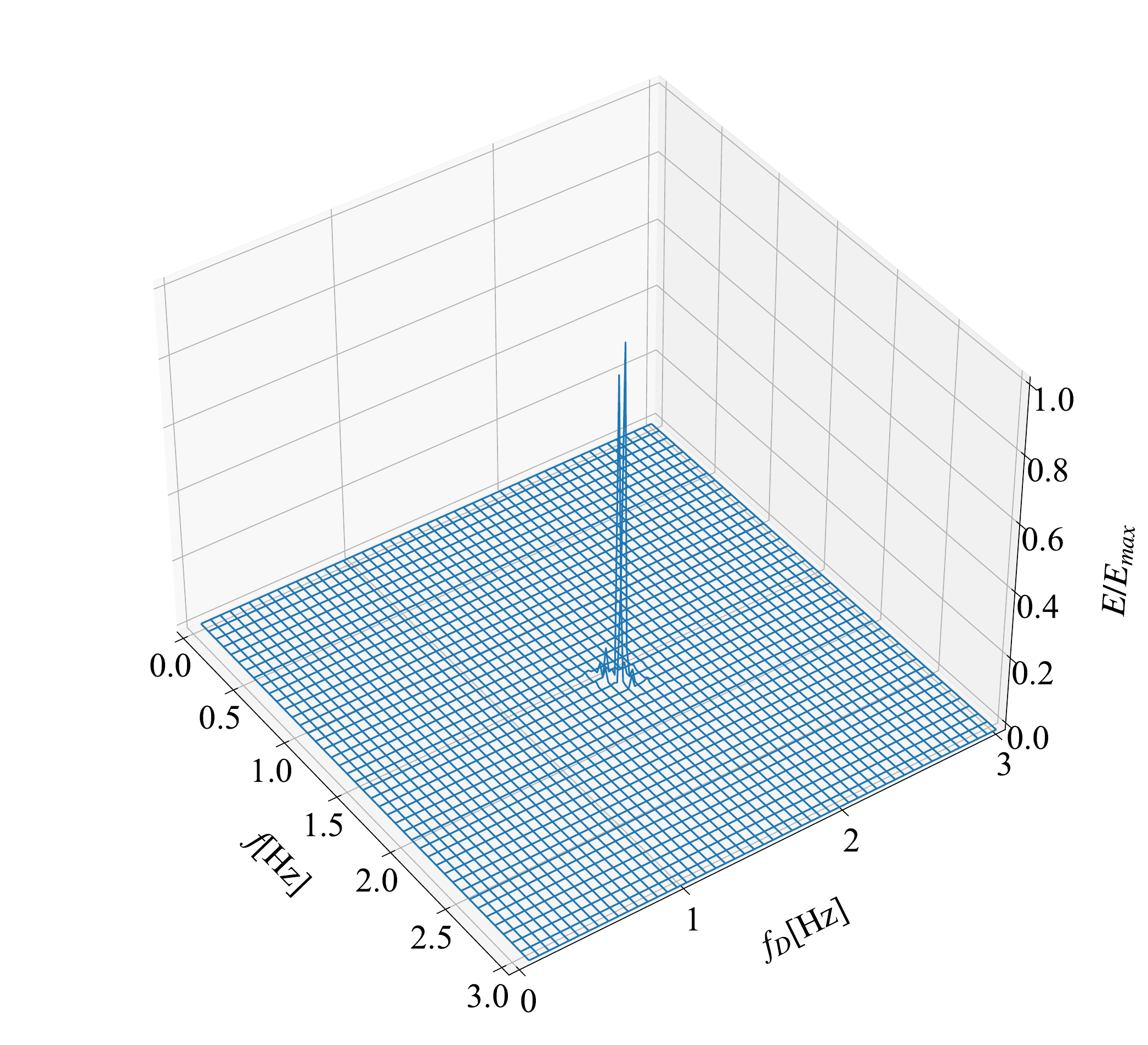}
\makebox[0.49\textwidth]{a)}
\includegraphics[trim={30 20 10 30},clip,width=0.84\linewidth]{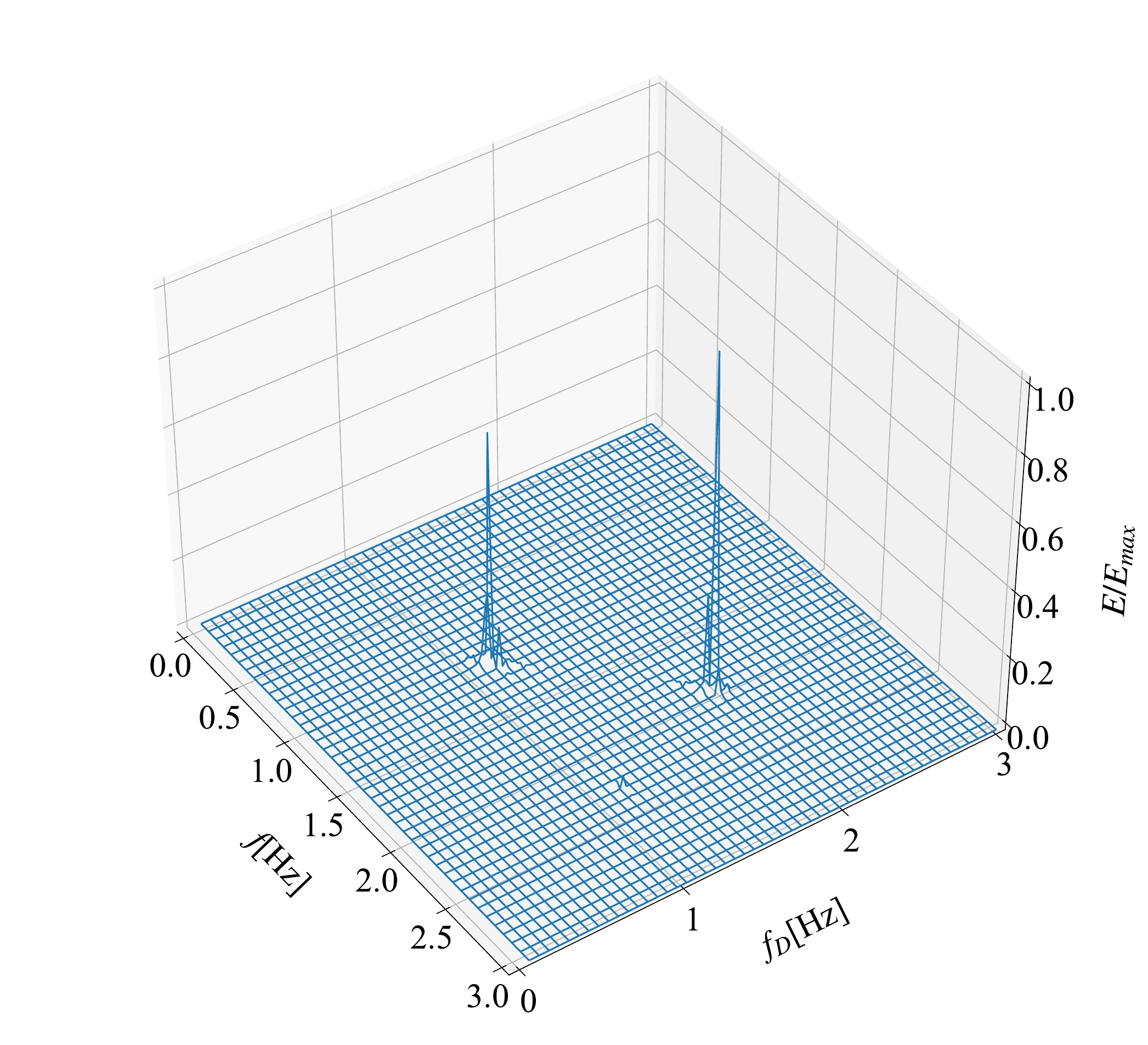}
\makebox[0.49\textwidth]{b)}
\includegraphics[trim={30 20 10 30},clip,width=0.84\linewidth]{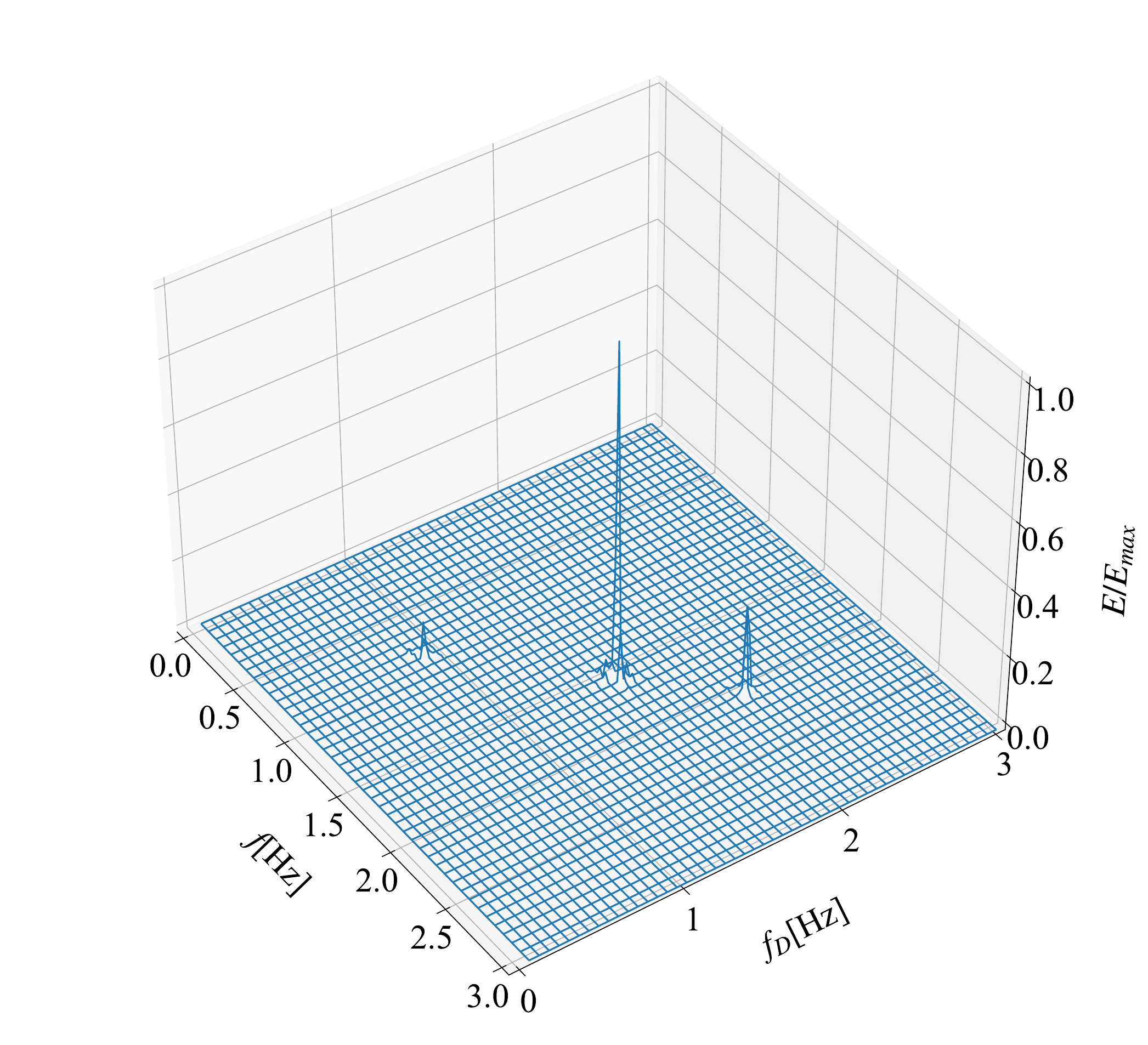}
\makebox[0.49\textwidth]{c)}
\caption{Normalized power spectra for several driving frequencies. In a) for two bubbles, b) three bubbles and c) four bubbles. The resonant frequency of the bubbles is $f=1.6$ Hz for the three cases. However, because of a change in the system size (number of bubbles) the system resonant frequencies appear at different positions along the diagonal line $f=f_{D}$. The number of peaks along the diagonal is $N-1$.}
\label{fig:1_fig5}
\end{figure}

In Fig. \ref{fig:1_fig5} (a), (b) and (c) we plot the power spectra of the aforementioned systems for all sampled driving frequencies. The main peaks appear along the diagonal $f = f_{D}$. The main contribution to the response is along this line. However in Figs. \ref{fig:1_fig5} (b) and (c) we can also see small peaks off the diagonal line, corresponding to higher harmonics of the smallest resonant peak $f_{1}$.

Clearly, increasing the number of bubbles ($N$) leads to the emergence of additional peaks ($N-1$), similar to what was shown in the comparison of the BI and MBP results for $N=10$ (Fig.~\ref{fig:1_fig3}), where the location of the peaks in $R$ remains in good agreement with the prediction from Eq.~\eqref{eq:1_17}.

\subsection{Bidisperse systems}\label{sec:1_4_2}
In this subsection we explore the effect of bidispersity on bubble systems using the MBP model. Since there are two bubble sizes we refer to the smaller size as the \textit{small} bubble and to the larger one as the \textit{big} bubble. We set the initial size of the small bubble equal to 0.5 m such that its resonant frequency is $f_s = 3.2$ Hz; the initial size of the big bubble is 3.5 m and its resonant frequency $f_b = 1.2$ Hz. So the average bubble size will be 2 m like in the monodisperse case.

There are many ways to arrange a bidisperse system, in this article we study two particular cases, a \textit{bidisperse alternating system} and a \textit{bidisperse stacked system}. In the first one, as the name says, the bubble sizes are alternating: ...-small-big-small-big-...; in the second case we place first all the bubbles of one size and then all the bubbles of the other size together. For both cases we use a driving amplitude of $A = 0.001$ m.

\subsubsection{Bidisperse alternating systems}

\begin{figure}
\includegraphics[width=\columnwidth]{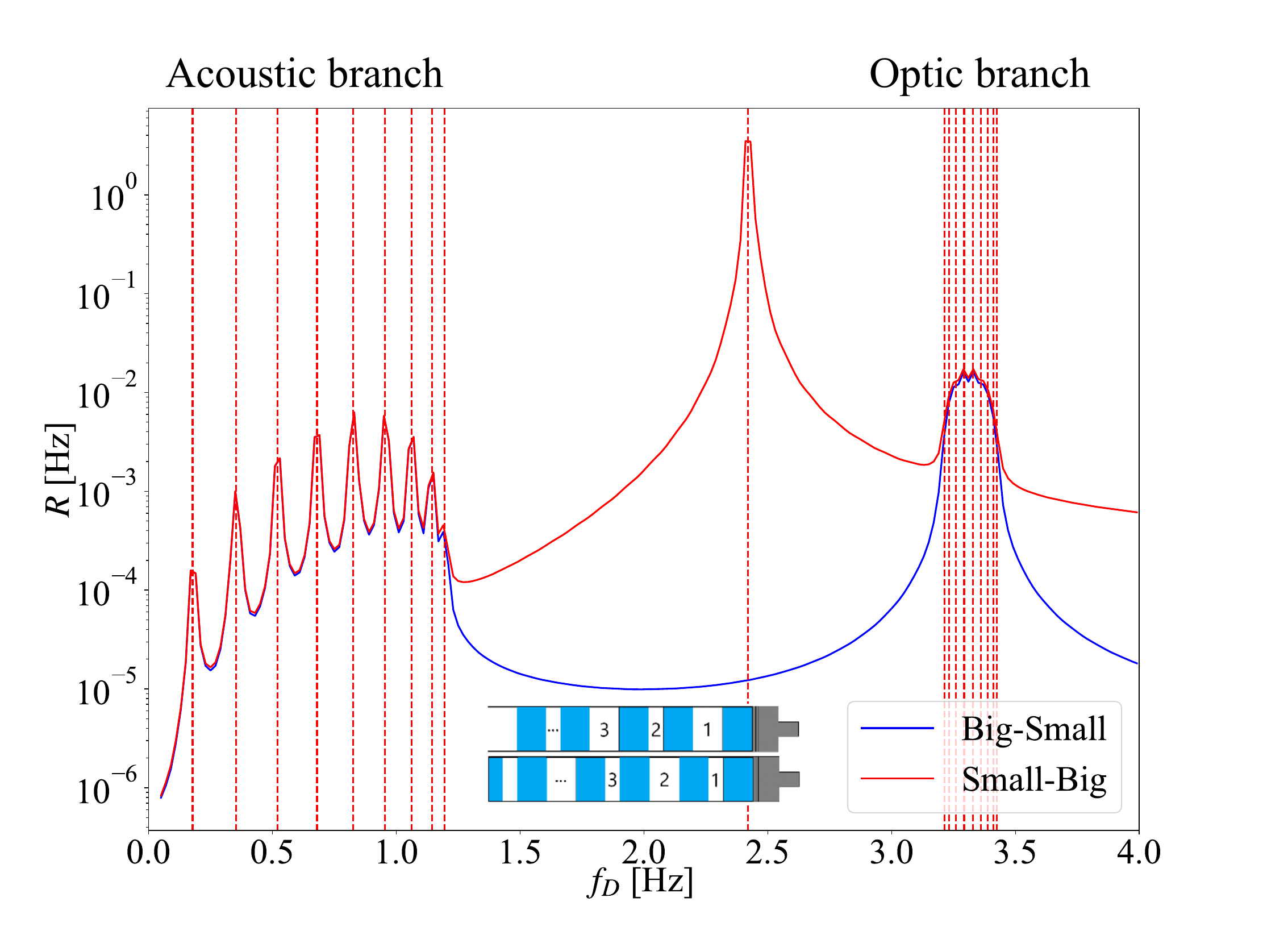}
\caption{Response curves, in blue for the alternating big-small system and in red for the small-big one. Both blue and red curve present the exact same peaks along the acoustic and optic branches as predicted by Eq.~\eqref{eq:1_19}. The small-big system however, has a very large response at the natural frequency of the pseudo-two-bubble system given by $f_{s,b} = 2.4$ Hz.
}\label{fig:1_fig6}
\end{figure}

We start by simulating the bidisperse alternating systems. Figure \ref{fig:1_fig6} shows the response against the driving frequency for two alternating setups with 20 bubbles each. In blue, the response of an alternating big-small system, where the first bubble next to the piston is big. In red, the alternating small-big one, where the first bubble is small. In the inset, next to the plot legend, one can see sketches of the setups.

It is possible to obtain an expression that predicts the position of each of the peaks in Fig \ref{fig:1_fig6}. for this alternating systems. This case resembles a problem in the classical theory of crystals being the bubble equivalent of a one-dimensional Bravais lattice with two ions, see, e.g,. \textcite{ashcroft1976solid}. Naturally the big and small bubbles correspond to the two ions with different \textit{stiffnesses}. Then the eigenfrequencies of the bidisperse alternating system are given by (see Appendix \ref{1_A_3} for a derivation)
\begin{equation}
f_n = \dfrac{1}{2 \pi}
\sqrt{
\dfrac{\omega_{s}^{2} + \omega_{b}^{2} }{2} \pm \dfrac{1}{2}
\sqrt{ \omega_{s}^{4} + \omega_{b}^{4} +
2 \omega_{s}^{2} \omega_{b}^{2}
\cos\left( \dfrac{n \pi}{N} \right) } }, \label{eq:1_19}
\end{equation}
where $\omega_{s}$ and $\omega_{b}$ are the natural angular frequencies of the small and big bubbles respectively given by their corresponding Eq.~\eqref{eq:1_8}, $N$ is the number of small (or big) bubbles 
and $n$ counts 
the eigenfrequencies 
going from 1 to $N-1$.

Equation \eqref{eq:1_19} foresees two different sets (or branches) of solutions, distinguished by the choice of sign in 
the $\pm$ symbol. The solutions with the plus sign inside the square root (close to $\sqrt{\omega_s^2 + \omega_b^2}/2\pi$) correspond to what is known in crystallography as the optic branch and the solutions with the minus sign (close to zero) are known as the acoustic branch. If we define a unit cell as the minimum building block from which the whole lattice can be built just by translation operations, then, the acoustic branch corresponds to the two bubbles (one big and one small) of the unit cell oscillating in phase. Accordingly, the optic branch corresponds to the two bubbles from the unit cell oscillating in anti-phase. In crystals, depending whether the system oscillates in the optic or acoustic branch will dictate how the lattice interacts with light. Here is not the case, but we keep the names for easy identification. In Fig. \ref{fig:1_fig6} we show in dotted red vertical lines the predicted peaks that agree nicely with the plotted response of both the big-small and small-big systems. The acoustic branch spawns from around 0 to 1.2 Hz and the optic branch rises around 3.2 Hz. However in this latter case the peaks are so close together that we are not able to resolve them given the resolution of the plot. 

Both 
plots in Fig. \ref{fig:1_fig6} are practically the same except for one 
salient feature. If the first bubble is small (red curve), the highest response is not in either of the branches but at $f_{D} = 2.4$ Hz. The reason for this is that, around that frequency, the system behaves almost like a two-bubble system of first one small and then one big bubble. Since 2.4 Hz is smaller than the small bubble resonant frequency, the first bubble follows the piston in phase, transmitting the signal unaffected to the second bubble. Then, since 2.4 Hz is higher than the big bubble resonant frequency, the second bubble acts as a low-pass filter that attenuates nearly all of the pressure signal. The motion in this system around 2.4 Hz is therefore almost exclusively confined to the first two bubbles. Then, following the same procedure that we used to obtain Eq.~\eqref{eq:1_7}, we can write the equation of motion for the big bubble (since the small bubble equation is just $\epsilon_1 = - A \sin( \omega_{D} t )$)
\begin{equation}
\ddot{\epsilon}_{2} +  \omega_{s,b}^{2} \epsilon_{2} \approx
\dfrac{1}{2} \omega_{s}^{2} \epsilon_{1}, \nonumber
\end{equation}
where the natural angular frequency of the pseudo-two-bubble system is
\begin{equation}
\omega_{s,b} = \sqrt{\dfrac{\omega_{s}^{2}+\omega_{b}^{2}}
{2}}. \label{eq:1_20}
\end{equation}
This expression gives us a resonant frequency of $f_{s,b} = \omega_{s,b}/ 2 \pi \approx 2.4$ Hz, just as seen in the red curve in Fig. \ref{fig:1_fig6}.

\begin{figure*}
\begin{center}
\includegraphics[trim={30 20 10 30},clip,width=0.49\textwidth]{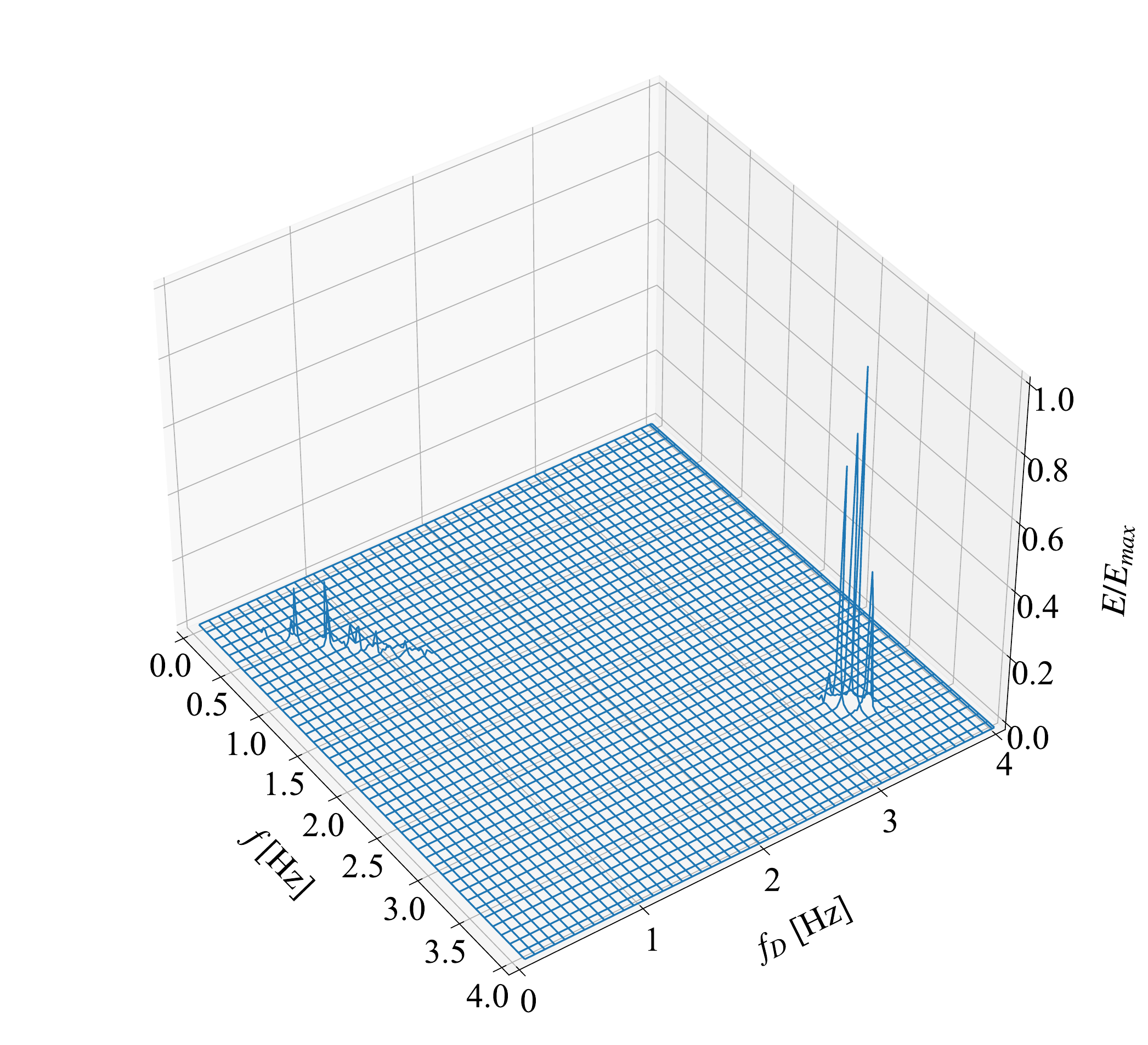}
\includegraphics[trim={30 20 10 30},clip,width=0.49\textwidth]{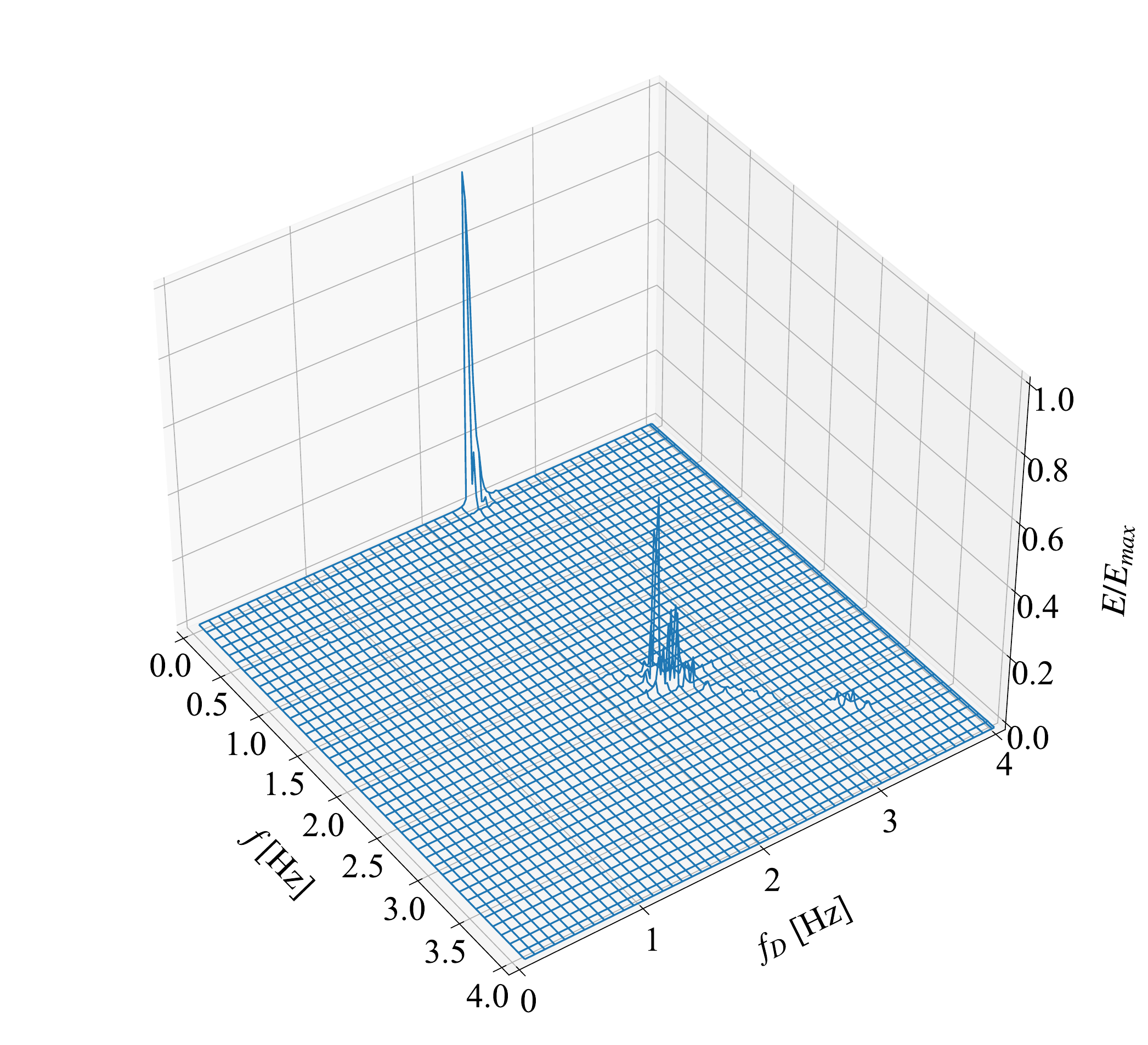}
\makebox[0.49\textwidth]{a)}
\makebox[0.49\textwidth]{b)}
\end{center}
\caption{Normalized power spectra for two bidisperse alternating systems, in a) the first bubble is big and in b) the first bubble is small. The size of the big bubble is 3.5 m with a resonant frequency of 1.2 Hz; the small bubble is 0.5 m in size with resonant frequency of 3.2 Hz. We see in a) that when $f_{D}$ is close to the resonant frequency of the big bubbles, the eigenfrequencies of the acoustic branch are excited. Correspondingly, when $f_{D}$ is around the small bubble resonant frequency, the optic branch resonates. In b) the response of both branches is imperceptible as compared to the big peaks close to $f_{D} = 2.4$ Hz. Around this driving frequency the peak with higher $f$ corresponds to the resonance of the pseudo-two-bubble system, the peak with very low frequency is due to beating caused by a small difference between $f_{D}$ and $f_{s,b}$.
}\label{fig:1_fig7} 
\end{figure*}

In Fig. \ref{fig:1_fig7} we plot the normalized power spectra for both alternating systems with driving frequencies between $f_{D} = 0.05$ and 4.0 Hz. Figure \ref{fig:1_fig7} (a) shows very well defined and separate resonant peaks along the diagonal close to each bubble resonant frequency at their corresponding branches. Figure \ref{fig:1_fig7} (b) explains why there is a high response around $f_{D} = 2.4$ Hz. There is a contribution to the response that comes from the peak along the diagonal at $f = f_{D} = 2.4$ Hz coming from $f_{s,b}$; another contribution appears off the diagonal and has a much lower frequency, in fact close to zero. This comes from what is known as \textit{beating}. When the driving frequency is close to but not \textit{exactly} $f_{s,b}$, the oscillations in the second bubble are modulated by a frequency given by half the difference between $f_{D}$ and $f_{s,b}$.

\subsubsection{Bidisperse stacked systems}

\begin{figure}
\includegraphics[width=\columnwidth]{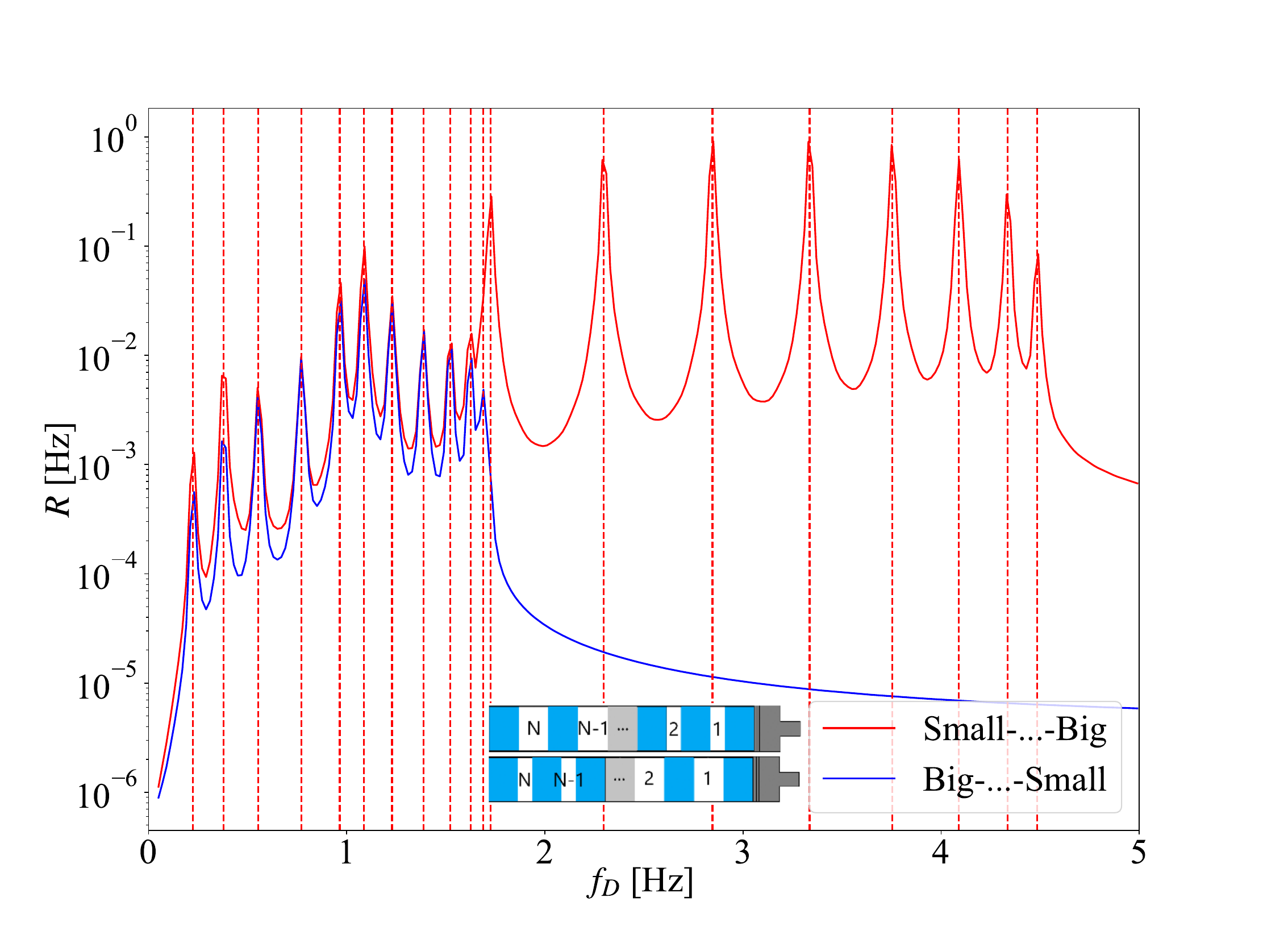}
\caption{Normalized response curves for bidisperse stacked systems, in red when all small bubbles are first, i.e., closest to the piston, and in blue when all big bubbles are first. We see that small bubbles allow the transmission of pressure signals over the entire sampled spectrum, as opposed to big bubbles that filter high frequencies. The small-...-big system presents coupling between the big and small bubble halves making the eigenfrequencies shift from their monodisperse counterpart.}\label{fig:1_fig8}
\end{figure}

The second bidisperse system we studied is the so called \textit{stacked} system. Here, all the bubbles of a particular size are arranged first next to the piston followed by all the bubbles of the other size. The same small and big sizes were used as in the alternating case and, again, there are $10$ bubbles of each size. Figure \ref{fig:1_fig8} shows in red the response of the stacked small-...-big setup, where all small bubbles are first; in blue, we observe the response of the stacked big-...-small setup, where all big bubbles are first. The inset shows sketches of these systems next to their corresponding names.

The red curve exhibits resonant peaks over the entire sampled driving frequency range. The wider higher peaks arise around the small bubble resonant frequency, whereas the smaller narrower peaks are closer to the big bubble resonant frequency. The blue curve shows only narrow peaks near the big bubble resonance, which is due to the filtering of signals with higher frequency than the natural one of the big bubbles, as we will explain later in this section.

Unlike the bidisperse alternating cases, the stacked system does not present branches, which might lead one to think incorrectly that the first half of the system behaves uncoupled from the second half. If this was the case, one could use Eq.~\eqref{eq:1_17} to find the eigenfrequencies of each half and predict the peaks positions. This, however, turns out not to be correct, since 
the equations of motion do present coupling at the connecting bubbles, where the system changes from one bubble size to the other. The eigenvalue problem that one must solve is then (see Appendix \ref{1_A_4} for a derivation)
\begin{equation}
\bm{B}\cdot \bm{\beta} = \lambda_{n} \bm{\beta} , \label{eq:1_21}
\end{equation}
where we have defined
\begin{equation}
\lambda_{n} = \omega_{n}^{2}  \label{eq:1_22}
\end{equation}
and the $(N-1)\times(N-1)$ matrix
\begin{equation}
\bm{B} = \begin{pmatrix}
    2a  &   -a     &     0     &           &          &          &         \\
   -a   &    2a    &    -a     &           &          &          &         \\
        &  \ddots  &  \ddots   &           &          &          &         \\
        &          &    -a     &   (a+b)   &    -b    &          &         \\
        &          &           &           &  \ddots  &  \ddots  &         \\       
        &          &           &           &    -b    &    2b    &   -b    \\
        &          &           &           &    0     &    -b    &   2b
  \end{pmatrix} , \label{eq:1_23}
\end{equation}
with $a = \omega_{s}^2/2$ and $b = \omega_{b}^2/2$. There does not appear to be a simple analytic solution to this problem and one may numerically solve for 
the eigenvalues of matrix $\bm{B}$ to obtain the natural angular frequencies of the system to compare to 
the frequency peaks emerging in Fig. \ref{fig:1_fig8}. This was done using Matlab (and Wolfram Mathematica for double check). The obtained eigenfrequencies are plotted as 
red dotted vertical lines in Fig. \ref{fig:1_fig8}. The big-...-small system presents the same eigenfrequencies as the small-...-big one up to the ones associated with the small bubbles, as can be seen in the figure.

As stated above, and unlike the monodisperse or bidisperse alternating systems, it is not possible to find analytic eigenfrequencies for this stacked case. However, there are other ways to describe theoretically what is observed in Fig. \ref{fig:1_fig8}. For this, we will treat both halves of the setup as two continuous 
media, i.e., the half-space with the big bubbles will correspond to a medium with lower average density and the half-space with small bubbles to higher density. It is well known that when an acoustic wave is incident to the interface between two media there will be transmission and reflection of the wave \cite{chapman1990normal}. Then, for the stacked setup it is possible to find transmission and reflection coefficients that describe how much of the pressure wave crosses or returns from the boundary between the two half-spaces.

To derive these coefficients, imagine a solitary pressure wave traveling from the first medium to the second; one may think about it as the first period of the signal after the piston starts moving. For simplicity, at first we assume that there is no attenuation of the pressure wave. By construction, the piston was placed on the right side of the system so the incident wave will travel to the left in the negative $x$ direction, as well as the transmitted wave; the reflected wave will thus travel to the right in positive $x$ direction. Then we can write the incident, reflected and transmitted pressure waves as 
\begin{equation}
\renewcommand{\arraystretch}{1.5}
\begin{array}{rl}
P_{i} &= A_{i} e^{i(-k_{1} x - \omega t)}
 , \\
P_{r} &= A_{r} e^{i(k_{1} x - \omega t)}
 , \\ 
P_{t} &= A_{t} e^{i(-k_{2} x - \omega t)}
 ,
\end{array} 
\label{eq:1_24}
\end{equation}
where $A_i$, $A_r$ and $A_t$ are just amplitudes of the incident, reflected and transmitted pressure waves; $k_1$ and $k_2$ are the wave numbers of medium 1 and 2 respectively, being the medium 1 the one next to the piston and medium 2 next to the  fixed wall.

The boundary conditions at the interface between the two media are
\begin{equation}
\renewcommand{\arraystretch}{2.1}
\begin{array}{rl}
P_{i} + P_{r} &= P_{t}
 , \\
\dfrac{1}{\rho_1} \left( \dfrac{\partial P_i}{\partial x} \right) +
\dfrac{1}{\rho_1} \left( \dfrac{\partial P_r}{\partial x} \right) &=
\dfrac{1}{\rho_2} \left( \dfrac{\partial P_t}{\partial x} \right) ,
\end{array} \label{eq:1_25}
\end{equation}
which just means that there are no pressure jumps and that the velocities between the media are the same at the boundary. If we define the transmission coefficient as $T_p = A_t / A_i$ and the reflection coefficient as $R_p = A_r / A_i$, we can use Eqs.  \ref{eq:1_24} and \ref{eq:1_25} to find
\begin{equation}
\renewcommand{\arraystretch}{2.1}
\begin{array}{rl}
T_p &= \dfrac{2 \rho_2 c_2}{\rho_1 c_1 + \rho_2 c_2}
 , \\
R_p &= \dfrac{\rho_2 c_2 - \rho_1 c_1}{\rho_1 c_1 + \rho_2 c_2}
,
\end{array} \label{eq:1_26}
\end{equation}
where $\rho_i = \rho_l \ell / (\ell+\delta x_{0,i} )$ is the density and $c_i = c_{0,i} \tilde{c_i}$ the sound speed of the corresponding medium $i = 1$ and 2, with $c_{0,i} = \omega_{i} (\ell+\delta x_{0,i} )$. Here, 
$\tilde{c_i}$ is given by Eq.~\eqref{eq:1_13} and naturally depends on $\tilde{\omega}_i = \omega/\omega_i$, where 
$\omega_i$ corresponds to the natural bubble frequency of medium $i = 1$ and 2 respectively. 
Inserting the above expressions in Eq.~\eqref{eq:1_26} we obtain
\begin{equation}
\renewcommand{\arraystretch}{2.1}
\begin{array}{rl}
T_p &= \dfrac{2 \omega_2 \tilde{c}_2}{\omega_1 \tilde{c}_1 + \omega_2\tilde{c}_2}
 , \\
R_p &= \dfrac{\omega_2 \tilde{c}_2 - \omega_1 \tilde{c}_1}{\omega_1 \tilde{c}_1 + \omega_2 \tilde{c}_2}\,\,,
.
\end{array} \label{eq:1_27}
\end{equation}
where $\tilde{c}_i$ depends both on the driving frequency $\omega$ and the natural frequency $\omega_i$ of each medium. See, e.g., \textcite{chapman1990normal} for further elaboration.

It is didactic to notice that 
\begin{equation}
T_p - R_p = 1 , \label{eq:1_28}
\end{equation}
for this non-attenuated case. In the case when we introduce attenuation (as it happens when the driving angular frequency is larger than $\sqrt{2}$ times the bubble natural frequency), the incident wave's amplitude decreases exponentially according to Eq.~\eqref{eq:1_48}. The derivation of the coefficients follows the same procedure and so we arrive to the attenuated transmission and reflection coefficients defined as
\begin{equation}
\renewcommand{\arraystretch}{2.1}
\begin{array}{rl}
\tilde{T}_p &= T_p e^{-N/\tilde{\xi}_1}
 , \\
\tilde{R}_p &= R_p e^{-N/\tilde{\xi}_1}
,
\end{array} \label{eq:1_29}
\end{equation}
where $N$ is the number of bubbles from medium 1 that the incident wave has to pass through to arrive at the interface and $\tilde{\xi_1}$ the dimensionless penetration depth of medium 1 given by
\begin{equation}
\tilde{\xi_1} =
    \begin{cases}
      \infty & \text{if $\tilde{\omega}_1 \leq \sqrt{2}$}\\
     - \left[\log\left(\tilde{\omega}_{1}^2(1-\sqrt{1-2\tilde{\omega}_{1}^{-2}})-1 \right)\right]^{-1} & \text{if $\tilde{\omega}_{1} \geq \sqrt{2}$}
    \end{cases} , \label{eq:1_30}
\end{equation}
see Appendix A for derivation. Equation \eqref{eq:1_28} is also transformed to
\begin{equation}
\tilde{T}_p - \tilde{R}_p = e^{-N/\tilde{\xi}_1} . \label{eq:1_31}
\end{equation}

Equations \ref{eq:1_29} predict a strong suppression of the transmitted and reflected waves depending on which is the first medium. On the one hand, for the small-...-big system, the small bubbles are present in medium 1 (i.e., closest to the driving piston) and the attenuation regime is reached when $f_D = 4.54$ Hz, way above the attenuation regime of the big bubbles medium. Therefore we are able to see peaks in the response all over the sampled driving frequencies up to 4.54 Hz. On the other hand, when medium 1 is the one with big bubbles (big-...-small), the attenuation regime is reached at $f_D = 1.71$ Hz and practically nothing of the pressure wave reaches medium 2. This is in agreement with what is observed in Fig. \ref{fig:1_fig8} blue curve. Even though both setups have the same eigenfrequencies obtained by solving Eq.~\eqref{eq:1_21}, attenuation suppresses the response of the small bubbles when big bubbles are located before them. In Appendix \ref{1_A_5} we show plots of $\tilde{T}_p$ and $\tilde{R}_p$ for the case 
$N = 10$. 

\begin{figure*}
\begin{center}
\includegraphics[trim={30 20 10 30},clip,width=0.49\textwidth]{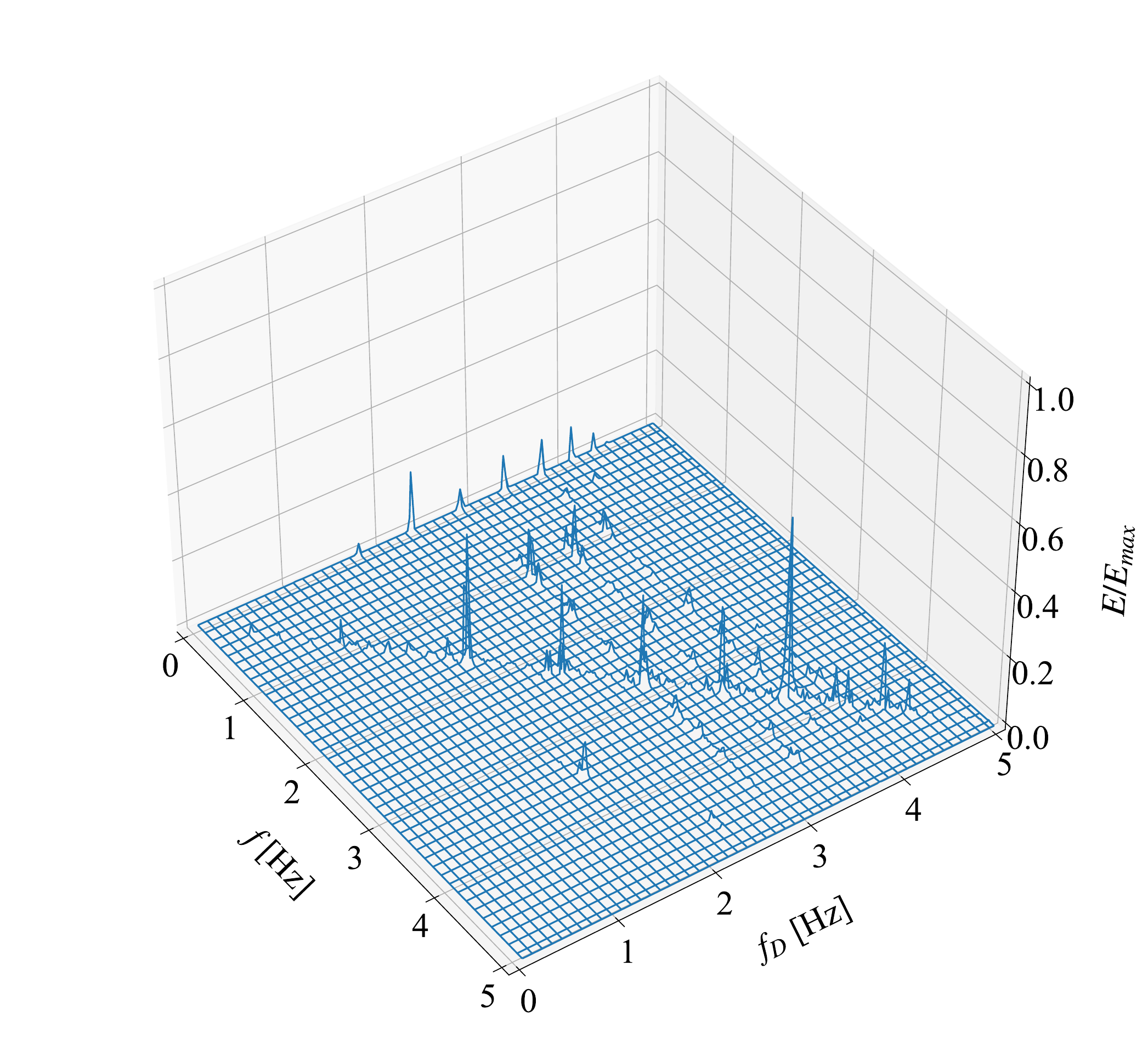}
\includegraphics[trim={30 20 10 30},clip,width=0.49\textwidth]{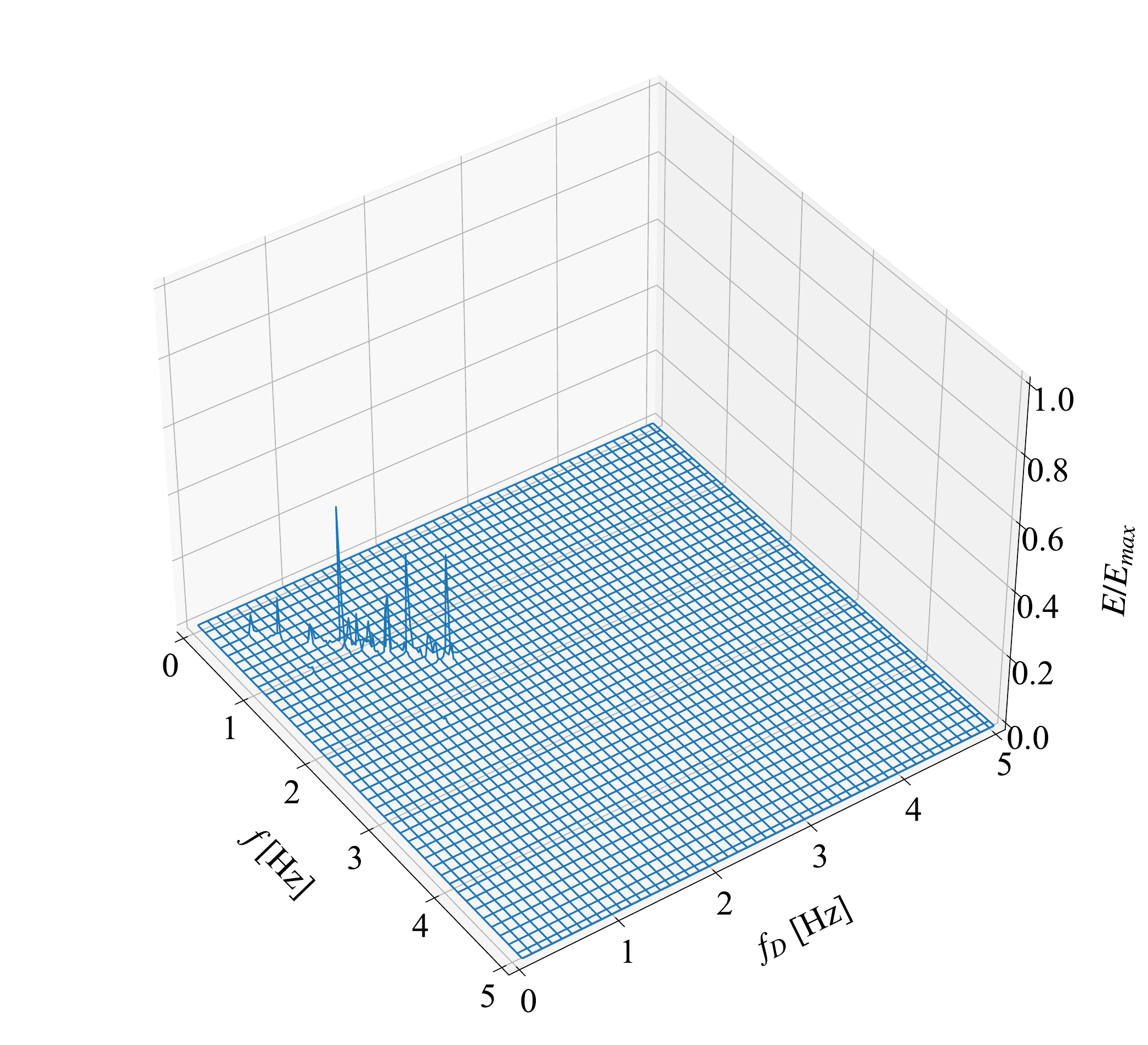}
\makebox[0.49\textwidth]{a)}
\makebox[0.49\textwidth]{b)}
\end{center}
\caption{Normalized power spectra for two bidisperse stacked systems. In a) all the small bubbles are placed first and then the big bubbles; in b) it is the other way around. The eigenfrequencies can be computed by solving numerically the eigenvalue problem as explained in the main text. We can appreciate in a) that small bubbles transmit pressure signals better since we get responses all over the power spectra. In b) we see that big bubbles filter the pressure signal since only eigenfrequencies associated to big bubbles resonate.}\label{fig:1_fig9} 
\end{figure*}

Figure \ref{fig:1_fig9} displays the normalized power spectra for the sampled driving frequencies between $f_{D} = 0.05$ and 4.0 Hz and $A = 0.01$ m for both stacked systems. Figure \ref{fig:1_fig9} (a) corresponds to the power spectrum for the case that 
all small bubbles are first. Here we notice that the main contributions to the response are along the diagonal. However, there are multiple small peaks along lines of constant $f_{D}$. These are caused by the nonlinear character of the equations of motion. The amplitude of the oscillation is one order of magnitude smaller than the small bubble size, but it is big enough to show this nonlinear behavior. To prove this, simulations with smaller driving amplitude were carried out. The results show similar peaks along the diagonal but the smaller peaks off this line were not present anymore, confirming our hypothesis.

On the other hand, in Fig. \ref{fig:1_fig9} (b), where the system is reversed and the large bubbles are first, we see that the main and almost only contribution to the response is along the diagonal. 
Above $f_D = 1.71$ Hz, all signals have been suppressed due to attenuation, in accordance to what was previously discussed. Moreover, there are 
practically no peaks outside of the diagonal since the driving amplitude is much smaller than the big bubble size and non-linear effects are reduced.

\subsection{Impulsively kicked system}\label{sec:1_4_3}

In this section we study the effect of an impulsive driving on a monodisperse system and the differences we observe with respect to the continuously driven case. The setup consists of $20$ monodisperse bubbles of initial sizes $\delta x_0 = 2$ m such that the bubble natural frequency is $f_0 = 1.6$ Hz. The piston is instantaneously displaced by an amplitude $A = 1$ m from its initial position, thus compressing the first bubble and subsequently the next ones in a chain reaction.

\begin{figure*}
\includegraphics[width=\textwidth]{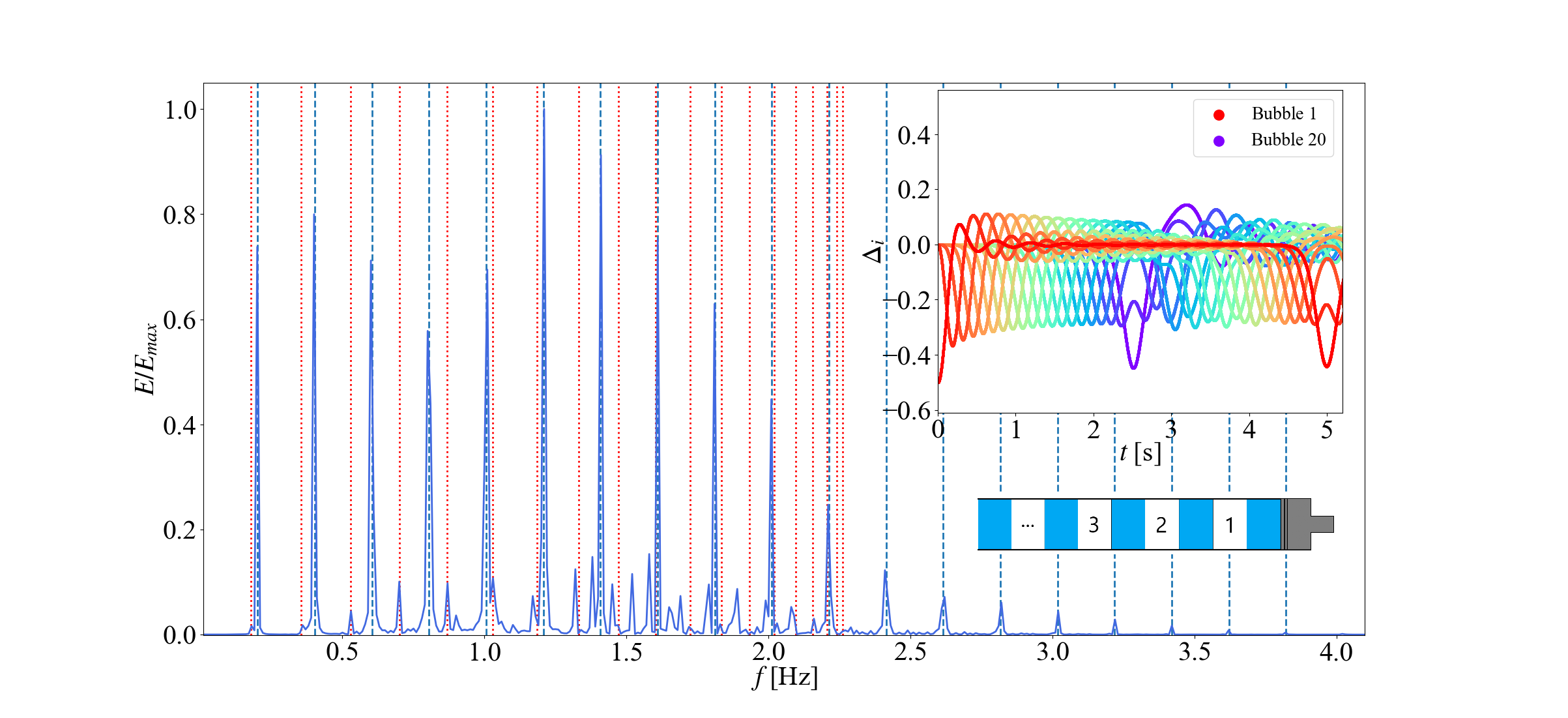}
\caption{Normalized power spectrum of an impulsively kicked monodisperse system of 20 bubbles with resonant frequency 1.6 Hz. In the inset we see the relative bubble volume signal from where the power spectrum is obtained. Since we are exciting all frequencies, we see many peaks at the harmonics of the system resonant frequency (dashed vertical blue lines). Note that these are not identical to the eigenfrequencies of the system (dotted vertical red lines) given by Eq.~\eqref{eq:1_17}.  
}\label{fig:1_fig10}
\end{figure*}
In this case there is no unique driving frequency. The \textit{kick} given by the piston excites all frequencies at the same time. We observe in Fig. \ref{fig:1_fig10} the normalized power spectrum for this setup. The peaks arise at the natural frequency of the system $f_\text{sys} = c_\text{imp}/L^*$, where $c_\text{imp}$ corresponds to the speed of the impulse and $L^*$ is equal to twice the distance the impulse travels from one end of the chain to the other, $L^* = 2[(N-1)\ell+N\delta x_0]$.

The inset in Fig. \ref{fig:1_fig10} shows the relative bubble volume signal where the chain reaction can be observed. The first bubble compresses (red curve), therefore perturbing the second bubble and so on all the way to the last bubble (purple curve). Then, when the last bubble has reached maximum compression, it bounces off the fixed wall and the signal travels back to the first bubble.

When comparing Fig.~\ref{fig:1_fig10} to the response $R$ of a harmonically driven $20$-bubble system, it is clear that the impulsively-driven case is dominated by the frequency dictated by the system size $f_\text{sys}$ and its higher harmonics (blue dashed vertical lines), the peaks of which overpower the ones corresponding to the eigenfrequencies of the system, which are distinctly different (red dotted vertical lines). This can be understood from the fact that, in spite of dispersion being present for frequencies close to $\sqrt{2}\omega_0$ the pressure pulse will as a whole continue to bounce back and forth between the boundaries of the system, leading to a strong presence of $f_\text{sys}$ in the power spectrum. Note that the power spectrum will change the longer one allows the signal to bounce back and forth into the system, with dispersion increasingly spreading the signal.

It is observed that the impulse speed accurately matches $c_\text{imp} = \frac{\pi}{4} c_0$ for an integration time of 100 seconds. Notice that this does not correspond to the propagation velocity expected from the group velocity (Eq.~\eqref{eq:1_groupvelocity} derived in Appendix \ref{1_A_1}) by solving
\begin{equation}
\frac{L^* \omega}{2\pi} = c_g(\omega)\,\,, 
\end{equation}
where the quantity $L^*$ equals twice the system size (defined above), or to solving a similar equation for the phase velocity for that matter (Eq.~\ref{eq:1_13}). Possibly, this is connected to the fact that 
these expressions were obtained for an infinite chain of bubbles. We speculate that both the impulsive way of driving the system and its finite size, i.e., the boundary conditions at the right piston and left fixed wall, play an important role in the speed of the impulse. However, the above mentioned value for $c_\text{imp}$ was tested for systems ranging from 10 to 100 bubbles for same integration times, providing a high degree of accuracy at predicting the location of the peaks.

\subsection{Polydisperse systems}\label{sec:1_4_4}

We now turn from a bidisperse to a fully disordered system, where, as a last set of simulations, we numerically solve the equations of motion for a continuously driven random polydisperse systems. Each setup consists of 50 bubbles of random initial sizes between 0.5 and 3.5 m, drawn from a uniform distribution. For every run, the integration time is computed such that the pressure signal can travel back and forth the whole system once, taking a typical wave speed of $c = 20$ m/s.

\begin{figure}
\includegraphics[width=\columnwidth]{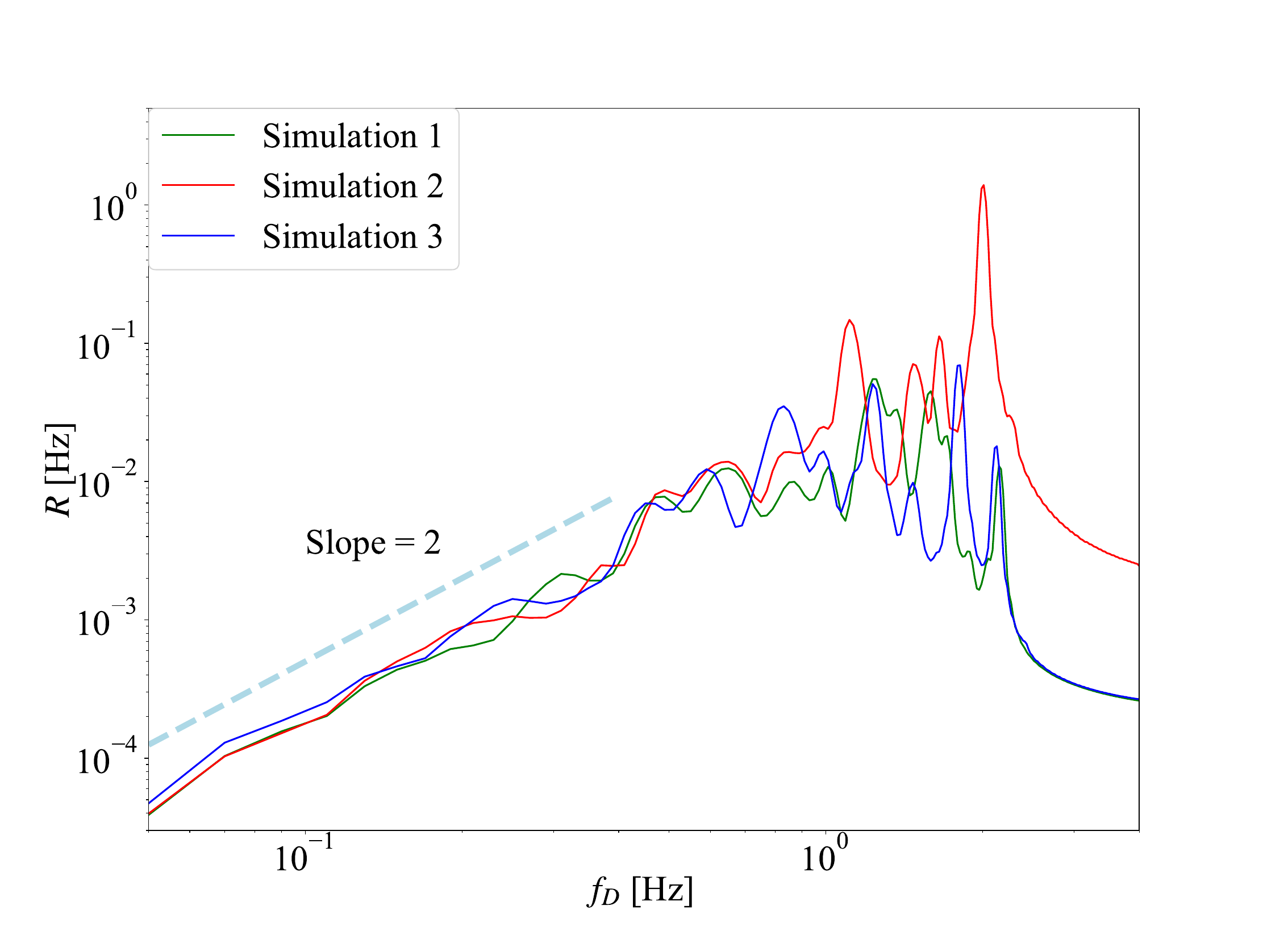}
\caption{Response curves for three different polydisperse systems with 50 bubbles of random sizes between 0.5 and 3.5 m. The integration time was sufficient to let the pressure signal travel back and forth once through the entire system. We notice that the three simulations present different peaks located at random positions. }\label{fig:1_fig11}
\end{figure}

In Fig. \ref{fig:1_fig11}, we see the response of three particular random realizations 
with sampled driving frequencies between $f_{D} = 0.05$ and 4.0 Hz. As opposed to previous sections, the peaks now arise at random locations since each setup is different. Nevertheless, peaks tend to appear around the average bubble resonant frequency $f_0 \approx 1.6$ Hz and, in principle once the individual bubble sizes are fixed, may be computed from solving a matrix eigenvalue problem similar to, e.g., that of Eq.~\eqref{eq:1_23} but now with different coefficients in each position on the tri-diagonal. 

It is observed that, for small driving frequencies, the response appears 
to follow a power law with exponent two, as shown by the dashed blue line in Fig. \ref{fig:1_fig11}. However, after more careful analysis, it is concluded that the 
exponent is similar but not exactly equal to two. Under the hypothesis 
that a polydisperse system behaves similar to a monodisperse one for driving frequencies smaller than the respective eigenfrequencies, we derive in Appendix \ref{1_A_3nw} that 
the response of a monodisperse system goes as a constant plus a sum with $N$ terms (corresponding to the number of bubbles, which are proportional to the square of the driving frequency (Eq.~\eqref{eq:RDeltafinal}). Staying away from the eigenfrequencies of the system, where the response diverges, we therefore expect and observe an intermediate region 
displaying a power law with exponent close to two.

\begin{figure}
\includegraphics[width=\columnwidth]{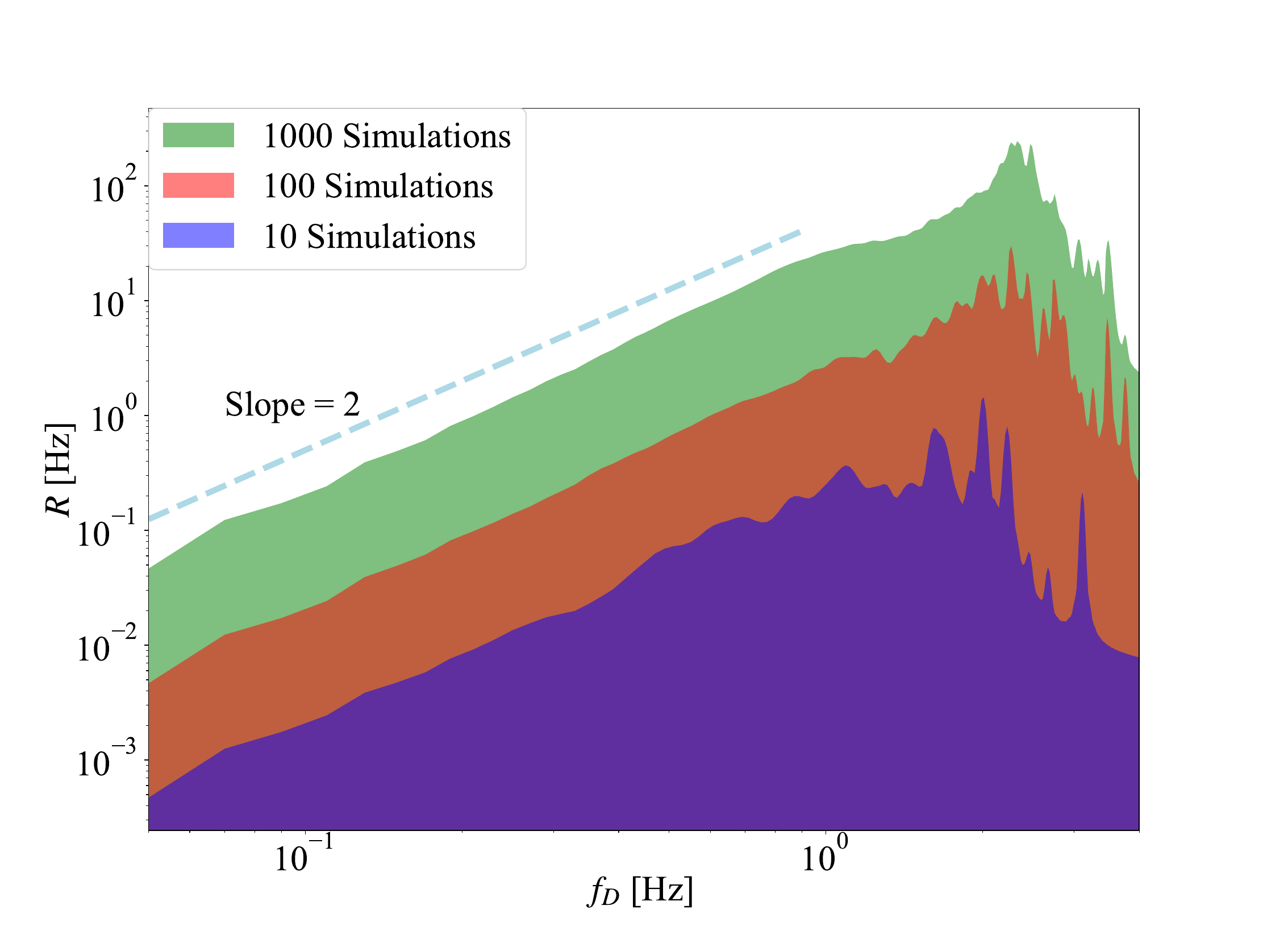}
\caption{Response curves for polydisperse systems. In purple we added the typical response curves of 10 different simulations, in orange for 100 simulations and in green for 1000. On each run we simulate 50 bubbles with random sizes between 0.5 and 3.5 m. The integration time was sufficient for the pressure signal to travel back and forth once. We see that the maximum is located approximately at the average resonant frequency of the bubbles, $f=2.4$ Hz . On the left side, we appreciate the power law with exponent 2 for the response at low frequencies, which is addressed in the main text. 
}\label{fig:1_fig12}
\end{figure}

Figure \ref{fig:1_fig12} shows the added response of $10$, $100$ and $1,000$ realizations of the simulation plotted in purple, orange and green colors respectively. This means, for example, that we added the response of 10 different simulations in order to obtain the purple plot. The responses were not normalized in order to avoid overlapping between curves. We notice that the case of $10$ simulations still presents pronounced random peaks, as were also present in Fig.~\ref{fig:1_fig11}. As we added more responses, the peaks smooth out until there are almost no discernible peaks in the total response, like in the green curve. As a result, it is easier to observe the power-law behavior with an exponent close to two for small driving frequencies in Fig. \ref{fig:1_fig12} as compared to Fig. \ref{fig:1_fig11}. This is due to the smoothing that occurs when adding multiple random responses.

\section{Concluding remarks}\label{sec:1_5}

We studied the transmission of a pressure signal through a confined bubble array. Generally, we found the system to be unexpectedly rich, where much of the richness appears to be already present in the linearized system consisting of coupled harmonic oscillators, a system that we wrongly believed to be fully explored and understood.

More specifically, we simulated a three-dimensional axisymmetrical cylinder filled with arranged monodisperse bubbles using a Boundary Integral (BI) code. We were able to match the bubble volume change signal with the result of the Multiple Bagnold Problem (MBP) model. A good agreement was obtained for initial times, when the system behaves quasi-linearly, and qualitatively good agreement for longer times. We were able to identify each of the peaks arising at the power spectrum for this monodisperse case for both BI and MBP simulations. The first peak corresponds to the driving frequency and the consecutive ones emerge at the eigenfrequencies of the system. We derived an equation that predicts the value of the eigenfrequencies with excellent agreement with the MBP peaks and also good agreement with the BI peaks up to the fifth eiqenfrequency. 

Using the MBP model, an expression for the sound speed was obtained by analyzing an infinite monodisperse system. Two regimes where found: When the driving frequency is smaller than $\sqrt{2}$ times 
the 
natural frequency of a single bubble, the pressure signal is transmitted without attenuation and with a driving-frequency-dependent phase shift. When the driving frequency is however higher than $\sqrt{2}$ times the bubble frequency, the pressure wave is attenuated and the phase shift becomes constant and equal to $\pi$.

We explored the effect of a range of driving frequencies on diverse MBP systems. When the bubbles are monodisperse, the response exhibits a number of peaks equal to the number of bubbles minus one, arising exactly at the eigenfrequencies predicted by Eq.~\eqref{eq:1_17}. The peaks arise around the bubble natural frequency. When the number of bubbles is odd, the middle eigenfrequency matches the bubble natural frequency and the response is enhanced. This corresponds to the largest peak in the frequency spectrum. When the number of bubbles is even, there is no peak at the bubble frequency but the two largest peaks emerge to the left and right of it. 

Subsequently, two different types of bidisperse systems were studied, called \textit{alternating} and \textit{stacked}, respectively. The alternating big-small and small-big system responses show peaks in two groups, corresponding to the acoustic and optic branches of Eq.~\eqref{eq:1_19} respectively. The big-small system shows suppression of the response in between the branches. Qualitatively, this happens due to the first big bubble acting as a filter that attenuates driving frequencies much higher than its natural one. Once $f_{D}$ is close to the natural frequency of the small bubble, the optic branch resonates. Unlike the big-small system, the alternating small-big system exhibits the biggest response at the natural frequency of the pseudo-two bubble system, which lies between the two branches. This is again a consequence of the position of the big bubble. Since the first bubble is small, the signal is transmitted non-attenuated to the second bubble, but since this one is big, it attenuates the signal before it reaches the third bubble.

The small-...-big stacked setup, where the small bubbles are closest to the driving piston, displays peaks over the entire driving frequency range, with narrower peaks near the big bubble's resonant frequency and wider peaks near the small bubble's resonant frequency with no suppression of the response for higher frequencies. Big and small bubble responses do not arise independently but are coupled due to the interface connecting both media. Thus, their eigenfrequencies do not coincide with their monodisperse counterparts; it is possible nevertheless to predict them by solving the corresponding eigenvalue problem numerically. The obtained eigenfrequencies exactly match the peaks observed in the response curve. The big-...-small system response only arises close to the big bubble resonant frequency, since big bubbles do not allow pressure signals with frequencies in their attenuation regime i.e., larger than $\sqrt{2}f_b$) to propagate through the system. The eigenfrequencies of the big-...-small system coincide with those of the small-...-big one, until the above suppression becomes dominant. Bidisperse stacked systems behave like two media connected by an interface. As such, we were able to deduce transmission and reflection coefficients that, once attenuation is taken into account, decay exponentially towards zero when the driving frequency is in the attenuation regime.

Next, we simulated the effect of an impulsive driving on a monodisperse system. The power spectrum, in addition to small peaks coming from the eigenvalue spectrum, shows a strong resonance peak of the system as a whole, the frequency of which corresponds to the pressure pulse moving back and forth between piston and wall, including almost as strong higher harmonics corresponding to this frequency. The relative bubble volume signal shows the chain reaction of bubble deformation and provides a graphical representation of the pressure signal traveling back and forth through the whole setup. 

The last set of simulations corresponds to random polydisperse systems. As more responses from different runs are added together, the peaks in the total response get smoothed until no high outlier peak is visible. For small driving frequencies, the total response presents a power law with exponent close to two, which becomes more evident as more responses from different random realizations are summed.

Finally, we come to the relevance of this study for possible applications, even if we have only been scratching the surface of the full complexity of the propagation of a pressure pulse through an inhomogeneous bubbly liquid, and therefore the following is necessarily speculative. Turning first to impact on a bubbly liquid, as may happen in an LNG container, it is clear that the frequencies corresponding to the system as a whole and its higher harmonics is likely to be dominant in the response, as we have observed in the impulsively driven system. Nevertheless, segregation of bubble sizes is likely to happen in such a system, e.g., due to the effect of buoyancy as a result of which larger bubbles will rise faster to the surface. Such a region of larger bubbles may act as a reflector for pressure waves similar to what was observed in the bidisperse stacked case that we studied. In any case, presence of bubbles in an impacting sloshing wave may be strongly affected by bubbles and the resulting decrease of the local speed of sound in the liquid.

Another area where our research may be of relevance is that of contrast agents used in ultrasound imaging, where usually one aims at creating a large response by selecting the bubbles to be as monodisperse as possible. From our results it is becoming clear that if multiple bubbles are confined to a relatively small area in space (e.g., as would be the case in an artery) they may in fact have multiple eigenmodes close to the expected resonant frequency. On the one hand, this may deteriorate the expected response, but on the other the presence of additional peaks in the spectrum may also provide information about the local structure of the object of interest that would not be directly accessible.

A final application may be found in the area of bubble curtains, where a jet of bubbles of finite small thickness is used to shield the sea fauna from a noisy event such as the creation of a foundation for a marine wind park. A theory like the one presented here could be used for predicting the attenuation of sound within the bubble curtain.

\begin{acknowledgements}
This publication is part of the Vici project IMBOL (Project No. 17070) which is partly financed by the Dutch Research Council (NWO). 
\end{acknowledgements}

\nocite{*}
\bibliographystyle{apsrev4-2}
\bibliography{TPSCBAbib}

\providecommand{\noopsort}[1]{}\providecommand{\singleletter}[1]{#1}%
\begin{thebibliography}{34}%
\makeatletter
\providecommand \@ifxundefined [1]{%
 \@ifx{#1\undefined}
}%
\providecommand \@ifnum [1]{%
 \ifnum #1\expandafter \@firstoftwo
 \else \expandafter \@secondoftwo
 \fi
}%
\providecommand \@ifx [1]{%
 \ifx #1\expandafter \@firstoftwo
 \else \expandafter \@secondoftwo
 \fi
}%
\providecommand \natexlab [1]{#1}%
\providecommand \enquote  [1]{``#1''}%
\providecommand \bibnamefont  [1]{#1}%
\providecommand \bibfnamefont [1]{#1}%
\providecommand \citenamefont [1]{#1}%
\providecommand \href@noop [0]{\@secondoftwo}%
\providecommand \href [0]{\begingroup \@sanitize@url \@href}%
\providecommand \@href[1]{\@@startlink{#1}\@@href}%
\providecommand \@@href[1]{\endgroup#1\@@endlink}%
\providecommand \@sanitize@url [0]{\catcode `\\12\catcode `\$12\catcode
  `\&12\catcode `\#12\catcode `\^12\catcode `\_12\catcode `\%12\relax}%
\providecommand \@@startlink[1]{}%
\providecommand \@@endlink[0]{}%
\providecommand \url  [0]{\begingroup\@sanitize@url \@url }%
\providecommand \@url [1]{\endgroup\@href {#1}{\urlprefix }}%
\providecommand \urlprefix  [0]{URL }%
\providecommand \Eprint [0]{\href }%
\providecommand \doibase [0]{https://doi.org/}%
\providecommand \selectlanguage [0]{\@gobble}%
\providecommand \bibinfo  [0]{\@secondoftwo}%
\providecommand \bibfield  [0]{\@secondoftwo}%
\providecommand \translation [1]{[#1]}%
\providecommand \BibitemOpen [0]{}%
\providecommand \bibitemStop [0]{}%
\providecommand \bibitemNoStop [0]{.\EOS\space}%
\providecommand \EOS [0]{\spacefactor3000\relax}%
\providecommand \BibitemShut  [1]{\csname bibitem#1\endcsname}%
\let\auto@bib@innerbib\@empty
\bibitem [{\citenamefont {Kieffer}(1977)}]{kieffer1977sound}%
  \BibitemOpen
  \bibfield  {author} {\bibinfo {author} {\bibfnamefont {S.~W.}\ \bibnamefont
  {Kieffer}},\ }\href@noop {} {\bibfield  {journal} {\bibinfo  {journal}
  {Journal of Geophysical research}\ }\textbf {\bibinfo {volume} {82}},\
  \bibinfo {pages} {2895} (\bibinfo {year} {1977})}\BibitemShut {NoStop}%
\bibitem [{\citenamefont {Kundu}\ \emph {et~al.}(2016)\citenamefont {Kundu},
  \citenamefont {Cohen},\ and\ \citenamefont {Dowling}}]{kundu1990fluid}%
  \BibitemOpen
  \bibfield  {author} {\bibinfo {author} {\bibfnamefont {P.~K.}\ \bibnamefont
  {Kundu}}, \bibinfo {author} {\bibfnamefont {I.~M.}\ \bibnamefont {Cohen}},\
  and\ \bibinfo {author} {\bibfnamefont {D.~R.}\ \bibnamefont {Dowling}},\
  }\href@noop {} {\emph {\bibinfo {title} {Fluid mechanics}}}\ (\bibinfo
  {publisher} {Academic press San Diego},\ \bibinfo {year} {2016})\BibitemShut
  {NoStop}%
\bibitem [{\citenamefont {Pozrikidis}(2016)}]{pozrikidis2016fluid}%
  \BibitemOpen
  \bibfield  {author} {\bibinfo {author} {\bibfnamefont {C.}~\bibnamefont
  {Pozrikidis}},\ }\href@noop {} {\emph {\bibinfo {title} {Fluid dynamics:
  theory, computation, and numerical simulation}}}\ (\bibinfo  {publisher}
  {Springer},\ \bibinfo {year} {2016})\BibitemShut {NoStop}%
\bibitem [{\citenamefont {Rayleigh}(1917)}]{rayleigh1917viii}%
  \BibitemOpen
  \bibfield  {author} {\bibinfo {author} {\bibfnamefont {L.}~\bibnamefont
  {Rayleigh}},\ }\href@noop {} {\bibfield  {journal} {\bibinfo  {journal} {The
  London, Edinburgh, and Dublin Philosophical Magazine and Journal of Science}\
  }\textbf {\bibinfo {volume} {34}},\ \bibinfo {pages} {94} (\bibinfo {year}
  {1917})}\BibitemShut {NoStop}%
\bibitem [{\citenamefont {Minnaert}(1933)}]{minnaert1933xvi}%
  \BibitemOpen
  \bibfield  {author} {\bibinfo {author} {\bibfnamefont {M.}~\bibnamefont
  {Minnaert}},\ }\href@noop {} {\bibfield  {journal} {\bibinfo  {journal} {The
  London, Edinburgh, and Dublin Philosophical Magazine and Journal of Science}\
  }\textbf {\bibinfo {volume} {16}},\ \bibinfo {pages} {235} (\bibinfo {year}
  {1933})}\BibitemShut {NoStop}%
\bibitem [{\citenamefont {Plesset}(1949)}]{plesset1949dynamics}%
  \BibitemOpen
  \bibfield  {author} {\bibinfo {author} {\bibfnamefont {M.~S.}\ \bibnamefont
  {Plesset}},\ }\href@noop {} {\bibfield  {journal} {\bibinfo  {journal}
  {Journal of applied mechanics}\ }\textbf {\bibinfo {volume} {16}},\ \bibinfo
  {pages} {277} (\bibinfo {year} {1949})}\BibitemShut {NoStop}%
\bibitem [{\citenamefont {Wijngaarden}(1972)}]{wijngaarden1972one}%
  \BibitemOpen
  \bibfield  {author} {\bibinfo {author} {\bibfnamefont {L.~v.}\ \bibnamefont
  {Wijngaarden}},\ }\href@noop {} {\bibfield  {journal} {\bibinfo  {journal}
  {Annual review of fluid Mechanics}\ }\textbf {\bibinfo {volume} {4}},\
  \bibinfo {pages} {369} (\bibinfo {year} {1972})}\BibitemShut {NoStop}%
\bibitem [{\citenamefont {Plesset}\ and\ \citenamefont
  {Prosperetti}(1977)}]{plesset1977bubble}%
  \BibitemOpen
  \bibfield  {author} {\bibinfo {author} {\bibfnamefont {M.~S.}\ \bibnamefont
  {Plesset}}\ and\ \bibinfo {author} {\bibfnamefont {A.}~\bibnamefont
  {Prosperetti}},\ }\href@noop {} {\bibfield  {journal} {\bibinfo  {journal}
  {Annual review of fluid mechanics}\ }\textbf {\bibinfo {volume} {9}},\
  \bibinfo {pages} {145} (\bibinfo {year} {1977})}\BibitemShut {NoStop}%
\bibitem [{\citenamefont {Prosperetti}\ \emph {et~al.}(1986)\citenamefont
  {Prosperetti}, \citenamefont {Lezzi} \emph {et~al.}}]{prosperetti1986bubble}%
  \BibitemOpen
  \bibfield  {author} {\bibinfo {author} {\bibfnamefont {A.}~\bibnamefont
  {Prosperetti}}, \bibinfo {author} {\bibfnamefont {A.}~\bibnamefont {Lezzi}},
  \emph {et~al.},\ }\href@noop {} {\bibfield  {journal} {\bibinfo  {journal}
  {J. FLUID MECH., 1986,}\ }\textbf {\bibinfo {volume} {168}},\ \bibinfo
  {pages} {457} (\bibinfo {year} {1986})}\BibitemShut {NoStop}%
\bibitem [{\citenamefont {Brennen}(2014)}]{brennen2014cavitation}%
  \BibitemOpen
  \bibfield  {author} {\bibinfo {author} {\bibfnamefont {C.~E.}\ \bibnamefont
  {Brennen}},\ }\href@noop {} {\emph {\bibinfo {title} {Cavitation and bubble
  dynamics}}}\ (\bibinfo  {publisher} {Cambridge University Press},\ \bibinfo
  {year} {2014})\BibitemShut {NoStop}%
\bibitem [{\citenamefont {Leighton}(2012)}]{leighton2012acoustic}%
  \BibitemOpen
  \bibfield  {author} {\bibinfo {author} {\bibfnamefont {T.}~\bibnamefont
  {Leighton}},\ }\href@noop {} {\emph {\bibinfo {title} {The acoustic
  bubble}}}\ (\bibinfo  {publisher} {Academic press},\ \bibinfo {year}
  {2012})\BibitemShut {NoStop}%
\bibitem [{\citenamefont {Smereka}\ and\ \citenamefont
  {Banerjee}(1988)}]{smereka1988dynamics}%
  \BibitemOpen
  \bibfield  {author} {\bibinfo {author} {\bibfnamefont {P.}~\bibnamefont
  {Smereka}}\ and\ \bibinfo {author} {\bibfnamefont {S.}~\bibnamefont
  {Banerjee}},\ }\href@noop {} {\bibfield  {journal} {\bibinfo  {journal} {The
  Physics of fluids}\ }\textbf {\bibinfo {volume} {31}},\ \bibinfo {pages}
  {3519} (\bibinfo {year} {1988})}\BibitemShut {NoStop}%
\bibitem [{\citenamefont {d'Agostino}\ and\ \citenamefont
  {Brennen}(1989)}]{d1989linearized}%
  \BibitemOpen
  \bibfield  {author} {\bibinfo {author} {\bibfnamefont {L.}~\bibnamefont
  {d'Agostino}}\ and\ \bibinfo {author} {\bibfnamefont {C.~E.}\ \bibnamefont
  {Brennen}},\ }\href@noop {} {\bibfield  {journal} {\bibinfo  {journal}
  {Journal of Fluid Mechanics}\ }\textbf {\bibinfo {volume} {199}},\ \bibinfo
  {pages} {155} (\bibinfo {year} {1989})}\BibitemShut {NoStop}%
\bibitem [{\citenamefont {Ida}(2004)}]{ida2004investigation}%
  \BibitemOpen
  \bibfield  {author} {\bibinfo {author} {\bibfnamefont {M.}~\bibnamefont
  {Ida}},\ }\href@noop {} {\bibfield  {journal} {\bibinfo  {journal} {Journal
  of the Physical Society of Japan}\ }\textbf {\bibinfo {volume} {73}},\
  \bibinfo {pages} {3026} (\bibinfo {year} {2004})}\BibitemShut {NoStop}%
\bibitem [{\citenamefont {Ida}(2005)}]{ida2005avoided}%
  \BibitemOpen
  \bibfield  {author} {\bibinfo {author} {\bibfnamefont {M.}~\bibnamefont
  {Ida}},\ }\href@noop {} {\bibfield  {journal} {\bibinfo  {journal} {Physical
  Review E}\ }\textbf {\bibinfo {volume} {72}},\ \bibinfo {pages} {036306}
  (\bibinfo {year} {2005})}\BibitemShut {NoStop}%
\bibitem [{\citenamefont {Zeravcic}\ \emph {et~al.}(2011)\citenamefont
  {Zeravcic}, \citenamefont {Lohse},\ and\ \citenamefont
  {Van~Saarloos}}]{zeravcic2011collective}%
  \BibitemOpen
  \bibfield  {author} {\bibinfo {author} {\bibfnamefont {Z.}~\bibnamefont
  {Zeravcic}}, \bibinfo {author} {\bibfnamefont {D.}~\bibnamefont {Lohse}},\
  and\ \bibinfo {author} {\bibfnamefont {W.}~\bibnamefont {Van~Saarloos}},\
  }\href@noop {} {\bibfield  {journal} {\bibinfo  {journal} {Journal of fluid
  mechanics}\ }\textbf {\bibinfo {volume} {680}},\ \bibinfo {pages} {114}
  (\bibinfo {year} {2011})}\BibitemShut {NoStop}%
\bibitem [{\citenamefont {Bagnold}(1939)}]{bagnold1939interim}%
  \BibitemOpen
  \bibfield  {author} {\bibinfo {author} {\bibfnamefont {R.}~\bibnamefont
  {Bagnold}},\ }\href@noop {} {\bibfield  {journal} {\bibinfo  {journal}
  {Journal of the Institution of Civil Engineers}\ }\textbf {\bibinfo {volume}
  {12}},\ \bibinfo {pages} {202} (\bibinfo {year} {1939})}\BibitemShut
  {NoStop}%
\bibitem [{\citenamefont {Braeunig}\ \emph {et~al.}(2010)\citenamefont
  {Braeunig}, \citenamefont {Brosset}, \citenamefont {Dias}, \citenamefont
  {Ghidaglia} \emph {et~al.}}]{braeunig2010effect}%
  \BibitemOpen
  \bibfield  {author} {\bibinfo {author} {\bibfnamefont {J.-P.}\ \bibnamefont
  {Braeunig}}, \bibinfo {author} {\bibfnamefont {L.}~\bibnamefont {Brosset}},
  \bibinfo {author} {\bibfnamefont {F.}~\bibnamefont {Dias}}, \bibinfo {author}
  {\bibfnamefont {J.-M.}\ \bibnamefont {Ghidaglia}}, \emph {et~al.},\ }in\
  \href@noop {} {\emph {\bibinfo {booktitle} {The Twentieth International
  Offshore and Polar Engineering Conference}}}\ (\bibinfo {organization}
  {International Society of Offshore and Polar Engineers},\ \bibinfo {year}
  {2010})\BibitemShut {NoStop}%
\bibitem [{\citenamefont {Ancellin}\ \emph {et~al.}(2012)\citenamefont
  {Ancellin}, \citenamefont {Ghidaglia}, \citenamefont {Brosset} \emph
  {et~al.}}]{ancellin2012influence}%
  \BibitemOpen
  \bibfield  {author} {\bibinfo {author} {\bibfnamefont {M.}~\bibnamefont
  {Ancellin}}, \bibinfo {author} {\bibfnamefont {J.-M.}\ \bibnamefont
  {Ghidaglia}}, \bibinfo {author} {\bibfnamefont {L.}~\bibnamefont {Brosset}},
  \emph {et~al.},\ }in\ \href@noop {} {\emph {\bibinfo {booktitle} {The
  Twenty-second International Offshore and Polar Engineering Conference}}}\
  (\bibinfo {organization} {International Society of Offshore and Polar
  Engineers},\ \bibinfo {year} {2012})\BibitemShut {NoStop}%
\bibitem [{\citenamefont {Oguz}\ and\ \citenamefont
  {Prosperetti}(1993)}]{oguz1993dynamics}%
  \BibitemOpen
  \bibfield  {author} {\bibinfo {author} {\bibfnamefont {H.~N.}\ \bibnamefont
  {Oguz}}\ and\ \bibinfo {author} {\bibfnamefont {A.}~\bibnamefont
  {Prosperetti}},\ }\href@noop {} {\bibfield  {journal} {\bibinfo  {journal}
  {Journal of Fluid Mechanics}\ }\textbf {\bibinfo {volume} {257}},\ \bibinfo
  {pages} {111} (\bibinfo {year} {1993})}\BibitemShut {NoStop}%
\bibitem [{\citenamefont {Bergmann}\ \emph {et~al.}(2009)\citenamefont
  {Bergmann}, \citenamefont {Van Der~Meer}, \citenamefont {Gekle},
  \citenamefont {Van Der~Bos},\ and\ \citenamefont
  {Lohse}}]{bergmann2009controlled}%
  \BibitemOpen
  \bibfield  {author} {\bibinfo {author} {\bibfnamefont {R.}~\bibnamefont
  {Bergmann}}, \bibinfo {author} {\bibfnamefont {D.}~\bibnamefont {Van
  Der~Meer}}, \bibinfo {author} {\bibfnamefont {S.}~\bibnamefont {Gekle}},
  \bibinfo {author} {\bibfnamefont {A.}~\bibnamefont {Van Der~Bos}},\ and\
  \bibinfo {author} {\bibfnamefont {D.}~\bibnamefont {Lohse}},\ }\href@noop {}
  {\bibfield  {journal} {\bibinfo  {journal} {Journal of Fluid Mechanics}\
  }\textbf {\bibinfo {volume} {633}},\ \bibinfo {pages} {381} (\bibinfo {year}
  {2009})}\BibitemShut {NoStop}%
\bibitem [{\citenamefont {Gekle}\ \emph {et~al.}(2009)\citenamefont {Gekle},
  \citenamefont {Snoeijer}, \citenamefont {Lohse},\ and\ \citenamefont {van~der
  Meer}}]{gekle2009approach}%
  \BibitemOpen
  \bibfield  {author} {\bibinfo {author} {\bibfnamefont {S.}~\bibnamefont
  {Gekle}}, \bibinfo {author} {\bibfnamefont {J.~H.}\ \bibnamefont {Snoeijer}},
  \bibinfo {author} {\bibfnamefont {D.}~\bibnamefont {Lohse}},\ and\ \bibinfo
  {author} {\bibfnamefont {D.}~\bibnamefont {van~der Meer}},\ }\href@noop {}
  {\bibfield  {journal} {\bibinfo  {journal} {Physical Review E}\ }\textbf
  {\bibinfo {volume} {80}},\ \bibinfo {pages} {036305} (\bibinfo {year}
  {2009})}\BibitemShut {NoStop}%
\bibitem [{\citenamefont {Gekle}\ and\ \citenamefont
  {Gordillo}(2011)}]{gekle2011compressible}%
  \BibitemOpen
  \bibfield  {author} {\bibinfo {author} {\bibfnamefont {S.}~\bibnamefont
  {Gekle}}\ and\ \bibinfo {author} {\bibfnamefont {J.~M.}\ \bibnamefont
  {Gordillo}},\ }\href@noop {} {\bibfield  {journal} {\bibinfo  {journal}
  {International journal for numerical methods in fluids}\ }\textbf {\bibinfo
  {volume} {67}},\ \bibinfo {pages} {1456} (\bibinfo {year}
  {2011})}\BibitemShut {NoStop}%
\bibitem [{\citenamefont {Bouwhuis}\ \emph {et~al.}(2012)\citenamefont
  {Bouwhuis}, \citenamefont {van~der Veen}, \citenamefont {Tran}, \citenamefont
  {Keij}, \citenamefont {Winkels}, \citenamefont {Peters}, \citenamefont
  {van~der Meer}, \citenamefont {Sun}, \citenamefont {Snoeijer},\ and\
  \citenamefont {Lohse}}]{bouwhuis2012maximal}%
  \BibitemOpen
  \bibfield  {author} {\bibinfo {author} {\bibfnamefont {W.}~\bibnamefont
  {Bouwhuis}}, \bibinfo {author} {\bibfnamefont {R.~C.}\ \bibnamefont {van~der
  Veen}}, \bibinfo {author} {\bibfnamefont {T.}~\bibnamefont {Tran}}, \bibinfo
  {author} {\bibfnamefont {D.~L.}\ \bibnamefont {Keij}}, \bibinfo {author}
  {\bibfnamefont {K.~G.}\ \bibnamefont {Winkels}}, \bibinfo {author}
  {\bibfnamefont {I.~R.}\ \bibnamefont {Peters}}, \bibinfo {author}
  {\bibfnamefont {D.}~\bibnamefont {van~der Meer}}, \bibinfo {author}
  {\bibfnamefont {C.}~\bibnamefont {Sun}}, \bibinfo {author} {\bibfnamefont
  {J.~H.}\ \bibnamefont {Snoeijer}},\ and\ \bibinfo {author} {\bibfnamefont
  {D.}~\bibnamefont {Lohse}},\ }\href@noop {} {\bibfield  {journal} {\bibinfo
  {journal} {Physical review letters}\ }\textbf {\bibinfo {volume} {109}},\
  \bibinfo {pages} {264501} (\bibinfo {year} {2012})}\BibitemShut {NoStop}%
\bibitem [{\citenamefont {Bouwhuis}\ \emph {et~al.}(2013)\citenamefont
  {Bouwhuis}, \citenamefont {Winkels}, \citenamefont {Peters}, \citenamefont
  {Brunet}, \citenamefont {van~der Meer},\ and\ \citenamefont
  {Snoeijer}}]{bouwhuis2013oscillating}%
  \BibitemOpen
  \bibfield  {author} {\bibinfo {author} {\bibfnamefont {W.}~\bibnamefont
  {Bouwhuis}}, \bibinfo {author} {\bibfnamefont {K.~G.}\ \bibnamefont
  {Winkels}}, \bibinfo {author} {\bibfnamefont {I.~R.}\ \bibnamefont {Peters}},
  \bibinfo {author} {\bibfnamefont {P.}~\bibnamefont {Brunet}}, \bibinfo
  {author} {\bibfnamefont {D.}~\bibnamefont {van~der Meer}},\ and\ \bibinfo
  {author} {\bibfnamefont {J.~H.}\ \bibnamefont {Snoeijer}},\ }\href@noop {}
  {\bibfield  {journal} {\bibinfo  {journal} {Physical Review E}\ }\textbf
  {\bibinfo {volume} {88}},\ \bibinfo {pages} {023017} (\bibinfo {year}
  {2013})}\BibitemShut {NoStop}%
\bibitem [{\citenamefont {Hendrix}\ \emph {et~al.}(2015)\citenamefont
  {Hendrix}, \citenamefont {Bouwhuis}, \citenamefont {van~der Meer},
  \citenamefont {Lohse},\ and\ \citenamefont
  {Snoeijer}}]{hendrix2015universal}%
  \BibitemOpen
  \bibfield  {author} {\bibinfo {author} {\bibfnamefont {M.~H.}\ \bibnamefont
  {Hendrix}}, \bibinfo {author} {\bibfnamefont {W.}~\bibnamefont {Bouwhuis}},
  \bibinfo {author} {\bibfnamefont {D.}~\bibnamefont {van~der Meer}}, \bibinfo
  {author} {\bibfnamefont {D.}~\bibnamefont {Lohse}},\ and\ \bibinfo {author}
  {\bibfnamefont {J.~H.}\ \bibnamefont {Snoeijer}},\ }\href@noop {} {\bibfield
  {journal} {\bibinfo  {journal} {arXiv preprint arXiv:1502.02869}\ } (\bibinfo
  {year} {2015})}\BibitemShut {NoStop}%
\bibitem [{\citenamefont {Van~der Meer}\ \emph {et~al.}(2007)\citenamefont
  {Van~der Meer}, \citenamefont {Dollet}, \citenamefont {Voormolen},
  \citenamefont {Chin}, \citenamefont {Bouakaz}, \citenamefont {de~Jong},
  \citenamefont {Versluis},\ and\ \citenamefont {Lohse}}]{van2007microbubble}%
  \BibitemOpen
  \bibfield  {author} {\bibinfo {author} {\bibfnamefont {S.~M.}\ \bibnamefont
  {Van~der Meer}}, \bibinfo {author} {\bibfnamefont {B.}~\bibnamefont
  {Dollet}}, \bibinfo {author} {\bibfnamefont {M.~M.}\ \bibnamefont
  {Voormolen}}, \bibinfo {author} {\bibfnamefont {C.~T.}\ \bibnamefont {Chin}},
  \bibinfo {author} {\bibfnamefont {A.}~\bibnamefont {Bouakaz}}, \bibinfo
  {author} {\bibfnamefont {N.}~\bibnamefont {de~Jong}}, \bibinfo {author}
  {\bibfnamefont {M.}~\bibnamefont {Versluis}},\ and\ \bibinfo {author}
  {\bibfnamefont {D.}~\bibnamefont {Lohse}},\ }\href@noop {} {\bibfield
  {journal} {\bibinfo  {journal} {The Journal of the Acoustical Society of
  America}\ }\textbf {\bibinfo {volume} {121}},\ \bibinfo {pages} {648}
  (\bibinfo {year} {2007})}\BibitemShut {NoStop}%
\bibitem [{\citenamefont {Versluis}\ \emph {et~al.}(2010)\citenamefont
  {Versluis}, \citenamefont {Goertz}, \citenamefont {Palanchon}, \citenamefont
  {Heitman}, \citenamefont {van~der Meer}, \citenamefont {Dollet},
  \citenamefont {de~Jong},\ and\ \citenamefont
  {Lohse}}]{versluis2010microbubble}%
  \BibitemOpen
  \bibfield  {author} {\bibinfo {author} {\bibfnamefont {M.}~\bibnamefont
  {Versluis}}, \bibinfo {author} {\bibfnamefont {D.~E.}\ \bibnamefont
  {Goertz}}, \bibinfo {author} {\bibfnamefont {P.}~\bibnamefont {Palanchon}},
  \bibinfo {author} {\bibfnamefont {I.~L.}\ \bibnamefont {Heitman}}, \bibinfo
  {author} {\bibfnamefont {S.~M.}\ \bibnamefont {van~der Meer}}, \bibinfo
  {author} {\bibfnamefont {B.}~\bibnamefont {Dollet}}, \bibinfo {author}
  {\bibfnamefont {N.}~\bibnamefont {de~Jong}},\ and\ \bibinfo {author}
  {\bibfnamefont {D.}~\bibnamefont {Lohse}},\ }\href@noop {} {\bibfield
  {journal} {\bibinfo  {journal} {Physical review E}\ }\textbf {\bibinfo
  {volume} {82}},\ \bibinfo {pages} {026321} (\bibinfo {year}
  {2010})}\BibitemShut {NoStop}%
\bibitem [{\citenamefont {Florencio~Jr}\ and\ \citenamefont
  {Lee}(1985)}]{florencio1985exact}%
  \BibitemOpen
  \bibfield  {author} {\bibinfo {author} {\bibfnamefont {J.}~\bibnamefont
  {Florencio~Jr}}\ and\ \bibinfo {author} {\bibfnamefont {M.~H.}\ \bibnamefont
  {Lee}},\ }\href@noop {} {\bibfield  {journal} {\bibinfo  {journal} {Physical
  Review A}\ }\textbf {\bibinfo {volume} {31}},\ \bibinfo {pages} {3231}
  (\bibinfo {year} {1985})}\BibitemShut {NoStop}%
\bibitem [{\citenamefont {Taghizadeh}\ \emph {et~al.}(2021)\citenamefont
  {Taghizadeh}, \citenamefont {Shrivastava},\ and\ \citenamefont
  {Luding}}]{taghizadeh2021stochastic}%
  \BibitemOpen
  \bibfield  {author} {\bibinfo {author} {\bibfnamefont {K.}~\bibnamefont
  {Taghizadeh}}, \bibinfo {author} {\bibfnamefont {R.~K.}\ \bibnamefont
  {Shrivastava}},\ and\ \bibinfo {author} {\bibfnamefont {S.}~\bibnamefont
  {Luding}},\ }\href@noop {} {\bibfield  {journal} {\bibinfo  {journal}
  {Materials}\ }\textbf {\bibinfo {volume} {14}},\ \bibinfo {pages} {1815}
  (\bibinfo {year} {2021})}\BibitemShut {NoStop}%
\bibitem [{Note1()}]{Note1}%
  \BibitemOpen
  \bibinfo {note} {Note that bubbles are made large to minimize the influence
  of surface tension, which is present in the BI simulations for stability
  reasons and set to the value of air-water. With the exception of surface
  tension, all results are scalable.}\BibitemShut {Stop}%
\bibitem [{\citenamefont {Ashcroft}\ \emph {et~al.}(1976)\citenamefont
  {Ashcroft}, \citenamefont {Mermin} \emph {et~al.}}]{ashcroft1976solid}%
  \BibitemOpen
  \bibfield  {author} {\bibinfo {author} {\bibfnamefont {N.~W.}\ \bibnamefont
  {Ashcroft}}, \bibinfo {author} {\bibfnamefont {N.~D.}\ \bibnamefont
  {Mermin}}, \emph {et~al.},\ }\href@noop {} {\bibinfo {title} {Solid state
  physics}} (\bibinfo {year} {1976})\BibitemShut {NoStop}%
\bibitem [{\citenamefont {Chapman}\ and\ \citenamefont
  {Ward}(1990)}]{chapman1990normal}%
  \BibitemOpen
  \bibfield  {author} {\bibinfo {author} {\bibfnamefont {D.~M.}\ \bibnamefont
  {Chapman}}\ and\ \bibinfo {author} {\bibfnamefont {P.~D.}\ \bibnamefont
  {Ward}},\ }\href@noop {} {\bibfield  {journal} {\bibinfo  {journal} {The
  Journal of the Acoustical Society of America}\ }\textbf {\bibinfo {volume}
  {87}},\ \bibinfo {pages} {601} (\bibinfo {year} {1990})}\BibitemShut
  {NoStop}%
\bibitem [{\citenamefont {Guardiola}\ and\ \citenamefont
  {Ros}(1982)}]{guardiola1982numerical}%
  \BibitemOpen
  \bibfield  {author} {\bibinfo {author} {\bibfnamefont {R.}~\bibnamefont
  {Guardiola}}\ and\ \bibinfo {author} {\bibfnamefont {J.}~\bibnamefont
  {Ros}},\ }\href@noop {} {\bibfield  {journal} {\bibinfo  {journal} {Journal
  of Computational Physics}\ }\textbf {\bibinfo {volume} {45}},\ \bibinfo
  {pages} {374} (\bibinfo {year} {1982})}\BibitemShut {NoStop}%
\end{thebibliography}%

\appendix
\section{Analysis of the linearized infinite MBP model}\label{1_A_1}
The linearized Multiple Bagnold model in an infinite chain of identical bubbles is described by the following set of equations: 
\begin{equation}
\ddot{\epsilon}_k + \omega_0^2\epsilon_k = \tfrac{1}{2}\omega_0^2\left(\epsilon_{k-1} + \epsilon_{k+1}\right)\qquad\textrm{for}\,\,k \in \mathbb{Z}\,,
\label{eq:1_33}
\end{equation}
with $\omega_0^2 = 2 \gamma P_0/(\rho_l \ell \delta x_0)$.

Now, if the system is excited with a certain driving frequency $\omega_D$, then (after transients have died out) the response at each position $k$ is expected to happen with the same frequency, and can be written as
\begin{equation}
\epsilon_k(t) = A_k e^{i(\omega_D t + \phi_k)}\qquad\textrm{for}\,\,k \in \mathbb{Z}\,,
\label{eq:1_34}
\end{equation}
where $A_k$ and $\phi_k$ are the (real) amplitude and phase at position $k$. Inserting Eqs. \ref{eq:1_34} in \ref{eq:1_33}, we obtain 
\begin{equation}
\begin{split}
& (-\omega_D^2+\omega_0^2)A_k e^{\left[i(\omega_Dt + \phi_k)\right]} = \\ & \tfrac{1}{2}\omega_0^2\left(A_{k+1} e^{\left[i(\omega_Dt + \phi_{k+1})\right]} + A_{k-1} e^{\left[i(\omega_Dt + \phi_{k-1})\right]}\right)\,,
\end{split}
\end{equation}
or, dividing by $A_k \exp\left[i(\omega_Dt + \phi_k)\right]$:
\begin{equation}
\begin{split}
& \omega_0^2 - \omega_D^2 = \\ & \tfrac{1}{2}\omega_0^2\left(\frac{A_{k+1}}{A_k} e^{\left[i(\phi_{k+1} - \phi_k)\right]} + \frac{A_{k-1}}{A_k} e^{\left[i(\phi_{k-1}-\phi_k)\right]}\right) = \\ & \tfrac{1}{2}\omega_0^2\left(\alpha_{k}e^{i\Delta\phi_{k}} + \frac{1}{\alpha_{k-1}} e^{-i\Delta\phi_{k-1}}\right)\,,
\end{split}
\label{eq:1_36}
\end{equation}
where we have defined the phase shift between consecutive bubbles $k$ and $k+1$ as $\Delta\phi_k \equiv \phi_{k+1} - \phi_k$ and the amplitude ratio as $\alpha_k \equiv A_{k+1}/A_k$. Because of translation invariance, position $k$ and $k-1$ are indistinguishable and therefore $\Delta\phi_{k} = \Delta\phi_{k-1} \equiv \Delta\phi$ and  $\alpha_{k} = \alpha_{k-1} \equiv \alpha$, with which Eq.~\eqref{eq:1_36} leads to 
\begin{equation}
\omega_0^2 - \omega_D^2 = \tfrac{1}{2}\omega_0^2\left(\alpha e^{i\Delta\phi} + \frac{1}{\alpha} e^{-i\Delta\phi}\right)\,.
\label{eq:1_37}
\end{equation}

Since the left hand side of Eq.~\eqref{eq:1_37} is real, the imaginary part of the right hand side must be zero
\begin{equation}
\left(\alpha  - \frac{1}{\alpha}\right)\sin(\Delta\phi) = 0 \; \Rightarrow \; \alpha = 1 \,\,\textrm{or}\,\,\sin(\Delta\phi) = 0\,,
\end{equation}
the latter of which leads to $\Delta\phi = n\pi$ for $n \in \mathbb{Z}$. It is easy to realize that the only sensible solution leading to a wave traveling in the positive $k$-direction is given by $\Delta\phi = \pi$. (Note that $\Delta\phi = -\pi$ travels towards negative $k$ and $\Delta\phi = 0$ does not travel at all.) 

Let us first examine the case that $\alpha = 1$. Inserting in Eq.~\eqref{eq:1_37} now leads to
\begin{equation}
\begin{split}
& \omega_0^2 - \omega_D^2 = \tfrac{1}{2}\omega_0^2\left(e^{i\Delta\phi} + e^{-i\Delta\phi}\right) = \omega_0^2 \cos(\Delta\phi) \\ \Rightarrow\qquad & \Delta\phi = \arccos(1-\tilde{\omega}^2)\,,
\end{split}
\end{equation}
where we have defined $\tilde{\omega} \equiv \omega_D/\omega_0$. The above equation only has a solution if the argument of the inverse cosine is between $-1$ and $1$, which leads to the condition that $\tilde{\omega} \leq \sqrt{2}$, where we note that $\tilde{\omega} = \sqrt{2}$ corresponds to $\Delta\phi = \pi $. 

The second case is $\Delta\phi = \pi$. Inserting in Eq.~\eqref{eq:1_37} now gives
\begin{equation}
\begin{split}
& \omega_0^2 - \omega_D^2 = -\tfrac{1}{2}\omega_0^2\left(\alpha + \frac{1}{\alpha}\right) \\ \Rightarrow\qquad & \alpha^2 + 2(1  - \tilde{\omega}^2)\alpha + 1 = 0\,.
\end{split}
\label{eq:1_40}
\end{equation}
This quadratic equation has a solution if $D = 4(1-\tilde{\omega}^2)^2 - 4 = 4\tilde{\omega}^2(\tilde{\omega}^2-2) > 0$, which is true when $\tilde{\omega}\geq \sqrt{2}$. Here, the insertion of $\tilde{\omega}= \sqrt{2}$ in Eq.~\eqref{eq:1_40} leads to the solution $\alpha = 1$, and the general solution is given by
\begin{equation}
\alpha = \tilde{\omega}^2 - 1 \pm \sqrt{\tilde{\omega}^2(\tilde{\omega}^2-2)}\,\,,
\end{equation}
where for a positive result one realizes that the negative root is the one to take.

Summarizing both cases, we obtain that
\begin{equation}
\alpha =
    \begin{cases}
      1 & \text{if $\tilde{\omega} \leq \sqrt{2}$}\\
      \tilde{\omega}^2\left(1 - \sqrt{1-2\tilde{\omega}^{-2}}\right) -1 & \text{if $\tilde{\omega} \geq \sqrt{2}$}
    \end{cases}\qquad ,
\label{eq:1_42}
\end{equation}
and
\begin{equation}
\Delta\phi =
    \begin{cases}
      \arccos(1-\tilde{\omega}^2) & \text{if $\tilde{\omega} \leq \sqrt{2}$}\\
      \pi & \text{if $\tilde{\omega} \geq \sqrt{2}$}
    \end{cases}\qquad ,
\label{eq:1_43}
\end{equation}
from which we learn that for $\tilde{\omega} \leq \sqrt{2}$ the signal is transmitted without attenuation ($\alpha = 1$) and with a frequency-dependent phase, whereas for $\tilde{\omega} \geq \sqrt{2}$ there is a strong attenuation of the signal and the phase shift becomes constant ($\Delta\phi = \pi$). 

Now we will relate the phase shift $\Delta\phi$ and attenuation factor $\alpha$ to quantities with a clearer physical interpretation, namely the wave speed $c$, wave length $\lambda$ and the attenuation length $\xi$. Starting with the wave speed, we realize that a phase shift $\Delta\phi$ corresponds to a time shift $\Delta t = \Delta\phi/\omega_D$. Since this time shift happens at a distance $\Delta x = \ell + \delta x_0$, we immediately obtain the wave speed as
\begin{equation}
c = \frac{\Delta x}{\Delta t} = \frac{(\ell + \delta x_0)\,\omega_D}{\Delta\phi} = \omega_0\,(\ell + \delta x_0)\frac{\tilde{\omega}}{\Delta\phi}\,.
\label{eq:1_44}
\end{equation}
In the right hand side, the last factor is dimensionless, and the first part is dimensional. Defining the typical wave speed $c_0$ as
\begin{equation}
c_0 \equiv \omega_0\,(\ell + \delta x_0) = \sqrt{\frac{2\gamma P_0}{\rho_l\nu(1-\nu)}}\,,
\end{equation}
where we have defined the gas fraction in the system $\nu = \delta x_0/(\ell + \delta x_0)$, this leads to the following expression for the dimensionless wave speed $\tilde{c} \equiv c/c_0$
\begin{equation}
\tilde{c} =
    \begin{cases}
      \frac{\tilde{\omega}}{\arccos(1-\tilde{\omega}^2)} & \text{if $\tilde{\omega} \leq \sqrt{2}$}\\
      \frac{\tilde{\omega}}{\pi} & \text{if $\tilde{\omega} \geq \sqrt{2}$}
    \end{cases}\qquad .
\label{eq:1_46}
\end{equation}
The wave length $\lambda = 2\pi/k$ can now be computed from the identity $c = \omega/k$ as 
\begin{equation}
\tilde{\lambda} =
    \begin{cases}
      \frac{1}{\arccos(1-\tilde{\omega}^2)} & \text{if $\tilde{\omega} \leq \sqrt{2}$}\\
      \dfrac{1}{\pi} & \text{if $\tilde{\omega} \geq \sqrt{2}$}
    \end{cases}\qquad , \label{eq:1_46_lambda}
\end{equation}
where we have defined $\tilde{\lambda} \equiv \lambda/\lambda_{0} = \tilde{c}/\tilde{\omega}$ with $\lambda_0 \equiv 2 \pi c_0/\omega_0$. Finally, in the attenuated regime ($\tilde{\omega}\geq \sqrt{2}$) we can compute a typical penetration depth $\xi$ by noting that the attenuation must be exponential in a translationally invariant system:
\begin{equation}
\alpha = \exp\left[-\frac{\Delta x}{\xi}\right] \,, \label{eq:1_48}
\end{equation}
which when equating with expression Eq.~\eqref{eq:1_42} leads to
\begin{equation}
\tilde{\xi} =
    \begin{cases}
      \infty & \text{if $\tilde{\omega} \leq \sqrt{2}$}\\
     - \left[\log\left(\tilde{\omega}^2(1-\sqrt{1-2\tilde{\omega}^{-2}})-1 \right)\right]^{-1} & \text{if $\tilde{\omega} \geq \sqrt{2}$}
    \end{cases}\qquad ,
\end{equation}
defining $\tilde{\xi} = \xi / \Delta x$.

The expression given by Eq.~\eqref{eq:1_46} corresponds to the rate at which the phase of a wave propagates, i.e., its phase velocity. In order to find an equation for the group velocity, one has to remember the definition
\begin{equation}
c_g = \frac{\partial \omega}{\partial k} \,,
\end{equation}
where $k$ is the wave number. We can rewrite this in terms of the wave length
\begin{equation}
c_g = - \frac{2 \pi}{\lambda^2}
\left( \frac{\partial \lambda}{\partial \omega} \right)^{-1} \,,
\end{equation}
which after carrying out the derivative making use of Eq.~\eqref{eq:1_46_lambda}, and noticing that frequencies in the attenuated regime ($\tilde{\omega} \geq \sqrt{2})$ do not last long, we arrive to the expression for the group velocity
\begin{equation}
c_g = \frac{c_0}{2}\sqrt{ 2 - \left( \frac{\omega}{\omega_0} \right)^2 }, \label{eq:1_groupvelocity}
\end{equation}
for any angular frequency $\omega$. 

It is important to notice that in a dispersive medium, as it is the case of an infinite chain of identical bubbles, an impulsive wave may initially contain many frequencies, a wave package. However, frequencies in the attenuated regime will almost immediately die out and frequencies in the unattenuated regime will travel at different speeds, thus distorting the wave package as it propagates.

\section{Computation of the eigenfrequencies of the monodisperse MBP}\label{1_A_2}

As described in the main text, the peaks in the response arise at the resonant frequencies of the system. In order to predict their exact location we can start thinking of the Multiple Bagnold Problem model as spring-mass system, where each liquid patch corresponds to a node with equivalent mass $m = \rho_{l} \ell A$, with $A$ an arbitrary cross section of the liquid patch. The equivalent rest length of the springs will be $\ell+\delta x_{0}$ with spring constant $k = \gamma P_{0} A/\delta x_{0}$. In this equivalent spring-mass system, the displacement of the $i$-th liquid patch $\epsilon_{i}$ corresponds to the motion of the $i$-th node. Since there are $N$ liquid patches, there are $N$ nodes respectively. Taking into account that the first liquid patch is driven and stuck to the piston, we will consider systems where $N \geq 2$ to have at least one degree of freedom. We may now look for any of the eigenmodes for which resonance occurs.

Next we will assume a wave speed $c(\omega)$ (although the $\omega$ dependence of the sound speed was already shown in Eq.~\eqref{eq:1_46} for an infinite system), where using well-known relations we have
\begin{equation}
c(\omega) = \dfrac{\omega}{k} = \dfrac{\omega \lambda}{2 \pi} . \label{eq:1_50}
\end{equation}
The equivalent mass-spring system has a length
\begin{equation}
L = N (\ell + \delta x_{0}). \label{eq:1_51}
\end{equation}

Resonance implies a much larger oscillation amplitude than the driving node $\epsilon_{1} \ll \epsilon_{i}$ for $i \geq 2$. Resonance occurs for standing waves where both endpoints are nodes, for which
\begin{equation}
L = n \dfrac{1}{2} \lambda , \label{eq:1_52}
\end{equation}
with $n \in \mathbb{N}$ the eigenmode number. At the same time, it is not possible to have structures smaller than twice the distance between nodes
\begin{equation}
\dfrac{1}{2} \lambda > (\ell + \delta x_{0}) . \label{eq:1_53}
\end{equation}
We substitute $\lambda$ from Eq.~\eqref{eq:1_52} into Eq.~\eqref{eq:1_50} and solve for $\omega$ to obtain
\begin{equation}
\omega_{n} = \dfrac{n \pi}{N} \dfrac{c}{(\ell + \delta x_{0})} , \label{eq:1_54}
\end{equation}
where we have renamed $\omega = \omega_{n}$ to explicitly point out that the angular frequency of the eigenmodes depends on the eigenmode number $n$. Substituting the expression for the wave speed Eq.~\eqref{eq:1_44} into this last equation and defining $\tilde{\omega}_{n} = \omega_{n} / \omega_{0}$ we obtain
\begin{equation}
\tilde{\omega}_{n} = \dfrac{n \pi}{N} \dfrac{\tilde{\omega}}{\Delta \phi} . \label{eq:1_55}
\end{equation}
Resonance occurs when $\tilde{\omega}_{n} = \tilde{\omega}$ for $\tilde{\omega} \leq \sqrt{2}$ since for $\tilde{\omega} > \sqrt{2}$ the transmitted signal is attenuated. Then using the expression for $\Delta \phi$ (Eq.~\eqref{eq:1_43}) and solving for $\tilde{\omega}_{n}$ we arrive at
\begin{equation}
\tilde{\omega}_{n} = \sqrt{1- \cos\left( \dfrac{n \pi}{N} \right)} , \label{eq:1_56}
\end{equation}
which predicts the angular eigenfrequencies of the system. Notice that combining the condition \ref{eq:1_53} with Eqs. \ref{eq:1_51} and \ref{eq:1_52} we obtain that $n < N$, i.e., $n = 1, ... , N-1$, the number of eigenmodes (peaks in the response curves) is less than the number of liquid patches (bubbles), as it shown in the response curves in the main text.

Translating $\tilde{\omega}_{n}$ to a frequency we obtain an expression that tells us exactly where the peaks will arise in the response curves
\begin{equation}
f_{n} = \dfrac{\omega_{0}}{2 \pi} \sqrt{1- \cos\left( \dfrac{n \pi}{N} \right)} . \label{eq:1_57}
\end{equation}

This first approach has the disadvantage that it has to rely on a result obtained for an infinite bubble system, namely the sound speed $c$. However, it is possible to arrive to the same expression in a more general and elegant (although slightly more complex) way.

From the linearized equations of the MBP we know that
\begin{equation}
\begin{array}{ll}
\ddot{\epsilon}_{2} +  \omega_{0}^{2} \epsilon_{2} &=
\tfrac{1}{2} \omega_{0}^{2} \epsilon_{3} \,\,\,\left[+\,\, \tfrac{1}{2} \omega_{0}^{2} \epsilon_{1}\right], \\
&\vdots \\ 
\ddot{\epsilon}_{i} + \omega_{0}^{2} \epsilon_{i} &=
\tfrac{1}{2} \omega_{0}^{2} \left( \epsilon_{i-1} + \epsilon_{i+1}
\right) , \\
&\vdots \\ 
\ddot{\epsilon}_{N} + \omega_{0}^{2} \epsilon_{N} &=
\tfrac{1}{2} \omega_{0}^{2} \epsilon_{N-1} ,
\end{array} \label{eq:1_58}
\end{equation}
where the last term of the first equation is put between square brackets considering that the equation for the first node is driven with a small amplitude such that $\epsilon_{1} \ll \epsilon_{i}$ for $i \geq 2$, as it should be for the resonance case, and may consequently be neglected. Next, we will assume a solution of the form
\begin{equation}
\epsilon_{i}(t) = \beta_{i} e^{i \omega_{n} t} , \label{eq:1_59}
\end{equation}
where $\beta_{i}$ are amplitudes. After substitution and rearranging we arrive to the set of equations
\begin{equation}
\begin{alignedat}{3}
2 \beta_{2} &- \beta_{3} &&=
2 \dfrac{\omega_{n}^{2}}{\omega_{0}^{2}}  \beta_{2}&& , \\ 
&\vdots \\ 
- \beta_{i-1} &+ 2 \beta_{i} - \beta_{i+1} &&=
2 \dfrac{\omega_{n}^{2}}{\omega_{0}^{2}}  \beta_{i}&& , \\
&\vdots \\ 
- \beta_{N-1} &+ 2 \beta_{N} &&=
2 \dfrac{\omega_{n}^{2}}{\omega_{0}^{2}}  \beta_{N}&& ,
\end{alignedat} \label{eq:1_60}
\end{equation}
which can be written in the compact form
\begin{equation}
\bm{A}\cdot \bm{\beta} = \lambda_{n} \bm{\beta} , \label{eq:1_61}
\end{equation}
where we have defined
\begin{equation}
\lambda_{n} = 2 \dfrac{\omega_{n}^{2}}{\omega_{0}^{2}} \label{eq:1_62}
\end{equation}
and the $(N-1)\times(N-1)$ matrix
\begin{equation}
\bm{A} = \begin{pmatrix}
    2  &   -1    &     0     &           &         \\
   -1  &    2    &    -1     &           &         \\
       & \ddots  &  \ddots   & \ddots    &         \\
       &         &    -1     &     2     &  -1     \\
       &         &     0     &    -1     &   2
  \end{pmatrix} . \label{eq:1_63}
\end{equation}

Equation \eqref{eq:1_61} is an eigenvalue problem for the matrix $\bm{A}$ with
\begin{equation}
\bm{A} = - \text{tridiag}(1,-2,1)  \label{eq:1_64}
\end{equation}
a well-studied tridiagonal matrix that appears often in numerical analysis, see for instance Guardiola \& Ros \cite{guardiola1982numerical}. It has the properties that the eigenvalues are given by
\begin{equation}
\lambda_{n} = 4 \sin^{2}\left( \dfrac{n \pi}{2 N} \right) \label{eq:1_65} 
\end{equation}
and the eigenvectors are the columns of the matrix
\begin{equation}
R_{n m} = \sqrt{\dfrac{2}{N}} \sin\left( \dfrac{n m \pi}{N} \right) \label{eq:1_66}
\end{equation}
for $n,m = 1, 2, ... , N-1$. Equating Eqs. \ref{eq:1_62} and \ref{eq:1_65} we arrive to
\begin{equation}
\omega_{n} = \sqrt{2} \omega_{0} \sin\left( \dfrac{n \pi}{2 N} \right) \label{eq:1_67}
\end{equation}
or, equivalently,
\begin{equation}
\tilde{\omega}_{n} = \sqrt{2} \sin\left( \dfrac{n \pi}{2 N} \right) . \label{eq:1_68}
\end{equation}
Remembering the trigonometric identity $\cos(x) = 1 - 2 \sin^{2} (x/2)$ we realize that Eq.~\eqref{eq:1_68} is exactly the same as Eq.~\eqref{eq:1_56}, proving that both approaches are equivalent.

\section{Computation of the frequency response of the monodisperse MBP}\label{1_A_3nw}

Incidentally, Eq.~\eqref{eq:1_58}, including the driving term, can also be used to compute the frequency response of the monodisperse system, noting that in matrix notation it can be written as
\begin{equation}
\ddot{\bm{\epsilon}} + \tfrac{1}{2}\omega_0^2\bm{A}\cdot\bm{\epsilon} = \epsilon_1(t)\bm{e}_1\,,\label{eq:evmatrix}
\end{equation}
where $\bm{A}$ is the $(N-1)\times (N-1)$-matrix defined in Eq.~\eqref{eq:1_63} and introducing the $(N-1)$-vectors $\bm{\epsilon} = (\epsilon_2,...,\epsilon_N)$ and $\bm{e}_1 = (1,0,...,0)$. Writing the driving term $\epsilon_1(t)$ as
\begin{equation}
\epsilon_1(t) = A e^{i\omega_Dt}\,,
\end{equation}
and assuming a response in the system of the same frequency $\omega_D$, i.e., $\bm{\epsilon} = A\bm{\beta}\exp[i\omega_Dt]$, we obtain after insertion in Eq.~\eqref{eq:evmatrix} that
\begin{equation}
\left[-\omega_D^2 \bm{1} + \tfrac{1}{2}\omega_0^2\bm{A}\right]\cdot\bm{\beta} = \bm{e}_1\,,\label{eq:evmatrixfourier}
\end{equation}
where $\bm{1}$ is the unit matrix. Note that we have included the amplitude $A$ in the definition to render the $\bm{\beta}$ dimensionless. Now, since $\bm{A}$ is a non-singular, symmetric matrix, there exists an orthonormal basis of eigenvectors $\hat{\bm{a}}_n$, which is formed by the columns of the matrix $\bm{R}$ defined in Eq.~\eqref{eq:1_66}. This implies that we may expand the vector $\bm{\beta}$ in this basis as $\bm{\beta} = \sum_{m}\beta_m \hat{\bm{a}}_m$ in Eq.~\eqref{eq:evmatrixfourier} , and since $\omega_0^2\bm{A}\cdot\hat{\bm{a}}_m = 2\omega_m^2\hat{\bm{a}}_m$ we obtain  
\begin{equation}
\sum_{m=1}^{N-1} (\omega_m^2-\omega_D^2) \beta_m \hat{\bm{a}}_m = \bm{e}_1\,.\label{eq:evmatrixbasis}
\end{equation}
By taking the dot product with eigenvector $\hat{\bm{a}}_n$ and using the orthonormality condition $\hat{\bm{a}}_m\cdot\hat{\bm{a}}_n = \delta_{mn}$ we obtain the coefficients
\begin{equation}
\beta_n = \frac{(\hat{\bm{a}}_n)_1}{(\omega_n^2-\omega_D^2)} = \sqrt{\frac{2}{N}}\frac{\sin(n\pi/N)}{(\omega_n^2-\omega_D^2)}\,,\label{eq:beta_n}
\end{equation}
where $(\hat{\bm{a}}_n)_1 = \bm{e}_1\cdot\hat{\bm{a}}_n = \sqrt{2/N}\sin(n\pi/N)$ is the first element of the vector $\hat{\bm{a}}_n$.
With this we have found the frequency response $R_\epsilon$ of the system as
\begin{equation}
R_\epsilon(\omega_D) = \bm{\beta}\cdot\bm{\beta} = \sum_{n=1}^{N-1} \frac{2\sin^2(n\pi/N)}{N(\omega_n^2-\omega_D^2)^2}\,.\label{eq:Repsilon}
\end{equation}
Using $\sin(x) = 2\sin(x/2)\cos(x/2)$ the sine term in the nominator may also be written in terms of the eigenfrequencies $\omega_n$ leading to
\begin{equation}
R_\epsilon(\omega_D) =  \sum_{n=1}^{N-1} \frac{\tilde{\omega}_n^2(2-\tilde{\omega}_n^2)}{2N(\tilde{\omega}_n^2-\tilde{\omega}_D^2)^2}\,.\label{eq:Repsilon}
\end{equation}
Unfortunately the above expression is not directly comparable to the frequency responses reported in this article, since they have been computed using the change of bubble volume $\Delta_k$ as an input, which are related to the $\epsilon_k$ by Eq.~\eqref{eq:1_14}, leading to $\delta x_0 \Delta_k = \epsilon_k - \epsilon_{k+1}$ (except for the last value for which $\delta x_0 \Delta_N = \epsilon_N$). This may be written as a matrix equation as well
\begin{equation}
\delta x_0 \bm{\Delta} = \bm{F}\cdot\bm{\epsilon}\,, \label{eq:matrixF}
\end{equation}
where the $(N-1)$-vector $\bm{\Delta}$ is defined as $\bm{\Delta} = (\Delta_2,\Delta_3,...,\Delta_N)$ and the $(N-1) \times (N-1)$-matrix $\bm{F}$ is minus the first forward difference matrix:
\begin{equation}
\bm{F} = \begin{pmatrix}
    1  &   -1    &     0     &           &         \\
       &    1    &    -1     &           &         \\
       & \ddots  &  \ddots   & \ddots    &         \\
       &         &     0     &     1     &  -1     \\
       &         &     0     &     0     &   1
  \end{pmatrix} . \label{eq:FFDmatrix}
\end{equation}
This matrix is non-singular and its inverse $\bm{F}^{-1}$ may easily be checked to be the upper triangular matrix where the elements of both the diagonal and the upper triangle are equal to 1, whereas the lower triangle is filled with 0. Since one may wonder about $\Delta_1$ not being included in $\bm{\Delta}$ and the subsequent calculation, it is good to note that 
\begin{equation}
\sum_{k=1}^N \Delta_k = \frac{\epsilon_1(t)}{\delta x_0}\label{eq:Deltaidentity}
\end{equation}
which implies that, once $\bm{\Delta}$ is known, we may compute $\Delta_1$ directly from the above identity, corroborating that also after transformation to the $N$ bubble sizes, the dimensionality of our problem does not change and remains $N-1$. Multiplying the matrix equation \eqref{eq:evmatrix} from the left with $\bm{F}$ directly leads to
\begin{equation}
\ddot{\bm{\Delta}} + \tfrac{1}{2}\omega_0^2(\bm{F}\cdot\bm{A}\cdot\bm{F}^{-1}))\cdot\bm{\Delta} = \epsilon_1(t)(\bm{F}\cdot\bm{e}_1)\,,\label{eq:Deltaevmatrix}
\end{equation}
from which we immediately conclude that the eigenfrequencies of the system remain the same, since the eigenvalues of the matrix $\bm{D} \equiv \bm{F}\cdot\bm{A}\cdot\bm{F}^{-1}$ are identical to those of $\bm{A}$. However, since it is easy to check that the eigenvectors $\bm{F}\cdot\hat{\bm{a}}_n$ of the (non-symmetrical) matrix $\bm{D}$ are not orthogonal, there is no straightforward manner to directly obtain the mode amplitudes of $\bm{\Delta}$ in a similar approach as was done for $\bm{\epsilon}$. We will therefore compute the amplitudes from those obtained for $\beta$ (cf. Eq.~\eqref{eq:beta_n}) . Letting $\Delta_k(t) = A\delta x_0 d_k \exp[i\omega_Dt]$ leads to the relation $(\delta x_0)\bm{d} = \bm{F}\cdot\bm{\beta}$ between the coefficients $\bm{d} = (d_2,d_3,...,d_N)$ and $\bm{\beta}$. Here, again, we include the prefactor $A\delta x_0$ in the definition to make the $d_k$ dimensionless.
Now, in the end what we want to obtain is
\begin{equation}
\bm{d}^T\!\!\cdot\!\bm{d} = (\bm{F}\!\cdot\!\bm{\beta})^T\!\!\cdot\!\bm{F}\!\cdot\!\bm{\beta} =  \bm{\beta}^T\!\!\cdot\!(\bm{F}^T\!\cdot\!\bm{F}\!\cdot\!\bm{\beta}) \,,\label{eq:d_squared}
\end{equation}
where $(...)^T$ stands for the transposed vector cq. matrix. Now, incidentally, the matrix $\bm{F}^T\!\!\cdot\!\bm{F}$ is almost equal to $\bm{A}$. More precisely, $\bm{F}^T\!\!\cdot\!\bm{F} = \bm{A} - \bm{E}_{11}$, where $\bm{E}_{11}$ is the matrix containing zeros everywhere except for one 1 on position in the upper left corner. Clearly, $\bm{E}_{11}$ is symmetric $\bm{E}_{11} = \bm{E}_{11}^T$, and $\bm{E}_{11}\!\cdot\!\bm{E}_{11} = \bm{E}_{11}$. Using these identities we may write  
\begin{equation}
\bm{d}^T\!\!\cdot\!\bm{d} = \bm{\beta}^T\!\!\cdot\!\bm{A}\!\cdot\!\bm{\beta} - (\bm{E}_{11}\!\cdot\!\bm{\beta})^T\!\!\cdot\!(\bm{E}_{11}\!\cdot\!\bm{\beta}) \,,\label{eq:d_squared}
\end{equation}
where the last term was written realizing that $\bm{\beta}^T\!\!\cdot\!\bm{E}_{11}\!\cdot\!\bm{\beta} = \bm{\beta}^T\!\!\cdot\!\bm{E}_{11}\!\cdot\!\bm{E}_{11}\!\cdot\!\bm{\beta} = (\bm{E}_{11}\!\cdot\!\bm{\beta})^T\!\!\cdot\!\bm{E}_{11}\!\cdot\!\bm{\beta}$. Now, expressing $\bm{\beta}$ in the orthogonal basis $\hat{\bm{a}}_n$ (i.e., $\bm{\beta} = \sum_n \beta_n\hat{\bm{a}}_n$) we obtain for the first term that
\begin{eqnarray}
\bm{\beta}^T\!\!\cdot\!\bm{A}\!\cdot\!\bm{\beta} &=& \sum_{m=1}^{N-1}\sum_{n=1}^{N-1} \beta_m\beta_n  \hat{\bm{a}}_m^T\!\!\cdot\!(\bm{A}\!\cdot\!\hat{\bm{a}}_m) \nonumber\\
&=& \sum_{n=1}^{N-1}2\tilde{\omega}_n^2\beta_n^2
  \,,\label{eq:firstterm}
\end{eqnarray}
and for the second term
\begin{eqnarray}
(\bm{E}_{11}\!\cdot\!\bm{\beta})^T\!\!\cdot\!(\bm{E}_{11}\!\cdot\!\bm{\beta}) &=& \sum_{m=1}^{N-1}\sum_{n=1}^{N-1} \beta_m\beta_n (\hat{\bm{a}}_m)_1(\hat{\bm{a}}_n)_1 \nonumber\\
&=& \left[\sum_{n=1}^{N-1}\beta_n(\hat{\bm{a}}_n)_1\right]^2 \,,\label{eq:secondterm}
\end{eqnarray}
with which we have found the following expression for $\bm{d}^T\!\!\cdot\!\bm{d}$: 
\begin{equation}
\bm{d}^T\!\!\cdot\!\bm{d} = \sum_{n=1}^{N-1}2\tilde{\omega}_n^2\beta_n^2 - \left[\sum_{n=1}^{N-1}\beta_n(\hat{\bm{a}}_n)_1\right]^2  
\,,\label{eq:d_squaredworkedout}
\end{equation}
To obtain the frequency response $R_\Delta$ we are still missing the amplitude of the first bubble $d_1$ which we will obtain directly from the definition $\delta x_0 \Delta_1 = \epsilon_1 - \epsilon_2$, leading to $d_1 = 1 - \bar{\beta}_1$. The bar in $\bar{\beta}_1$ is there to stress that we need to have the first component of $\bm{\beta}$ in the original basis instead of that of the basis of eigenvectors computed in Eq.~\eqref{eq:beta_n}, which are related by $\bar{\bm{\beta}} = \bm{R}^T\!\!\cdot\!\bm{\beta} = \bm{R}\!\cdot\!\bm{\beta}$, given the symmetry of $\bm{R}$. This then leads to
\begin{equation}
d_1 = 1-\bar{\beta}_1 = 1 -  \sum_{n=1}^{N-1} \beta_n (\hat{\bm{a}}_n)_1
  \,,\label{eq:d1}
\end{equation}
from which we directly obtain that
\begin{equation}
d_1^2 =  1 -  2\sum_{n=1}^{N-1} \beta_n (\hat{\bm{a}}_n)_1 + \left[\sum_{n=1}^{N-1}\beta_n(\hat{\bm{a}}_n)_1\right]^2
  \,.\label{eq:d1squared}
\end{equation}
To obtain $R_\Delta$ we now need to add $d_1^2$ and $\bm{d}^T\!\!\cdot\!\bm{d}$, i.e., we need to sum Eqs.~\eqref{eq:d_squaredworkedout} and~\eqref{eq:d1squared} in which we note that the second term of $\bm{d}^T\!\!\cdot\!\bm{d}$ is cancelled by the last term in $d_1^2$, leading to  
\begin{eqnarray}
R &=& d_1^2 + \bm{d}^T\!\!\cdot\!\bm{d}  \nonumber\\
&=& 1 + \sum_{n=1}^{N-1}2\tilde{\omega}_n^2\beta_n^2 - 2\sum_{n=1}^{N-1} \beta_n (\hat{\bm{a}}_n)_1
\,.\label{eq:RDeltasemifinal}
\end{eqnarray}
Using that $\beta_n = (\hat{\bm{a}}_n)_1/(\tilde{\omega}_n^2 - \tilde{\omega}_D^2)$ (Eq.~\eqref{eq:beta_n}) and $\sqrt{N/2}(\hat{\bm{a}}_n)_1 = \sin(n\pi/N) = \tilde{\omega}_n\sqrt{2-\tilde{\omega}_n^2}$ we thus find after some algebraic manipulation that
\begin{equation}
R = 1 + \frac{4}{N}\sum_{n=1}^{N-1}\frac{\tilde{\omega}_D^2\tilde{\omega}_n^2(2-\tilde{\omega}_n^2)}{(\tilde{\omega}_n^2-\tilde{\omega}_D^2)^2}
  \,,\label{eq:RDeltafinal}
\end{equation}

From Eq.~\eqref{eq:RDeltafinal} it is clear that, as the number of bubbles $N$ increases, the term containing the sum dominates over the 
the constant one (equal to one). Thus, it is no surprise that for small driving frequencies $\tilde{\omega}_D$ (staying away 
from the eigenfrequencies $\tilde{\omega}_n$ where the response diverges), 
the total response $R$ presents 
an intermediate region where the response converges to 
a power law of exponent two, following $\tilde{\omega}_D^2$ in the nominator of the sum.

\begin{figure}
\includegraphics[width=\columnwidth]{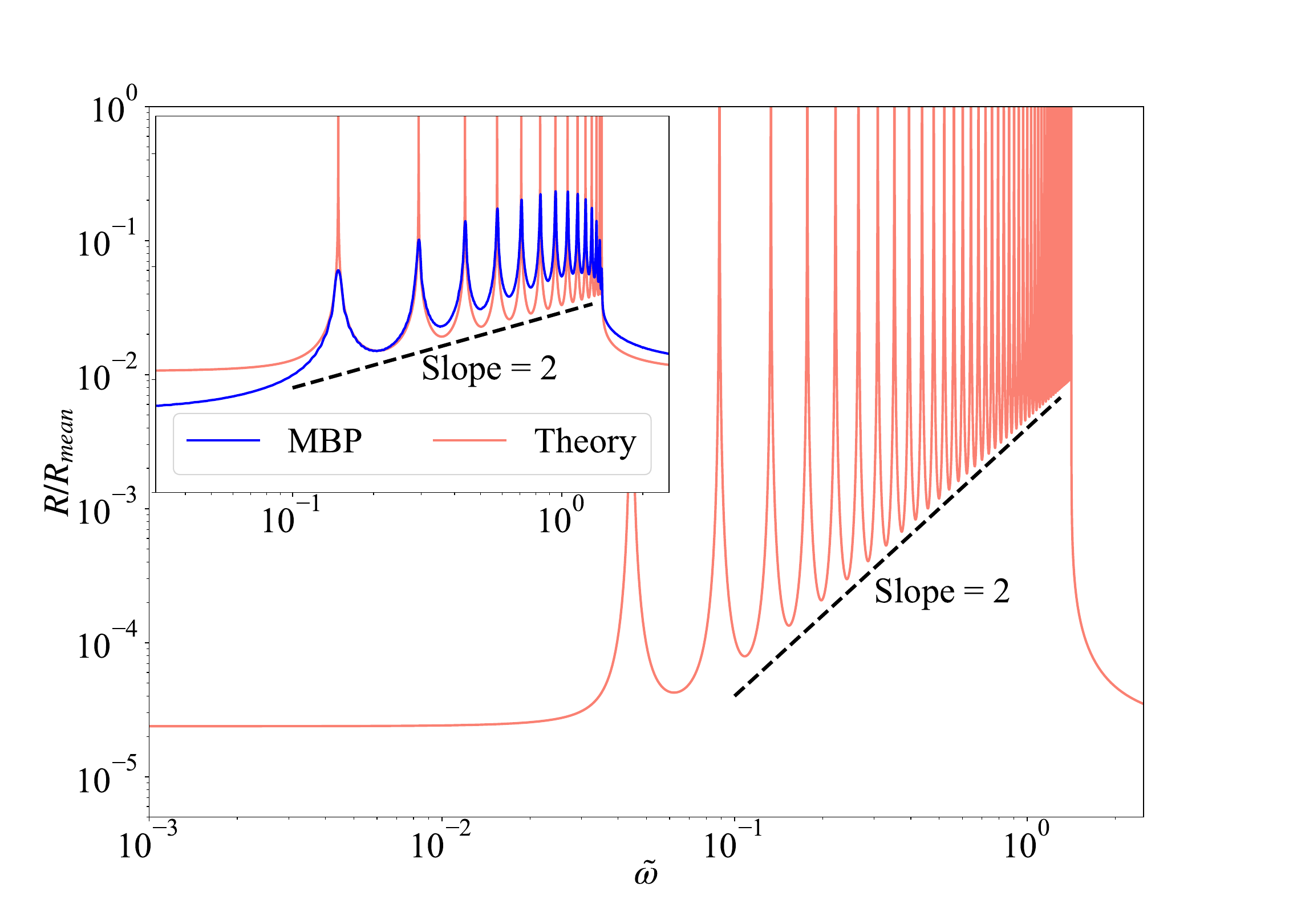}
\caption{Frequency response $R$ for a harmonically driven monodisperse system of $50$ bubbles with resonant frequency 1.6 Hz, normalized by an arbitrary value 
for visualization purposes. In the inset we see a comparison between the frequency response $R$ of the MBP model (blue curve), obtained by numerically solving 
Eq.~\eqref{eq:1_18}, and the theoretical result (red curve) obtained with Eq.~\eqref{eq:RDeltafinal}, for a harmonically driven monodisperse system of $15$ bubbles with resonant frequency 1.6 H. The power law with exponent two is plotted in a dashed black line for both plots respectively.
}\label{fig:1_fig_RvsOmtilde_appendix}
\end{figure}

In Fig. \ref{fig:1_fig_RvsOmtilde_appendix} we plot the response $R$ (red curve) computed from Eq.~\eqref{eq:RDeltafinal} for a harmonically driven monodisperse system of 50 bubbles (normalized by an arbitrary value for better visualization). Each peak corresponds to one of the multiple eigenfrequencies as predicted by Eq.~\eqref{eq:1_68}. More importantly, we show that the bottom part of the peaks approximately follows the above mentioned power law with exponent two (dashed black line). In the inset, we plot a comparison between the response $R$ obtained from Eq.~\eqref{eq:1_18} (blue curve) and that obtained from Eq.~\eqref{eq:RDeltafinal} (red curve), for a harmonically driven monodisperse system of 15 bubbles. Although both curves present peaks at the same eigenfrequencies, as expected by construction, the difference in normalization constants and the resolution of the blue curve do not allow them to overlap exactly. However, the approximate slope two at the bottom of the peaks is present for both curves.

\section{Computation of the MBP eigenfrequencies for bidisperse alternating systems}\label{1_A_3}

The alternating bidisperse system resembles a well known problem in the classical theory of crystals, when one considers a one dimensional Bravais lattice with two ions, see for instance Ashcroft \& Mermin \cite{ashcroft1976solid}.

In this case let us assume a long chain of bubbles of two sizes alternating in position, where we will call one (the smallest) size the \textit{small} and the other the \textit{big} size. In order to avoid confusing notation, let us call $u_n$ a small perturbation on the equilibrium position of the $n$th big bubble (and liquid patch) and $v_n$ for the adjacent small bubble (liquid patch). We will call $\omega_b$ and $\omega_s$ the natural frequencies of the big and small bubble respectively given by Eq.~\eqref{eq:1_8} but with initial sizes $\delta x_b(0)$ and $\delta x_s(0)$. Then, the equations of motion for the position $n$ do not take the form of Eq.~\eqref{eq:1_7} but rather
\begin{equation}
\begin{array}{ll}
2 \ddot{u}_{n} &=
- \omega_{b}^{2} ( u_n - v_n )
- \omega_{s}^{2} ( u_n - v_{n-1} ) , \\
\\
2 \ddot{v}_{n} &=
- \omega_{b}^{2} ( v_n - u_n )
- \omega_{s}^{2} ( v_n - u_{n+1} ) .
\end{array}
\end{equation}
Now, we will look for a solution of the form
\begin{equation}
u_n(t) = \alpha_n e^{i(k a - \omega_n t)}
\; , \;
v_n(t) = \beta_n e^{i(k a - \omega_n t)} ,
\end{equation}
where $\beta_n$ and $\alpha_n$ are just amplitudes, $a$ is equal to $2l + \delta x_s(0) + \delta x_b(0)$ and $k = 2 \pi n / a N$. This is naturally motivated by our previous analysis in monodisperse systems. After substitution we obtain the system
\begin{equation}
\begin{array}{ll}
\alpha_n ( 2 \omega_{n}^{2} - \omega_{b}^{2} - \omega_{s}^{2} ) +
\beta_n ( \omega_{b}^{2} e^{- i k a} + \omega_{s}^{2} )
&= 0 , \\
\\
\alpha_n ( \omega_{b}^{2} + \omega_{s}^{2} e^{i k a} ) +
\beta_n ( 2 \omega_{n}^{2} - \omega_{b}^{2} - \omega_{s}^{2} ) 
&= 0 .
\end{array}
\end{equation}
The pair of equations for each $n$ will have a non-trivial solution provided that the determinant of the amplitude coefficients vanishes. After expanding the determinant and solving for $\omega_n$ we arrive at an expression for the normal angular frequencies of oscillations for $n = 1, ... , N/2 -1$
\begin{equation}
\begin{array}{ll}
\omega_n &= 2\pi\, f_n\\
&= \sqrt{
\dfrac{\omega_{s}^{2} + \omega_{b}^{2} }{2} \pm \dfrac{1}{2}
\sqrt{ \omega_{s}^{4} + \omega_{b}^{4} +
2 \omega_{s}^{2} \omega_{b}^{2}
\cos\left( \dfrac{n \pi}{N} \right) } } .
\end{array}
\end{equation}
The two options indicated by the $\pm$-sign correspond to two possible sets of solutions or two \textit{branches}, as they call it in solid state physics. One is the optic branch ($+$) and the other the acoustic branch ($-$). In our case, since we are working with bubbles, the optic branch corresponds to two consecutive (one small and one big) bubbles vibrating in anti-phase and the acoustic branch to consecutive (one small and one big) bubbles vibrating in phase. 

\section{Computation of the MBP eigenfrequencies for bidisperse stacked systems}\label{1_A_4}

Once more we write the linearized equations of the MBP, taking into account that half the bubbles are small and the other half big
\begin{equation}
\renewcommand{\arraystretch}{2.1}
\begin{array}{rl}
\ddot{\epsilon}_{2} +  \omega_{s}^{2} \epsilon_{2} &=
\tfrac{1}{2} \; \omega_{s}^{2} \epsilon_{3} , \\
\ddot{\epsilon}_{3} + \omega_{s}^{2} \epsilon_{3} &=
\tfrac{1}{2} \; \omega_{s}^{2} \left( \epsilon_{2} + \epsilon_{4}
\right) , \\
&\vdots \\ 
\ddot{\epsilon}_{N/2} + \tfrac{1}{2} \left( \omega_{s}^{2} \epsilon_{N/2} 
+ \omega_{b}^{2} \epsilon_{N/2} \right) &=
\tfrac{1}{2} \left( \omega_{s}^{2} \epsilon_{N/2-1} + 
\omega_{b}^{2} \epsilon_{N/2+1} \right) , \\
&\vdots \\ 
\ddot{\epsilon}_{N-1} + \omega_{b}^{2} \epsilon_{N-1} &=
\tfrac{1}{2} \; \omega_{b}^{2} \left( \epsilon_{N-2} + \epsilon_{N}
\right) , \\
\ddot{\epsilon}_{N} + \omega_{b}^{2} \epsilon_{N} &=
\tfrac{1}{2} \; \omega_{b}^{2} \epsilon_{N-1} ,
\end{array}
\end{equation}
assuming that the equation for the first bubble is driven with a small amplitude such that $\epsilon_{1} \ll \epsilon_{i}$ for $i \geq 2$.  We assume as usual a solution of the form
\begin{equation}
\epsilon_{i}(t) = \beta_{i} e^{i \omega_{n} t} ,
\end{equation}
where $\beta_{i}$ are amplitudes. After substitution and rearranging in matrix form we arrive at
\begin{equation}
\bm{B}\cdot \bm{\beta} = \lambda_{n} \bm{\beta} ,
\end{equation}
where we have defined
\begin{equation}
\lambda_{n} = \omega_{n}^{2} 
\end{equation}
and the $(N-1)\times(N-1)$ matrix
\begin{equation}
\bm{B} = \begin{pmatrix}
    2a  &   -a     &     0     &           &          &          &         \\
   -a   &    2a    &    -a     &           &          &          &         \\
        &  \ddots  &  \ddots   &           &          &          &         \\
        &          &    -a     &   (a+b)   &    -b    &          &         \\
        &          &           &           &  \ddots  &  \ddots  &         \\       
        &          &           &           &    -b    &    2b    &   -b    \\
        &          &           &           &    0     &    -b    &   2b
  \end{pmatrix} ,
\end{equation}
with $a = \omega_{s}^2/2$ and $b = \omega_{b}^2/2$. It is easy to solve the eigenvalue problem of this matrix numerically using Matlab or Wolfram Mathematica to obtain its eigenvalues and therefore the eigenfrequencies of the system.

\section{Attenuated Transmission and Reflection Coefficients Plots}\label{1_A_5}

\begin{figure}
\begin{center}
\includegraphics[width=0.83\linewidth]{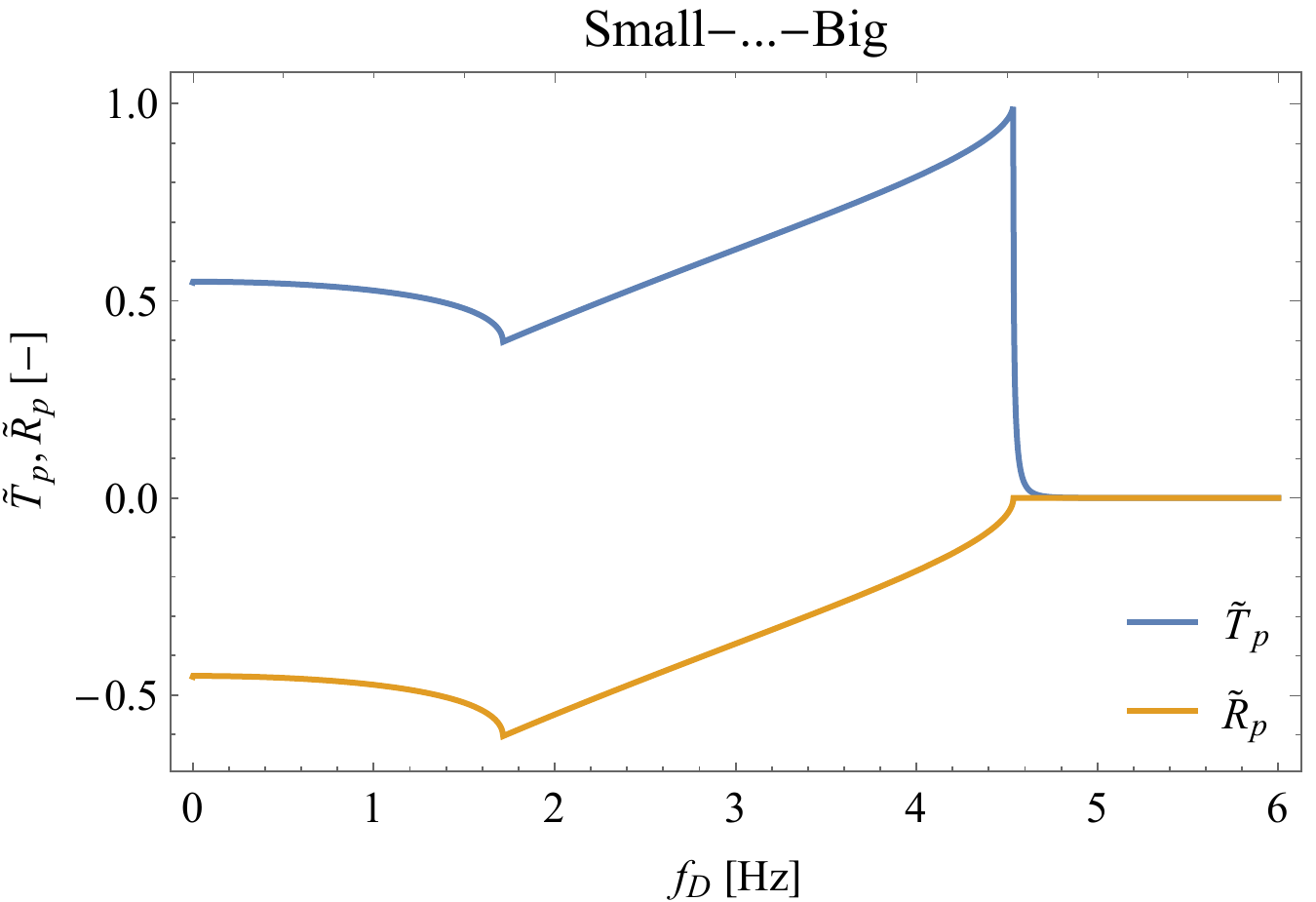}
\makebox[0.50\textwidth]{a)}
\includegraphics[width=0.83\linewidth]{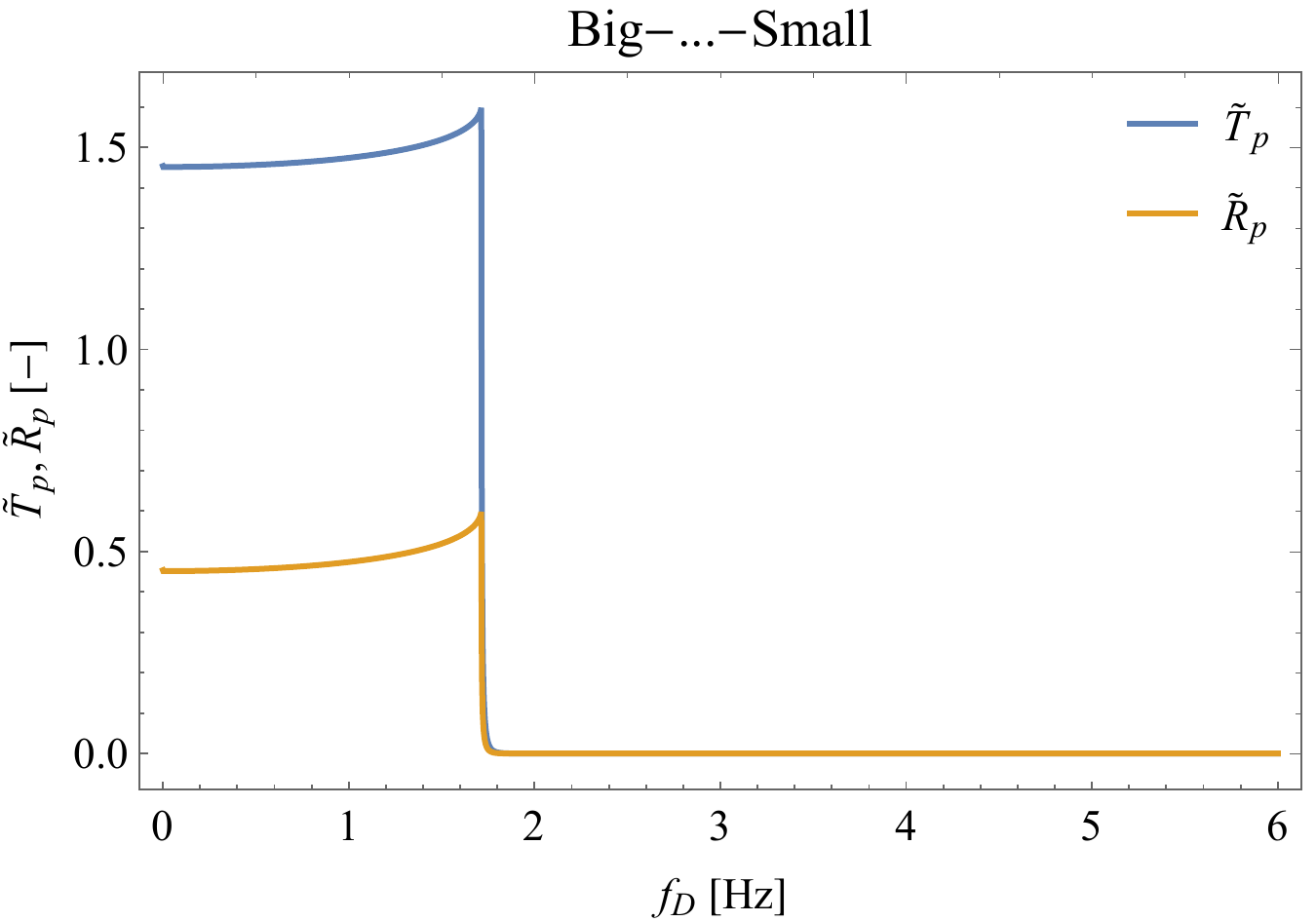}
\makebox[0.50\textwidth]{b)}
\end{center}
\label{1_A1}
\caption{Attenuated transmission and reflection coefficients for a) the small-...-big system and b) the big-...-small one. Both coefficients go to zero once $f_D$ reaches the attenuation regime of the first medium.}
\end{figure}

Writing Eq.~\eqref{eq:1_29} more explicitly
\begin{equation}
\renewcommand{\arraystretch}{2.1}
\begin{array}{rl}
\tilde{T}_p &= \dfrac{2 \omega_2 \tilde{c}_2}{\omega_1 \tilde{c}_1 + \omega_2\tilde{c}_2} e^{-N/\tilde{\xi}_1}
 , \\
\tilde{R}_p &= \dfrac{\omega_2 \tilde{c}_2 - \omega_1 \tilde{c}_1}{\omega_1 \tilde{c}_1 + \omega_2 \tilde{c}_2} e^{-N/\tilde{\xi}_1}
,
\end{array}
\end{equation}
we notice the dependencies of the attenuated coefficients on sound speed and natural bubble frequency. The bidisperse stacked cases described in the main text presents 10 bubbles of each size. Then, using the expressions found for bubble frequency, sound speed, penetration depth and $N = 10$ we can plot the coefficients as seen in Fig. \ref{1_A1}.

\end{document}